\documentclass[arxiv,usenatbib,iupartex-v23]{iupartex}

\usepackage{multicol}   
\usepackage{bm}		    
\usepackage{pdflscape}	
\usepackage{longtable}
\usepackage{booktabs}
\usepackage{soul}
\title[The ZP Constants for $Gaia$]{Spectroscopic Bolometric Corrections and Empirical Zero-point Constants of \textit{Gaia} Magnitudes, $G$, $G_{\rm BP}$, and $G_{\rm RP}$, from \textit{Gaia} XP Spectra}

\author[Eker et al.]{%
Z. Eker\,$^{1}$\orcid{0000-0003-1883-6255},
and
G. Y\"{u}cel\,$^{1\cc}$\orcid{0000-0002-9846-3788}
\affsep \\
$^1$Akdeniz University, Faculty of Science, Department of Space Sciences \& Technologies, 07058, Antalya, T\"{u}rkiye\\ 
}

\corres{G. Y\"{u}cel}{gokhannyucel@gmail.com}

\pubyear{2026}

\doiheader{XXXXXXX/PAR.20XX.00000}
\date{
	\pSubmit{00.00.0000}
	\pRevReq{00.00.0000}
	\pLastRevRec{00.00.0000}
	\pAccept{00.00.0000}
	\pPubOnl{00.00.0000}
}
\volume{4}
\volnumber{1}

\begin{document}

\defcitealias{Yucel2026}{Y26}

\maketitle
\begin{abstract}
The International Astronomical Union 2015 Resolution B2 (IAU2015GARB2) has ended the long-standing problem of zero-point constants of absolute and apparent bolometric magnitude scales and opened up a new window in fundamental astrophysics. The empirical zero-point constants of the bolometric corrections, $C_2(\xi)$, and the absolute/apparent magnitudes, $C_\xi/c_\xi$, for the $Gaia$ passbands were obtained from 88 $Gaia$ XP spectra, and absolute bolometric/filtered magnitudes. The individual  zero-point constants $\langle C_{\rm 2}\rangle$ of the bolometric corrections ($BC_\xi$) for each star  revealed weighted averages of  $\langle C_{\rm 2}(G)\rangle=0.8677\pm0.0109$ mag, $\langle C_{\rm 2}(G_{\rm BP})\rangle=1.0449\pm0.0116$ mag, and $\langle C_{\rm 2}(G_{\rm RP})\rangle=2.0510\pm0.0087$ mag. Furthermore, $C_{\rm Bol}=71.197425...$ mag and $c_{\rm Bol} =-18.997351...$mag announced by IAU2015GARB2, and using the definition of 
$C_{\rm 2}=C_{\rm Bol}-C_{\xi}=c_{\rm Bol}-c_{\xi}$, where the subscript $2$ indicate the wavelength ranges of two in which one is for bolometric and the other for one of the three filters, the zero-point constants of magnitudes for $Gaia$ filters as $C_{\rm G}=70.1525\pm0.0109$ mag and $c_{\rm G}=-19.8651\pm0.0105$ mag, $C_{\rm G_{\rm BP}}=70.1525\pm0.0116$ mag and $c_{\rm G_{\rm BP}}=-20.0423\pm0.0116$ mag, and $C_{\rm G_{\rm RP}}=69.1464\pm0.0087$ mag and $c_{\rm G_{\rm RP}}=-21.0484\pm0.0087$ mag, if $L_{\xi}$ and  $f_{\xi}$ are in SI units in case no extinctions. Lastly, spectroscopic $BC$s for $Gaia$ magnitudes of 88 stars and the spectroscopic $BC-T_{\rm eff}$ relation for each $Gaia$ filter are presented.
\end{abstract}



\section{Introduction}

Stars emit electromagnetic radiation at all wavelengths. Despite advances in technologies and theoretical knowledge, there is no telescope or detector to be able to collect all the photons from a star. Instead, collecting photons from a limited wavelength range, such as the visual band, is only for comparing the brightness of stars.  Existence of a different band besides the visual, such as the photographic or the blue magnitudes, had been noticed soon after usages of photographic plates in astronomical photometry \citep{Eker2025}, but bolometric magnitudes representing stellar brightness at all wavelengths first introduced by \cite{Eddington1926} in order to be used when expressing his famous mass$-$luminosity relation, which he was inspired it from empirically discovered mass$-$absolute visual magnitude relation discovered just few years ago independently by  \cite{Hertzsprung1923} and \cite{Russell1923}. Long after twelve years, \cite{Kuiper1938} gave the definition of bolometric correction for a star as the difference between its bolometric and visual magnitudes as

\begin{equation}
    BC = M_{\rm Bol} - M_{\rm V} = m_{\rm Bol}-V,
    \label{eq:1}
\end{equation}
because if the quantity $BC$ is added to its visual magnitude, one obtains its bolometric magnitude regardless of its brightness type being absolute or apparent in magnitude units.

\cite{Kuiper1938} originally suggested this definition for calibrating the stellar effective temperature ($T_{\rm eff}$) scale provided with observed stellar radii ($R$) from eclipsing binaries with known distances, and luminosities ($L$) calculated by

\begin{equation}
    M_{\rm Bol} = M_{\rm Bol, \odot} -2.5 \times \log{L/L_\odot},
    \label{eq:2}
\end{equation}
from absolute bolometric magnitudes ($M_{\rm Bol}$), where $M_{\rm Bol, \odot}$ and $L_\odot$ are constants representing solar absolute bolometric magnitude and luminosity since it is straight forward to obtain $M_{\rm Bol}$ as $M_{\rm Bol} = M_{\rm V} + BC$. Implying that the zero-point constant of the $BC$ scale is zero or a negative number because $L_{\rm V}=L\times10^{BC/2.5}$, where $L$ and $L_{\rm V}$ are the total and visual luminosities from $M_{\rm Bol}$ and $M_{\rm V}$ of a star and thus, if $BC>0$, $L_{\rm V}$ is unphysical, the definition of $BC$ from \cite{Kuiper1938} was ill posed by definiton to cause paradigms such ``{\it BC} of a star must always be negative,'' ``the bolometric magnitude of a star ought to be brighter than its $V$-magnitude,'' and ``the zero point of bolometric corrections are arbitrary'' \citep{Torres2010, Eker2021, Eker2021b, Eker2022}.

Apparently, the paradigms hide the ill-posed nature, such that the definition of $BC$ from \cite{Kuiper1938} was well accepted throughout the century. One of the good reasons could be that it was the only tool for calculating the mass and the luminosity of single stars by the help of the mass-luminosity relation for a long time \citep{Eker2025} and thus problems such: inconsistent $T_{\rm eff}-BC$ tables floated astronomical literature, there were cases with more than one $BC$ value assigned to a star, thus, inconstant $L$ were predicted due to multiple $BC$ assigned even if the stars have a unique $M_{\rm Bol}$, were inevitable \citep{Eker2025}. 

Part of the problem was Equation~(\ref{eq:2}) because there was no consensus on which values of $M_{\rm Bol, \odot}$ and $L_\odot$ must be used, thus the first step into the solution was taken by International Astronomical Union (IAU) by announcing a General Assembly Resolution B2, which mainly suggests replacing Equation~(\ref{eq:2}) by

\begin{equation}
    M_{\rm Bol} = -2.5 \times \log{L} + C_{\rm Bol},
    \label{eq:3}
\end{equation}
for solving the problems caused by the paradigms \citep{Eker2022, Eker2025} and standardize the predicted $L$ of stars in its XXIXth meeting in Honolulu in 2015 \citep{Mamajek2015} by giving the value of the zero-point constant of the absolute bolometric magnitude scale $C_{\rm Bol}=71.197$ $425$ ... if $L$ is in SI units. Because $L$ of a star ($L = 4 \pi R^2 \sigma T^4$) could also be defined as the received bolometric flux just above the atmosphere is multiplied by the surface area of the sphere with a radius equal to the distance of the star if no extinction, fixing the zero-point of absolute bolometric magnitudes must also fix the zero-point of the apparent bolometric magnitudes, too. Thus, the resolution also gave the zero-point constant of apparent bolometric magnitudes as $c_{\rm Bol}=-18.997$ $425$ ...  with the formula as

\begin{equation}
    m_{\rm Bol}=-2.5 \log{f_{\rm Bol}}+c_{\rm Bol},
    \label{eq:4}
\end{equation}
where $m_{\rm Bol}$ is the apparent bolometric magnitude, $f_{\rm Bol}$ is the received bolometric flux if it is in SI units on the telescope detector if there is no atmospheric and interstellar extinction, and $c_{\rm Bol}$ is the zero-point constant of the apparent bolometric magnitudes.

The IAU General Assembly did not prefer to fix the zero-point constant of the {\it BC} scale directly because fixing the zero-point constant of absolute bolometric magnitudes automatically fixes the zero-point constant of the {\it BC} scale also. This becomes obvious if Equation~(\ref{eq:3}) is adopted for absolute visual magnitudes as
\begin{equation}
    M_{\rm V}=-2.5 \log{L_{\rm V}}+C_{\rm V},
    \label{eq:5}
\end{equation}
and then subtracting it from Equation~(\ref{eq:3}) one obtains the following
\begin{equation}
    BC_{\rm V}=M_{\rm Bol}-M_{\rm V}=2.5\times\log{{\frac{L_{\rm V}}{L}}}+C_{\rm Bol}-C_{\rm V}, 
    \label{eq:6}
\end{equation}
where $L_V/L$ is the ratio of the visual to bolometric luminosity of the star and $C_{\rm Bol}-C_{\rm V}$ stands for the zero-point constant of the {\it BC} scale.  

If the same steps were applied to apparent magnitudes, one obtains 
\begin{equation}
    BC_{\rm V}=m_{\rm Bol}-V
    =2.5\times\log{{\frac{f_{\rm V}}{f_{\rm Bol}}}}+c_{\rm Bol}-c_{\rm V}, 
    \label{eq:7}
\end{equation}
where $f_{\rm V}/f_{\rm Bol}$ is the fraction of visual photons among the total that reach Earth from the star if there is no extinction. Assuming stars radiate isotropically, Equations (6) and (7) indicates that the zero-point constant of the $BC_{\rm V}$ scale is $C_2(V) = C_{\rm Bol} - C_{\rm V} = c_{\rm Bol} - c_{\rm V}$.

It is because old habits are very difficult to change, or it is because paradigm change is not so easy, that the resolution announced by the IAU 2015 General Assembly was ineffective or has not been noticed for a while. It took six years \cite{Eker2021} to be aware of it before everyone else and seven years to understand that it is a revolutionary document that provides ideas to discard all three paradigms mentioned above \citep{Eker2022}. It took ten years to publish a new generalized definition of $BC_\xi$, where the subscript $\xi$ stands for any photometric band, implying $L_{\rm \xi}=L\times10^{(BC_\xi-C_2(\xi))/2.5}$, which allows for a limited amount of positive $BC$, not only for the $V-$band, without violating the rule $L_{\rm \xi} < L$ in the cases where $C_2(\xi) > 0$, thus $BC_\xi$ values less than $C_2(\xi)$ greater than zero are possible \citep{Eker2025}. It took 11 years until \cite[hereafter Y26]{Yucel2026} to calculate spectroscopic $BC_{\rm V}$ values from 128 high resolution and high signal-to-noise ratio spectra and empirically determine the zero-point constant of the $BC_{\rm V}$ scale as $C_2(V)= 2.3653 \pm 0.0067$ and $C_2(V)= 2.3826 \pm 0.0076$ corresponding to filter transmission profiles of \cite{Bessell1990} and \cite{Landolt1992} respectively. 

In this study, we are motivated to extend availability of spectroscopic bolometric corrections and to fix the non-arbitrary zero-point constants of $BC$ scales for the $Gaia$ bands ($G$, $G_{\rm BP}$ and $G_{\rm RP}$) empirically by following the same method described by \citetalias{Yucel2026} using $Gaia$ XP spectra \citep{Gaia_DR3, DeAngeli2023} of the same stars used by \citetalias{Yucel2026}. 

\section{Data}

\begin{figure}[hb]
    \centering
    \includegraphics[width=1\linewidth]{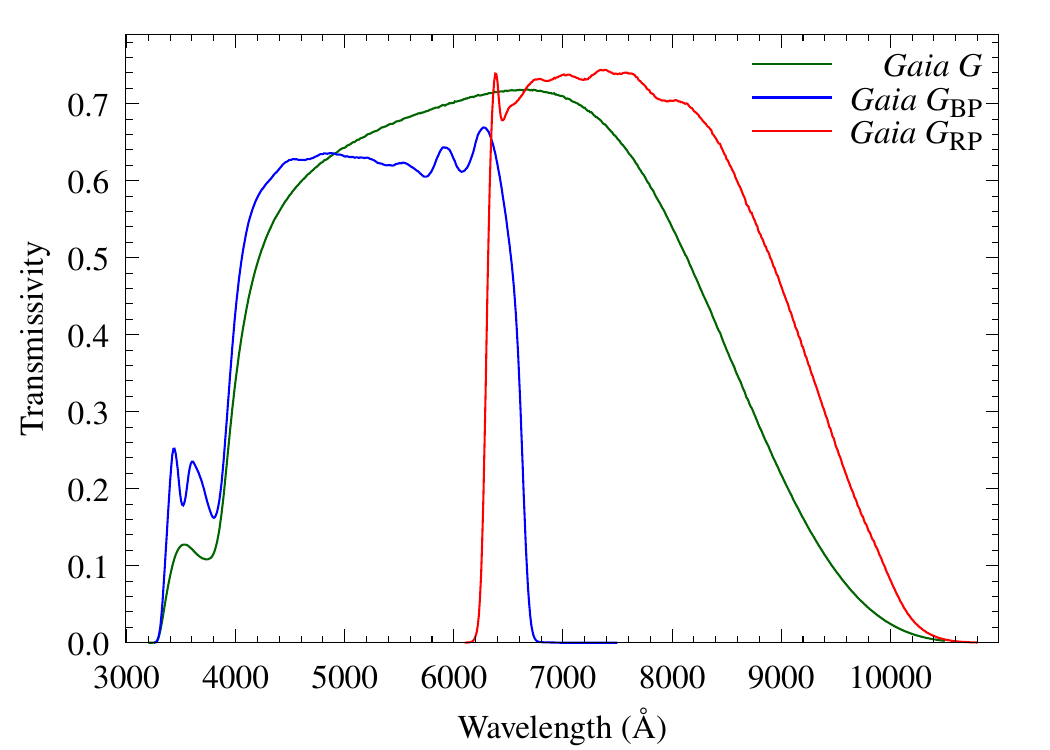}
    \caption{Transparency curves of $Gaia$ passbands from \protect\cite{Riello2021}.}
    \label{fig:gaia_filters}
\end{figure}

Measuring fractional luminosity $L_\xi /L$ of a star from its spectrum, where $\xi$ is one of the photometric bands of Gaia, is necessary but not enough to calculate the corresponding spectroscopic $BC_\xi$ value if the zero-point constant for the $BC_\xi$ scale is not available. The zero-point constant of  $BC_\xi$, symbolized as $C_2(\xi)$, must first be determined as was done for the spectroscopic values of $BC_{\rm V}$ by \citetalias{Yucel2026}. The zero-point constant of spectroscopic $BC$ values of any band requires determinations of $C_2(\xi)$ for a large group of stars using their absolute bolometric ($M_{\rm Bol}$) and filtered ($M_\xi$) magnitudes and fractional luminosities ($L_\xi/L$), which also requires accurate stellar parameters ($R$, $T_{\rm eff}$), photometric and astrometric data in addition to the spectra covering the wavelength ranges of the $Gaia$ bands ($G$, $G_{\rm BP}$ and $G_{\rm RP}$).  Then, the mean of the computed $C_2(\xi)$ values could be used in the formula as 
\begin{equation}
    {BC}_\xi=2.5\log{\frac{L_\xi}{L}+C_2(\xi)},
    \label{eq:8}
\end{equation}
to calculate the spectroscopic $BC_\xi$ from a fractional luminosity value $L_\xi /L$. 
The fractional luminosities, however, are obtained according to the formula
\begin{equation}
        \frac{L_\xi}{L}=\frac{\int_{0}^{\infty} S_\lambda (\xi)F_\lambda \,d\lambda}{\int_{0}^{\infty} F_\lambda \,d\lambda}=\frac{F_\xi}{\sigma {T_{\rm eff}}^4},
\label{eq:9}
\end{equation}
where $S_\lambda\left(\xi\right)$ is the function representing the transparency profile for one of the $Gaia$ bands ($G$, $G_{\rm BP}$, and $G_{\rm RP}$) and $F_\lambda$ is the surface flux per unit wavelength, therefore $F_\xi$ is the filtered flux out of the total flux, $\sigma {T_{\rm eff}}^4$, which is known as the bolometric flux. The transparency curves of the {\it Gaia} passbands used by the {\it Gaia} team in the production of DR3 photometric data are shown in Figure~\ref{fig:gaia_filters}.

\begin{figure*}
    \centering
    \includegraphics[width=0.95\linewidth]{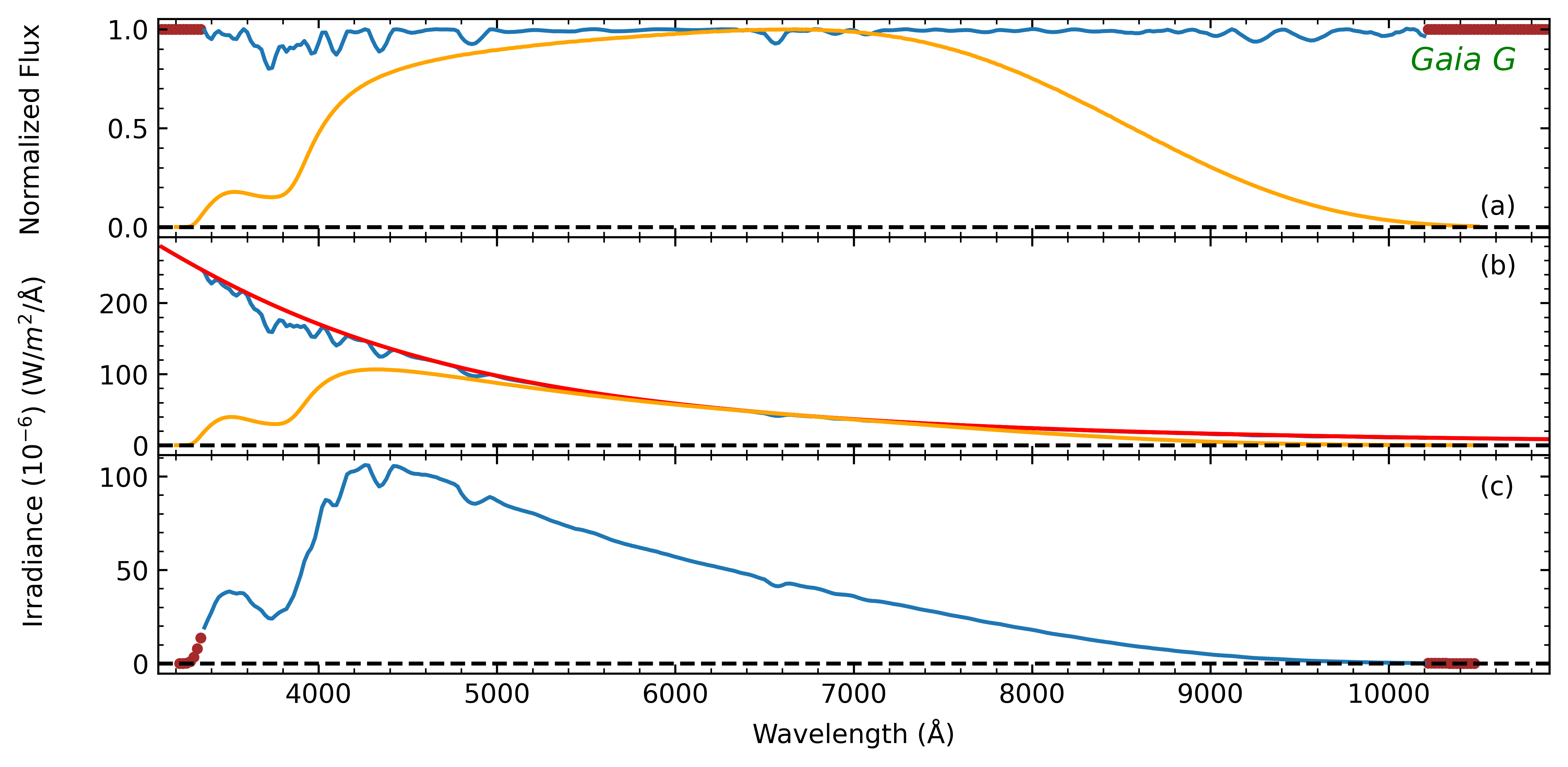} \\
    \includegraphics[width=0.95\linewidth]{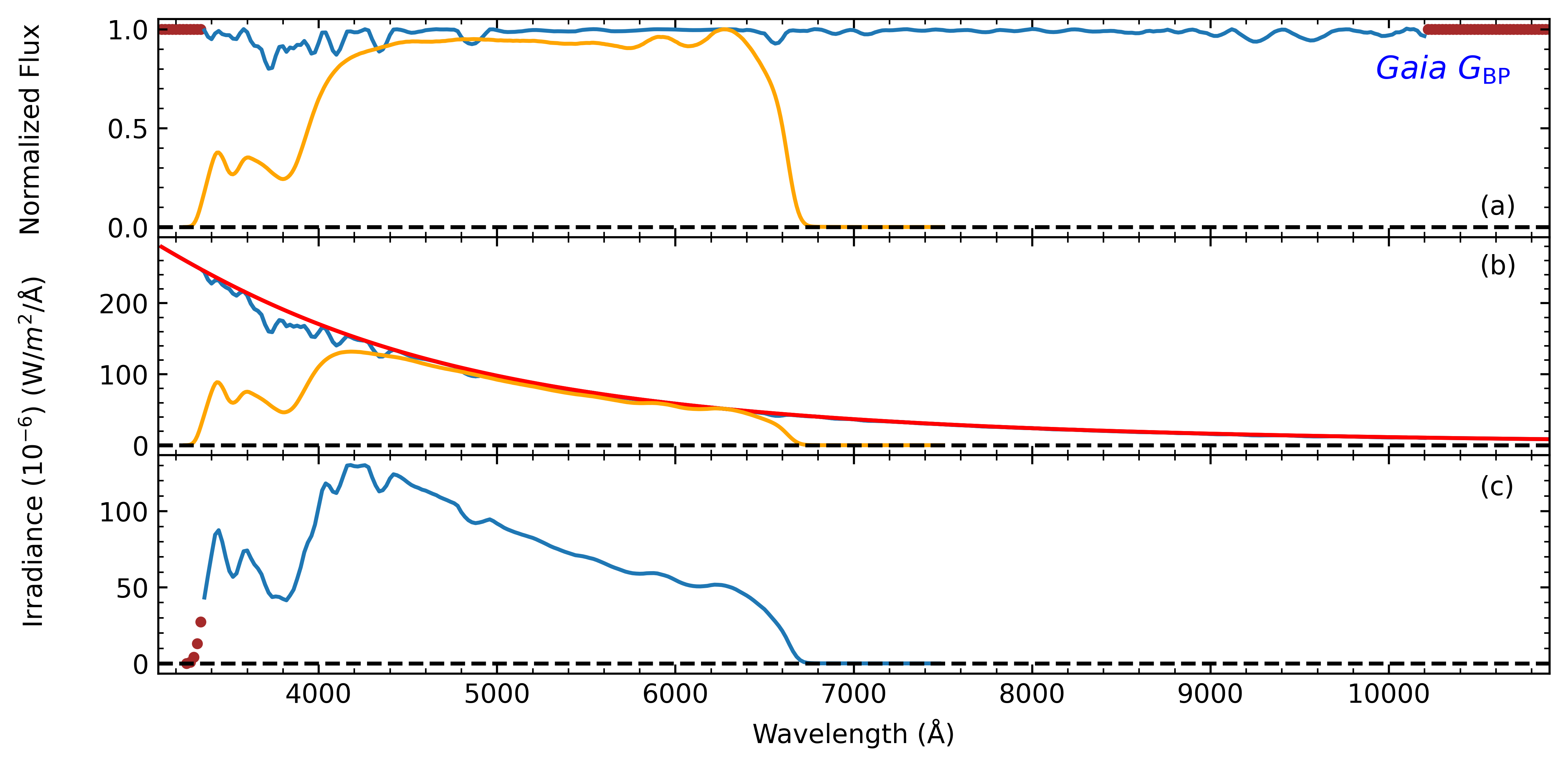}\\
    
    \includegraphics[width=0.95\linewidth]{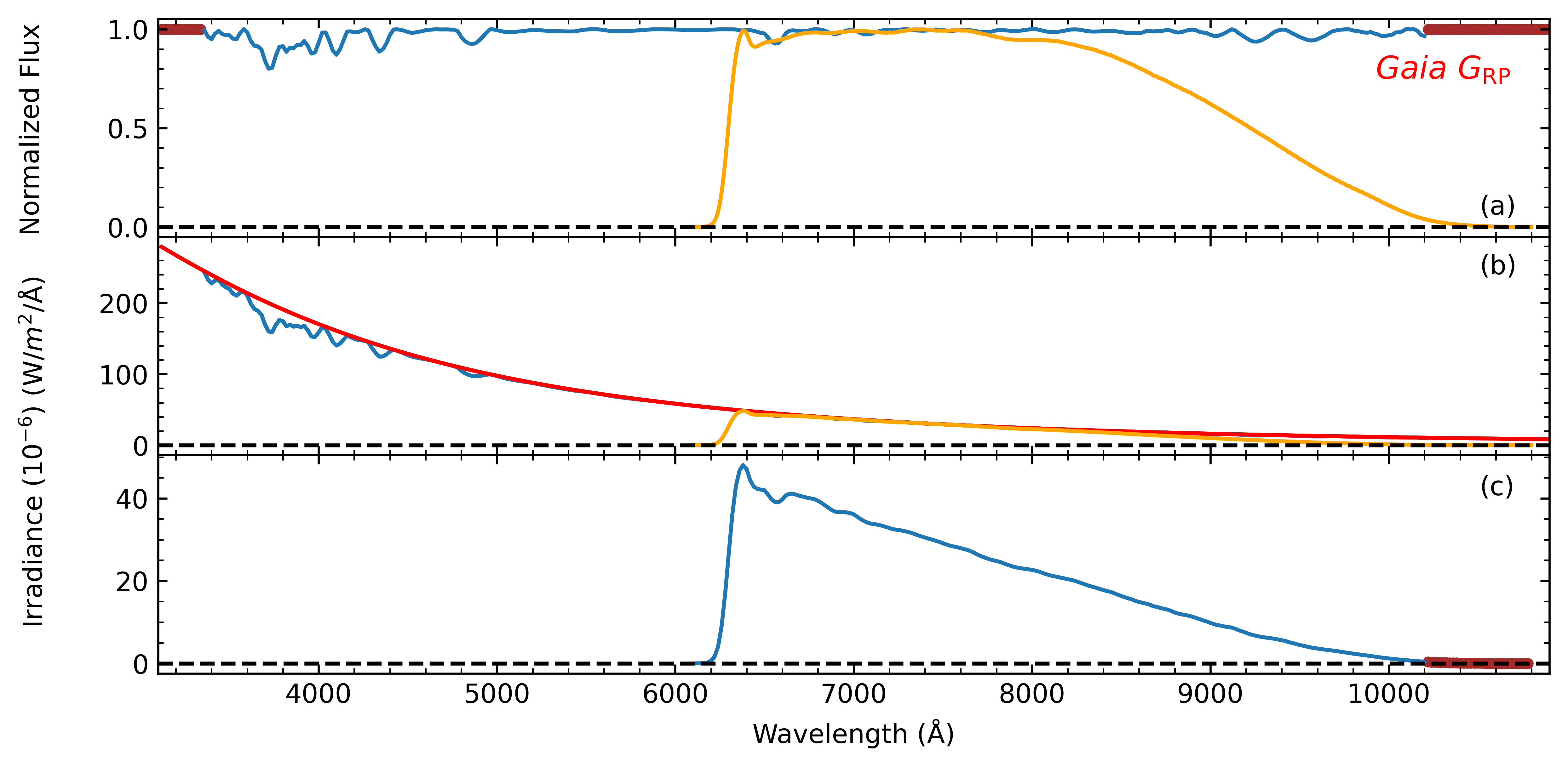}\\
    \caption{Normalized spectrum of HD\,29589 and \textit{Gaia} filter profiles (a), de-normalized flux and \textit{Gaia} filter spectra (b), and the convoluted spectrum of HD\,29589 (c) for $G$ (top panel), $G_{\rm BP}$ (middle panel), and $G_{\rm RP}$ (bottom panel), respectively.  Dividing the area under the convoluted spectrum by $\sigma {T_{\rm eff}}^4$ gives $L_{\xi}/L$ as 21.72, 19.87, and 6.27\% for $G$, $G_{\rm BP}$, and $G_{\rm RP}$ profile function, respectively.}  
    \label{fig:spec}
\end{figure*}

The steps of obtaining the filtered fluxes from the sample spectra are demonstrated in Figure~\ref{fig:spec} on a spectrum, which is from HD 29589. The spectrum and transparency curves must be normalized to one in the first step. This step is needed to remove the effects of interstellar extinction on the observed spectrum. The normalized spectrum and the transparency profiles of each band are shown in the top boxes of each panel of Figure~\ref{fig:spec}. 

The second step is for the de-normalisation of the spectrum and the transparency profiles. The de-normalization is done by multiplying both the spectrum and the profiles by a Planck function with a temperature equal to the effective temperature of the Star. The middle panels of each box in Figure~\ref{fig:spec} show the de-normalized spectrum and filter profiles for the star HD\,29589. 

The third step is convolution, that is, pixel-to-pixel multiplication of the spectra and the profiles to obtain the wavelength distribution of filtered star flux corresponding to $Gaia$ filters. The last step is to obtain integrated total fluxes by integrating the convoluted profiles. This can be done using any numerical integration technique. We have chosen Simpson's rule and applied it via \texttt{scipy.integrate.simpson} \citep{scipy}. Finally, fractional luminosities $L_{\rm G}/L$, $L_{\rm G_{BP}}/L$, and $L_{\rm G_{RP}}/L$ could be obtained by dividing each integrated quantity by the bolometric flux 
$\sigma\,{T_{\rm eff}}^4$ of the star.

\subsection{Spectroscopic Data}

Despite their low resolution ($\Delta\lambda\,=20$ \AA), \textit{Gaia} DR3 XP spectra have a great advantage that they are free of atmospheric telluric lines \citep{Griffin1973, Catanzaro1997, Griffin2000} because \textit{Gaia} satellite operated outside the Earth's atmosphere at the L2 point, which is 1.5 million km away from the Earth directly opposite side of the Sun; another advantage of them for this study is that their spectral range includes all the optical regions of stellar spectrum (3360 -- 10200 \AA) covering almost three of the {\it Gaia} bandpasses. 

The small missing parts of the observed spectra at the shortest and longest wavelengths have been noticed. The missing parts have been artificially completed by adding values one to the normalized spectra to extend them to fit the wavelength limits of the {\it Gaia} filters. The artificially extended parts of the normalized spectra and convoluted profiles are shown by dotted extensions in Figure~\ref{fig:spec}. This was done to minimize the negative effect of the missing part of the spectral coverage. The cases with and without artificial extensions are checked, and it is seen that the differences in the luminosity ratios are at the fourth digit after the decimal point, which are indeed negligible. 

\begin{table}
\renewcommand{\arraystretch}{0.95}
    \centering
    \scriptsize
    \caption{Spectroscopic data from the sample spectra with resolution ($\Delta\lambda=20$ \AA) covering spectral range of 3360--10200 \AA\,  
    }
    \label{speck}
    \begin{tabular}{llcrrrc}
    \toprule
    Order & Star & S/N & \multicolumn{1}{c}{$L_{\rm G}/L$} & \multicolumn{1}{c}{$L_{\rm G_{\rm BP}}/L$} & \multicolumn{1}{c}{$L_{\rm G_{\rm RP}}/L$} & err \\[1.5ex]
          &      &     &  \multicolumn{1}{c}{(\%)}    &      \multicolumn{1}{c}{(\%)}     &     \multicolumn{1}{c}{(\%)}      & (\%) \\
    \midrule
        01 & HD\,1279 & 254 & 24.700 & 22.230 & 7.415 & 0.557 \\ 
        02 & HD\,1404 & 196 & 40.309 & 32.439 & 15.472 & 0.722 \\ 
        03 & HD\,1439 & 352 & 37.822 & 31.436 & 13.641 & 0.402 \\ 
        04 & HD\,2729 & 295 & 22.422 & 20.456 & 6.525 & 0.479 \\ 
        05 & HD\,4778 & 211 & 37.407 & 31.009 & 13.577 & 0.670 \\ 
        06 & HD\,10362 & 183 & 21.950 & 20.067 & 6.334 & 0.773 \\ 
        07 & HD\,10700 & 168 & 41.980 & 25.290 & 25.313 & 0.842 \\ 
        08 & HD\,15318 & 307 & 32.796 & 28.184 & 10.975 & 0.461 \\ 
        09 & HD\,16440 & 257 & 12.682 & 12.022 & 3.302 & 0.550 \\ 
        10 & HD\,19736 & 204 & 20.782 & 19.065 & 5.946 & 0.693 \\ 
        11 & HD\,22879 & 332 & 46.515 & 31.058 & 24.549 & 0.426 \\ 
        12 & HD\,23300 & 276 & 22.160 & 20.220 & 6.448 & 0.512 \\ 
        13 & HD\,27295 & 316 & 30.388 & 26.538 & 9.818 & 0.448 \\ 
        14 & HD\,27563 & 238 & 24.129 & 21.740 & 7.220 & 0.594 \\ 
        15 & HD\,27778 & 238 & 19.328 & 17.826 & 5.457 & 0.594 \\ 
        16 & HD\,29138 & 147 & 7.626 & 7.432 & 1.836 & 0.962 \\ 
        17 & HD\,29335 & 278 & 23.402 & 21.170 & 6.955 & 0.509 \\ 
        18 & HD\,29589 & 279 & 21.718 & 19.865 & 6.265 & 0.507 \\ 
        19 & HD\,32040 & 173 & 29.800 & 26.013 & 9.590 & 0.817 \\ 
        20 & HD\,32115 & 254 & 45.569 & 33.670 & 20.462 & 0.557 \\ 
        21 & HD\,32309 & 288 & 34.188 & 28.968 & 11.804 & 0.491 \\ 
        22 & HD\,32612 & 322 & 11.780 & 11.170 & 3.073 & 0.439 \\ 
        23 & HD\,34078 & 150 & 4.046 & 3.986 & 0.945 & 0.943 \\ 
        24 & HD\,34310 & 207 & 32.671 & 27.970 & 11.011 & 0.683 \\ 
        25 & HD\,35299 & 251 & 7.681 & 7.469 & 1.863 & 0.563 \\ 
        26 & HD\,35693 & 216 & 39.795 & 32.150 & 15.064 & 0.655 \\ 
        27 & HD\,35912 & 200 & 11.507 & 10.927 & 2.983 & 0.707 \\ 
        28 & HD\,36285 & 283 & 9.634 & 9.286 & 2.401 & 0.500 \\ 
        29 & HD\,36430 & 264 & 12.173 & 11.514 & 3.199 & 0.536 \\ 
        30 & HD\,36512 & 364 & 3.472 & 3.451 & 0.787 & 0.389 \\ 
        31 & HD\,36591 & 451 & 5.870 & 5.776 & 1.384 & 0.314 \\ 
        32 & HD\,36960 & 458 & 4.941 & 4.878 & 1.148 & 0.309 \\ 
        33 & HD\,37077 & 275 & 45.896 & 33.813 & 20.728 & 0.514 \\ 
        34 & HD\,37356 & 255 & 8.955 & 8.631 & 2.226 & 0.555 \\ 
        35 & HD\,37744 & 208 & 7.551 & 7.291 & 1.860 & 0.680 \\ 
        36 & HD\,40967 & 349 & 17.416 & 16.202 & 4.794 & 0.405 \\ 
        37 & HD\,46189 & 165 & 15.822 & 14.824 & 4.280 & 0.857 \\ 
        38 & HD\,54764 & 264 & 12.467 & 11.731 & 3.314 & 0.536 \\ 
        39 & HD\,55856 & 265 & 9.741 & 9.412 & 2.420 & 0.534 \\ 
        40 & HD\,71155 & 275 & 37.147 & 30.730 & 13.488 & 0.514 \\ 
        41 & HD\,78556 & 208 & 34.357 & 29.177 & 11.727 & 0.680 \\ 
        42 & HD\,82106 & 180 & 38.643 & 20.855 & 25.944 & 0.786 \\ 
        43 & HD\,82621 & 260 & 40.347 & 32.551 & 15.382 & 0.544 \\ 
        44 & HD\,82943 & 236 & 45.735 & 30.384 & 24.267 & 0.599 \\ 
        45 & HD\,84937 & 254 & 47.587 & 33.813 & 22.954 & 0.557 \\ 
        46 & HD\,85503 & 178 & 35.221 & 17.988 & 24.943 & 0.795 \\ 
        47 & HD\,89021 & 115 & 34.938 & 27.507 & 14.293 & 1.230 \\ 
        48 & HD\,99922 & 207 & 40.437 & 32.061 & 15.971 & 0.683 \\ 
        49 & HD\,101364 & 379 & 45.972 & 30.002 & 24.970 & 0.373 \\ 
        50 & HD\,103095 & 324 & 43.912 & 26.704 & 26.114 & 0.436 \\ 
        51 & HD\,113226 & 128 & 38.743 & 21.191 & 25.385 & 1.105 \\ 
        52 & HD\,117176 & 490 & 44.567 & 27.790 & 25.655 & 0.289 \\ 
        53 & HD\,120198 & 171 & 35.228 & 29.641 & 12.351 & 0.827 \\ 
        54 & HD\,122563 & 279 & 39.710 & 21.967 & 26.249 & 0.507 \\ 
        55 & HD\,125924 & 351 & 10.293 & 9.860 & 2.602 & 0.403 \\ 
        56 & HD\,128167 & 487 & 46.487 & 32.511 & 22.785 & 0.290 \\ 
        57 & HD\,133208 & 127 & 38.937 & 21.684 & 25.552 & 1.114 \\ 
        58 & HD\,141714 & 403 & 43.427 & 26.167 & 25.994 & 0.351 \\ 
        59 & HD\,146233 & 252 & 45.438 & 29.499 & 24.770 & 0.561 \\ 
        60 & HD\,148379 & 262 & 15.372 & 14.287 & 4.264 & 0.540 \\ 
        61 & HD\,148688 & 196 & 10.358 & 9.882 & 2.664 & 0.722 \\ 
        62 & HD\,150117 & 188 & 33.065 & 27.873 & 11.588 & 0.752 \\ 
        63 & HD\,160762 & 269 & 13.608 & 12.745 & 3.663 & 0.526 \\ 
        64 & HD\,162570 & 262 & 44.631 & 32.895 & 19.969 & 0.540 \\ 
        65 & HD\,171301 & 204 & 27.408 & 23.907 & 8.864 & 0.693 \\ 
        66 & HD\,174959 & 195 & 20.435 & 18.600 & 5.948 & 0.725 \\ 
        67 & HD\,186427 & 517 & 45.442 & 29.367 & 24.877 & 0.274 \\ 
        68 & HD\,189319 & 153 & 29.146 & 12.851 & 23.470 & 0.924 \\ 
        69 & HD\,189957 & 178 & 3.778 & 3.736 & 0.869 & 0.795 \\ 
        70 & HD\,192263 & 198 & 40.103 & 22.270 & 26.103 & 0.714 \\ 
        71 & HD\,193183 & 247 & 10.804 & 10.357 & 2.759 & 0.573 \\ 
        72 & HD\,195556 & 171 & 14.600 & 13.676 & 3.910 & 0.827 \\ 
        73 & HD\,196740 & 389 & 19.051 & 17.529 & 5.391 & 0.364 \\ 
        74 & HD\,197512 & 247 & 7.808 & 7.509 & 1.944 & 0.573 \\ 
        75 & HD\,201091 & 201 & 34.419 & 16.626 & 25.562 & 0.704 \\ 
        76 & HD\,201092 & 199 & 32.101 & 15.090 & 24.543 & 0.711 \\ 
        77 & HD\,205139 & 187 & 5.672 & 5.462 & 1.395 & 0.756 \\ 
        78 & HD\,207538 & 209 & 3.937 & 3.878 & 0.924 & 0.677 \\ 
        79 & HD\,208266 & 231 & 6.691 & 6.446 & 1.677 & 0.612 \\ 
        80 & HD\,209419 & 326 & 19.571 & 17.920 & 5.688 & 0.434 \\ 
        81 & HD\,209975 & 207 & 3.555 & 3.531 & 0.810 & 0.683 \\ 
        82 & HD\,213087 & 246 & 6.818 & 6.604 & 1.696 & 0.575 \\ 
        83 & HD\,214263 & 201 & 10.426 & 10.002 & 2.642 & 0.704 \\ 
        84 & HD\,214994 & 261 & 39.231 & 32.006 & 14.594 & 0.542 \\ 
        85 & HD\,215191 & 245 & 9.289 & 8.969 & 2.314 & 0.577 \\ 
        86 & HD\,217014 & 284 & 45.331 & 29.247 & 24.946 & 0.498 \\ 
        87 & HD\,222173 & 290 & 29.241 & 25.530 & 9.423 & 0.488 \\ 
        88 & HD\,222762 & 247 & 25.809 & 22.914 & 7.974 & 0.573 \\
    \bottomrule
    \label{tab:1}
    \end{tabular}
\end{table}

Unfortunately, \textit{Gaia} XP spectra also suffered systemic errors that depend, mainly, on the color of the object, especially below 4000 \AA~ \citep{Montegriffo2023}. To overcome this problem, the correction proposed in \cite{Huang2024} is applied. In this study, 88 of 128 stars were used from the list of \citetalias{Yucel2026}, where only the stars had effective temperatures, radii, and spectra with S/N > 100, resolution: R > 20000, and the wavelength range that included the visual band.
The absence of some bright stars ($G \leq 6$ mag) in $Gaia$ DR3 is a known selection effect caused by detector saturation and processing limitations \citep{Gaia_DR3, Lindegren2021}. Searching for new stars with effective temperatures and radii was not intended, which would make us lose the advantage of using a list of stars with internally consistent parameters established by the SED models in  \citetalias{Yucel2026}.

Spectroscopic data of the sample stars used in this study are given in Table~\ref{speck} where the columns indicate order, star name, signal-to-noise ratio, fractional luminosities obtained by the method described above in percent units for the \textit{Gaia} bands $G$, $G_{\rm BP}$ and $G_{\rm RP}$, respectively, and the last column is for the relative errors according to,
\begin{equation}
    \frac{\Delta \left(L_{\xi}/L\right)}{\left(L_{\xi}/L\right)}=\sqrt{\left(\frac{\Delta L_{\xi}}{L_{\xi}}\right)^2+\left(\frac{\Delta L}{L}\right)^2}=\frac{\sqrt{2}}{S/N},
    \label{equ:10}
\end{equation}
the formula given by \citetalias{Yucel2026}.  

{
\small
\setlength{\tabcolsep}{3pt}
\renewcommand{\arraystretch}{0.55}
\begin{longtable}{llrclrrccrrc}

\caption{Stellar parameters for calculating absolute bolometric magnitudes of the sample stars according to the Stefan-Boltzmann law (credit to \citetalias{Yucel2026}).} \\
\toprule
    Order & Star & \multicolumn{1}{c}{$T_{\rm eff}$} & err & Reference & \multicolumn{1}{c}{$R$} & err & $L$ & $L$ & \multicolumn{1}{c}{err} & $M_{\rm Bol}$ & err \\
          &      & \multicolumn{1}{c}{(K)} & (\%) & & \multicolumn{1}{c}{($R_\odot$)} & (\%) & (W) & ($L_\odot$) & (\%) & \multicolumn{2}{c}{(mag)} \\
\midrule
\endfirsthead

\multicolumn{12}{l}%
{{{\bf\tablename\ \thetable{}} -- continued from previous page}} \\
\toprule
    Order & Star & \multicolumn{1}{c}{$T_{\rm eff}$} & err & Reference & \multicolumn{1}{c}{$R$} & err & $L$ & $L$ & \multicolumn{1}{c}{err} & $M_{\rm Bol}$ & err \\
          &      & \multicolumn{1}{c}{(K)} & (\%) & & \multicolumn{1}{c}{($R_\odot$)} & (\%) & (W) & ($L_\odot$) & (\%) & \multicolumn{2}{c}{(mag)} \\
\midrule
\endhead

\midrule
\multicolumn{12}{r}{{Continued on next page}} \\
\endfoot

\bottomrule
\endlastfoot

        01 & HD\,1279  & 13300 & 0.9 & \citet{Monier2023} & 5.73 & 5.0 & 3.54E+29 & 926 & 10.7 & -2.676 & 0.116 \\ 
        02 & HD\,1404  & 8840 & 2.0 & \citet{Hillen2012} & 2.06 & 5.8 & 8.89E+27 & 23 & 14.0 & 1.325 & 0.152 \\ 
        03 & HD\,1439  & 9640 & 1.3 & \citet{Royer2014} & 3.47 & 4.6 & 3.58E+28 & 94 & 10.5 & -0.188 & 0.114 \\ 
        04 & HD\,2729  & 14125 & 6.8 & \citet{SimonDiaz2017} & 3.27 & 4.5 & 1.47E+29 & 384 & 28.7 & -1.720 & 0.311 \\ 
        05 & HD\,4778  & 9605 & 1.0 & \citet{Espadons} & 2.10 & 5.6 & 1.30E+28 & 34 & 11.9 & 0.913 & 0.129 \\ 
        06 & HD\,10362 & 14300 & 2.1 & \citet{Martin2018} & 6.04 & 11.5 & 5.26E+29 & 1373 & 24.4 & -3.104 & 0.265 \\ 
        07 & HD\,10700 & 5397 & 0.3 & \citet{Soubiran2024} & 0.75 & 9.2 & 1.65E+26 & 0.431 & 18.5 & 5.654 & 0.200 \\ 
        08 & HD\,15318 & 10900 & 4.6 & \citet{Gebran2016} & 2.37 & 5.0 & 2.74E+28 & 72 & 20.9 & 0.104 & 0.227 \\ 
        09 & HD\,16440 & 19000 & 5.3 & \citetalias{Yucel2026} & 8.51 & 8.8 & 3.25E+30 & 8496 & 17.7 & -5.083 & 0.192 \\ 
        10 & HD\,19736 & 14727 & 1.4 & \citet{Huang2010} & 2.89 & 6.1 & 1.36E+29 & 355 & 13.4 & -1.636 & 0.146 \\ 
        11 & HD\,22879 & 5962 & 1.4 & \citet{Soubiran2024} & 0.98 & 5.0 & 4.21E+26 & 1.100 & 11.6 & 4.636 & 0.126 \\ 
        12 & HD\,23300 & 14207 & 3.0 &  \citet{Takeda2010} & 3.69 & 4.6 & 1.91E+29 & 498 & 15.1 & -2.004 & 0.164 \\ 
        13 & HD\,27295 & 11572 & 2.7 & \citet{Arentsen2019} & 1.78 & 5.6 & 1.96E+28 & 51 & 15.5 & 0.469 & 0.168 \\ 
        14 & HD\,27563 & 13461 & 0.7 & \citet{Huang2010} & 3.88 & 4.8 & 1.71E+29 & 446 & 10.1 & -1.883 & 0.110 \\ 
        15 & HD\,27778 & 15345 & 1.3 & \citet{Huang2010} & 3.35 & 8.6 & 2.15E+29 & 560 & 17.9 & -2.131 & 0.194 \\ 
        16 & HD\,29138 & 24090 & 1.0 & \citet{Borisov2023} & 14.01 & 2.1 & 2.28E+31 & 59553 & 5.8 & -7.197 & 0.063 \\ 
        17 & HD\,29335 & 13711 & 1.1 & \citet{Huang2010} & 2.94 & 3.0 & 1.06E+29 & 276 & 7.5 & -1.362 & 0.081 \\ 
        18 & HD\,29589 & 14400 & 1.4 & \citet{Niemczura2022} & 2.20 & 5.8 & 7.20E+28 & 188 & 12.9 & -0.946 & 0.140 \\ 
        19 & HD\,32040 & 11695 & 0.7 & \citet{Zorec2012} & 2.06 & 5.3 & 2.75E+28 & 72 & 10.9 & 0.099 & 0.118 \\ 
        20 & HD\,32115 & 7250 & 1.4 & \citet{Fossati2011} & 1.49 & 4.6 & 2.12E+27 & 5.54 & 10.8 & 2.881 & 0.117 \\ 
        21 & HD\,32309 & 10471 & 0.5 & \citet{Zorec2012} & 1.94 & 6.1 & 1.55E+28 & 41 & 12.4 & 0.718 & 0.134 \\ 
        22 & HD\,32612 & 19612 & 1.6 & \citet{Paunzen2005} & 3.40 & 1.5 & 5.89E+29 & 1540 & 7.0 & -3.229 & 0.076 \\ 
        23 & HD\,34078 & 31057 & 5.5 & \citet{Koleva2012} & 4.78 & 2.7 & 7.34E+30 & 19166 & 22.5 & -5.966 & 0.244 \\ 
        24 & HD\,34310 & 10864 & 0.7 & \citet{Zorec2012} & 2.35 & 7.1 & 2.66E+28 & 69.4 & 14.6 & 0.137 & 0.158 \\ 
        25 & HD\,35299 & 24000 & 0.8 & \citet{Nieva2011} & 3.47 & 2.6 & 1.38E+30 & 3595 & 6.1 & -4.149 & 0.066 \\ 
        26 & HD\,35693 & 9016 & 1.9 & \citet{Zorec2012} & 3.43 & 3.5 & 2.68E+28 & 70.0 & 10.1 & 0.128 & 0.110 \\ 
        27 & HD\,35912 & 19833 & 3.8 &\citet{Hourihane2023} &  3.19 & 4.3 & 5.43E+29 & 1420 & 17.4 & -3.140 & 0.189 \\ 
        28 & HD\,36285 & 21700 & 1.4 & \citet{Nieva2011} & 2.97 & 1.0 & 6.76E+29 & 1767 & 5.9 & -3.378 & 0.064 \\ 
        29 & HD\,36430 & 19300 & 1.0 & \citet{Nieva2011} & 3.77 & 2.1 & 6.82E+29 & 1781 & 5.9 & -3.386 & 0.064 \\ 
        30 & HD\,36512 & 33400 & 1.5 &  \citet{Nieva2011} & 5.17 & 2.1 & 1.15E+31 & 29936 & 7.3 & -6.450 & 0.079 \\ 
        31 & HD\,36591 & 27000 & 0.9 & \citet{Nieva2011} & 4.80 & 4.7 & 4.23E+30 & 11039 & 10.2 & -5.367 & 0.110 \\ 
        32 & HD\,36960 & 29000 & 1.2 & \citet{Nieva2011} & 5.00 & 3.6 & 6.10E+30 & 15926 & 8.6 & -5.765 & 0.093 \\ 
        33 & HD\,37077 & 7174 & 1.3 & \citet{Paunzen2006} & 5.68 & 2.3 & 2.95E+28 & 77.0 & 6.9 & 0.024 & 0.075 \\ 
        34 & HD\,37356 & 22370 & 2.0 & \citet{Cunha1994} & 4.97 & 2.0 & 2.13E+30 & 5572 & 9.0 & -4.625 & 0.097 \\ 
        35 & HD\,37744 & 24000 & 1.7 & \citet{Nieva2011} & 3.23 & 3.1 & 1.19E+30 & 3120 & 9.0 & -3.995 & 0.098 \\ 
        36 & HD\,40967 & 16258 & 1.3 & \citet{Paunzen2005} & 6.02 & 4.6 & 8.72E+29 & 2279 & 10.6 & -3.654 & 0.115 \\ 
        37 & HD\,46189 & 17010 & 1.8 & \citet{Paunzen2005} & 4.07 & 3.2 & 4.78E+29 & 1250 & 9.5 & -3.002 & 0.103 \\ 
        38 & HD\,54764 & 19000 & 5.3 & \citet{Lefever2007} & 18.77 & 9.7 & 1.58E+31 & 41374 & 28.7 & -6.802 & 0.311 \\ 
        39 & HD\,55856 & 21616 & 2.1 & \citet{Paunzen2005} & 5.78 & 3.6 & 2.52E+30 & 6571 & 11.0 & -4.804 & 0.119 \\ 
        40 & HD\,71155 & 9705 & 0.9 & \citet{Zorec2012} & 2.28 & 11.7 & 1.59E+28 & 41.6 & 23.7 & 0.691 & 0.257 \\ 
        41 & HD\,78556 & 10484 & 0.5 & \citet{Huang2010} & 5.02 & 2.0 & 1.05E+29 & 274 & 4.4 & -1.355 & 0.047 \\ 
        42 & HD\,82106 & 4836 & 0.3 & \citet{Soubiran2022} &  0.70 & 12.8 & 9.15E+25 & 0.239 & 25.6 & 6.294 & 0.278 \\ 
        43 & HD\,82621 & 8892 & 1.6 & \citet{Zorec2012} & 4.20 & 5.2 & 3.80E+28 & 99 & 12.2 & -0.252 & 0.133 \\ 
        44 & HD\,82943 & 5981 & 0.2 & \citet{Soubiran2022} & 1.18 & 5.9 & 6.10E+26 & 1.59 & 11.8 & 4.234 & 0.128 \\ 
        45 & HD\,84937 & 6484 & 1.6 & \citet{Soubiran2024} & 1.21 & 6.6 & 8.86E+26 & 2.31 & 14.7 & 3.829 & 0.159 \\ 
        46 & HD\,85503 & 4484 & 0.4 & \citet{Soubiran2022} & 12.03 & 9.6 & 2.02E+28 & 52.7 & 19.3 & 0.435 & 0.209 \\ 
        47 & HD\,89021 & 9280 & 2.2 & \citet{Hill1995} & 5.10 & 5.4 & 6.65E+28 & 174 & 13.9 & -0.859 & 0.150 \\ 
        48 & HD\,99922 & 8689 & 0.9 & \citet{Zorec2012} & 3.23 & 3.7 & 2.05E+28 & 53.6 & 8.2 & 0.417 & 0.089 \\ 
        49 & HD\,101364 & 5800 & 0.4 & \citet{Melendez2012} & 0.98 & 10.1 & 3.71E+26 & 0.969 & 20.3 & 4.774 & 0.221 \\ 
        50 & HD\,103095 & 5235 & 0.3 & \citet{Soubiran2024} & 0.55 & 14.3 & 7.93E+25 & 0.207 & 28.6 & 6.449 & 0.311 \\ 
        51 & HD\,113226 & 4950 & 0.2 & \citet{Soubiran2024} & 12.58 & 4.4 & 3.28E+28 & 86 & 8.8 & -0.091 & 0.096 \\ 
        52 & HD\,117176 & 5473 & 0.4 & \citet{Soubiran2024} & 1.91 & 4.1 & 1.13E+27 & 2.939 & 8.4 & 3.569 & 0.092 \\ 
        53 & HD\,120198 & 10174 & 1.0 & \citet{Espadons} & 2.10 & 5.6 & 1.64E+28 & 42.7 & 11.9 & 0.663 & 0.130 \\ 
        54 & HD\,122563 & 4642 & 0.8 & \citet{Soubiran2024} & 31.71 & 4.5 & 1.61E+29 & 421 & 9.5 & -1.820 & 0.103 \\ 
        55 & HD\,125924 & 21000 & 1.9 & \citet{Levenhagen2006} & 4.26 & 2.1 & 1.22E+30 & 3177 & 8.7 & -4.015 & 0.094 \\ 
        56 & HD\,128167 & 6575 & 1.0 & \citet{Soubiran2022} & 1.40 & 10.6 & 1.27E+27 & 3.314 & 21.5 & 3.439 & 0.233 \\ 
        57 & HD\,133208 & 5171 & 0.8 & \citet{dasilva2015} & 16.88 & 13.1 & 7.02E+28 & 183 & 26.3 & -0.919 & 0.286 \\ 
        58 & HD\,141714 & 5280 & 0.5 & \citet{Soubiran2022} & 7.21 & 4.8 & 1.39E+28 & 36.4 & 9.8 & 0.837 & 0.106 \\ 
        59 & HD\,146233 & 5824 & 0.5 & \citet{Soubiran2024} & 0.96 & 5.2 & 3.64E+26 & 0.952 & 10.5 & 4.793 & 0.114 \\ 
        60 & HD\,148379 & 17000 & 5.9 & \citet{Fraser2010} & 61.62 & 14.6 & 1.09E+32 & 285729 & 37.5 & -8.900 & 0.408 \\ 
        61 & HD\,148688 & 20700 & 4.8 & \citet{Fraser2010} & 39.52 & 12.5 & 9.89E+31 & 258350 & 31.6 & -8.791 & 0.343 \\ 
        62 & HD\,150117 & 10568 & 0.5 & \citet{Zorec2012} & 3.49 & 5.4 & 5.23E+28 & 137 & 10.9 & -0.599 & 0.119 \\ 
        63 & HD\,160762 & 18200 & 3.8 & \citet{Gordon2019} & 4.55 & 10.0 & 7.85E+29 & 2051 & 25.2 & -3.540 & 0.274 \\ 
        64 & HD\,162570 & 7396 & 0.7 & \citet{Zorec2012} & 3.03 & 4.6 & 9.49E+27 & 24.8 & 9.5 & 1.254 & 0.103 \\ 
        65 & HD\,171301 & 12153 & 1.6 & \citet{Adelman2017} & 2.24 & 7.0 & 3.78E+28 & 99 & 15.6 & -0.247 & 0.169 \\ 
        66 & HD\,174959 & 14681 & 1.4 & \citet{Cenarro2007} & 4.38 & 4.3 & 3.07E+29 & 802 & 10.2 & -2.520 & 0.110 \\ 
        67 & HD\,186427 & 5807 & 0.6 & \citet{Soubiran2024} & 1.18 & 8.4 & 5.42E+26 & 1.416 & 17.0 & 4.362 & 0.184 \\ 
        68 & HD\,189319 & 3904 & 0.8 & \citet{Soubiran2024} & 57.04 & 22.0 & 2.61E+29 & 681 & 44.1 & -2.343 & 0.478 \\ 
        69 & HD\,189957 & 32100 & 3.1 & \citet{Holgado2022} & 8.95 & 11.4 & 2.93E+31 & 76645 & 25.9 & -7.471 & 0.282 \\ 
        70 & HD\,192263 & 4967 & 0.3 & \citet{Soubiran2022} & 0.73 & 10.8 & 1.12E+26 & 0.293 & 21.6 & 6.072 & 0.235 \\ 
        71 & HD\,193183 & 20480 & 0.7 & \citet{Mugnes2015} & 30.01 & 10.5 & 5.46E+31 & 142698 & 21.2 & -8.146 & 0.230 \\ 
        72 & HD\,195556 & 17680 & 5.9 & \citet{Zorec2009} & 8.01 & 16.5 & 2.16E+30 & 5652 & 40.6 & -4.640 & 0.440 \\ 
        73 & HD\,196740 & 15427 & 1.8 & \citet{Wu2011} & 3.30 & 7.2 & 2.13E+29 & 556 & 16.1 & -2.122 & 0.175 \\ 
        74 & HD\,197512 & 23570 & 4.0 & \citet{Daflon2003} & 5.02 & 19.3 & 2.68E+30 & 7004 & 41.8 & -4.873 & 0.454 \\ 
        75 & HD\,201091 & 4398 & 0.8 & \citet{Soubiran2024} & 0.66 & 10.5 & 5.60E+25 & 0.146 & 21.2 & 6.826 & 0.230 \\ 
        76 & HD\,201092 & 4174 & 1.1 & \citet{Tabernero2022} & 0.53 & 24.1 & 2.98E+25 & 0.078 & 48.4 & 7.512 & 0.525 \\ 
        77 & HD\,205139 & 26810 & 7.9 & \citet{Wu2011} & 13.41 & 13.9 & 3.20E+31 & 83667 & 41.9 & -7.566 & 0.455 \\ 
        78 & HD\,207538 & 31400 & 1.6 & \citet{Aschenbrenner2023} & 6.18 & 20.0 & 1.28E+31 & 33395 & 40.5 & -6.569 & 0.440 \\ 
        79 & HD\,208266 & 24840 & 4.0 & \citet{Daflon2007} & 5.26 & 8.5 & 3.63E+30 & 9476 & 23.3 & -5.202 & 0.253 \\ 
        80 & HD\,209419 & 15020 & 1.5 & \citet{Mugnes2015} & 5.00 & 4.7 & 4.39E+29 & 1146 & 11.3 & -2.908 & 0.123 \\ 
        81 & HD\,209975 & 32983 & 4.3 & \citet{Prugniel2011} & 16.80 & 13.2 & 1.15E+32 & 300792 & 31.5 & -8.956 & 0.342 \\ 
        82 & HD\,213087 & 24870 & 0.5 & \citet{Mugnes2015} & 21.31 & 12.1 & 5.99E+31 & 156535 & 24.3 & -8.247 & 0.264 \\ 
        83 & HD\,214263 & 20860 & 0.7 & \citet{Mugnes2015} & 3.61 & 10.4 & 8.49E+29 & 2218 & 21.0 & -3.625 & 0.228 \\ 
        84 & HD\,214994 & 9209 & 8.1 & \citet{Hourihane2023} & 3.39 & 5.0 & 2.85E+28 & 74 & 34.1 & 0.061 & 0.370 \\ 
        85 & HD\,215191 & 22010 & 1.1 & \citet{Mugnes2015} & 4.05 & 7.3 & 1.33E+30 & 3469 & 15.3 & -4.111 & 0.166 \\ 
        86 & HD\,217014 & 5746 & 1.3 & \citet{Soubiran2024} & 1.13 & 5.3 & 4.77E+26 & 1.246 & 11.7 & 4.501 & 0.127 \\ 
        87 & HD\,222173 & 11800 & 4.2 & \citet{Bailey2013} & 5.57 & 2.3 & 2.08E+29 & 542 & 17.6 & -2.096 & 0.191 \\ 
        88 & HD\,222762 & 12828 & 0.8 & \citet{Huang2010} & 6.43 & 7.4 & 3.86E+29 & 1009 & 15.1 & -2.770 & 0.164 
        \label{tab:2}
\end{longtable}
}


\begin{figure}
    \centering
    \includegraphics[width=1\linewidth]{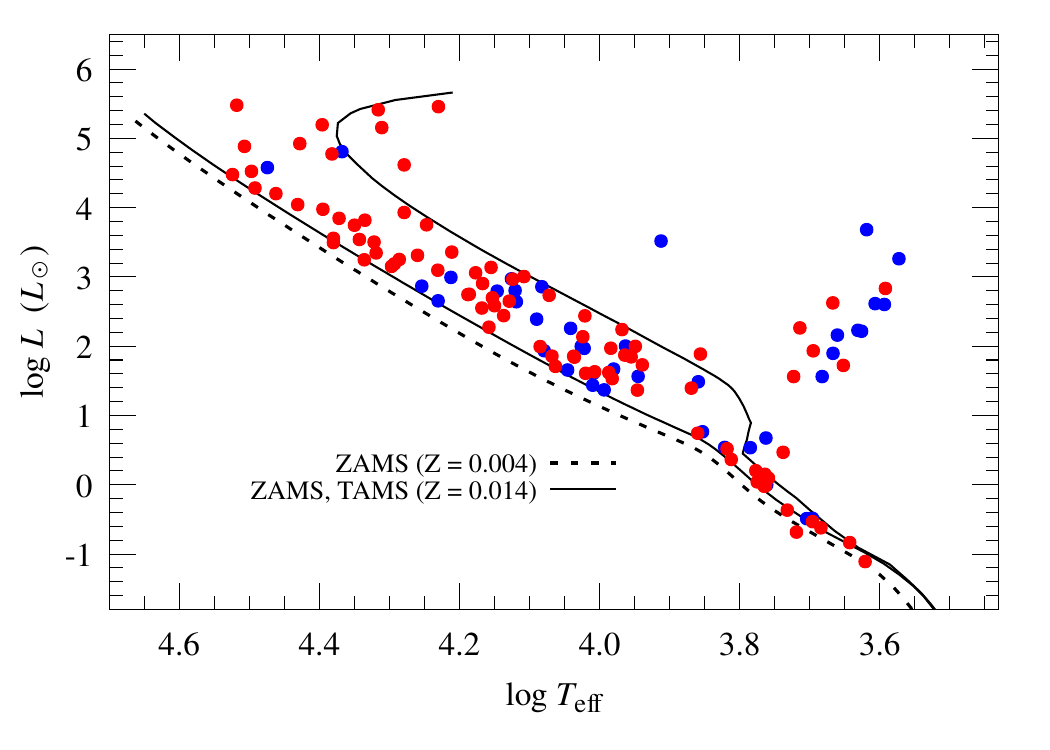}
    \caption{Distribution of the sample stars (filled red circles) on the H-R diagram. Filled blue circles are for the stars excluded from the list of \citetalias{Yucel2026}, because they do not have XP spectra. ZAMS and TAMS, according to PARSEC evolution models \citep{Bressan2012}.}
    \label{fig:hr}
\end{figure}

\subsection{Absolute Bolometric and Filtered Magnitudes}

Once, fractional luminosities (Table~\ref{speck}) are ready,  the absolute bolometric ($M_{\rm Bol}$) and filtered ($M_\xi$) magnitudes of the sample stars are needed to empirically determine the zero-point constants of the $BC_\xi$ scales for the first time in order to use Equation~(\ref{eq:8}) to calculate the spectroscopic $BC_\xi$. Absolute bolometric magnitudes from \citetalias{Yucel2026}, which were calculated from accurate stellar parameters ($R$, $T_{\rm eff}$) according to the Stefan-Boltzmann law, are presented in Table~\ref{tab:2} with their propagated uncertainties. The distribution of the sample stars used in this study in the H-R diagram is presented in Figure~\ref{fig:hr}, where the missing ones are also indicated to show a nearly uniform distribution of the sample covering almost all effective temperatures and radii, mainly main-sequence and some giants and subgiants. Therefore, the zero-point constants $C_2(\xi)$ would be assumed to have no bias, so that they would represent all stars on the H-R diagram.

Nevertheless, we had to calculate the absolute filtered magnitudes $M_\xi$ of the sample stars according to the following equation 
\begin{equation}
    M_\xi = \xi + 5\log{\varpi}+5-A_\xi,
    \label{eq:11}
\end{equation}
where the subscript $\xi$ indicates one of the $Gaia$ filters $G$, $G_{BP}$, and $G_{RP}$. However, the value indicated by $\xi$ is for the apparent magnitudes, $\varpi$ is the trigonometric parallax in units of arc seconds, and $A_\xi$ is interstellar extinction in magnitude units. We have used the same SED method of \citetalias{Yucel2026} to calculate $A_\xi$ as 
\begin{equation}
        A_\xi=2.5\times\log\frac{\int_{0}^{\infty} S_\lambda (\xi)f_\lambda^0 \,d\lambda}{\int_{0}^{\infty} S_\lambda (\xi)f_\lambda \,d\lambda},
        \label{eq:12}
\end{equation}
where $f_\lambda^0$ and $f_\lambda$ are monochromatic fluxes reaching above the Earth's atmosphere without and with interstellar extinctions, respectively, while $S_\lambda (\xi)$ is the transparency function for the filter $\xi$. Astrometric and photometric data for the calculation of absolute filtered magnitudes from the \textit{Gaia} DR3 database and the computed absolute filtered magnitudes corrected for interstellar extinctions are presented together with propagated uncertainties in Table~\ref{tab:3}.       


\begin{landscape}
\scriptsize
\setlength{\tabcolsep}{3pt}
\renewcommand{\arraystretch}{0.55}
\begin{longtable}{clr@{\hspace{0.4cm}}ccr@{\hspace{0.4cm}}ccr@{\hspace{0.4cm}}ccr}
\caption{Astrometric and photometric data with observational uncertainties for calculating the absolute filtered magnitudes and propagated errors.} \\
\toprule
Order & Star & \multicolumn{1}{c}{\,\,$\varpi$} & $G$ & $A_{\rm G}$  & \multicolumn{1}{c}{$M_{\rm G}$}  & $G_{\rm BP}$ & $A_{\rm G_{\rm BP}}$ & \multicolumn{1}{c}{$M_{\rm G_{\rm BP}}$}  & $G_{\rm RP}$  & $A_{\rm G_{\rm RP}}$  & \multicolumn{1}{c}{$M_{\rm G_{\rm RP}}$}  \\
&  & \multicolumn{1}{c}{(mas)} &  \multicolumn{1}{c}{(mag)} &  \multicolumn{1}{c}{(mag)} &   \multicolumn{1}{c}{(mag)} &  \multicolumn{1}{c}{(mag)} &   \multicolumn{1}{c}{(mag)} &   \multicolumn{1}{c}{(mag)} &  \multicolumn{1}{c}{(mag)} &  \multicolumn{1}{c}{(mag)} &  \multicolumn{1}{c}{(mag)} \\
\midrule
\endfirsthead

\multicolumn{12}{l}%
{{\normalsize{\bf\tablename\ \thetable{}} -- continued from previous page}} \\
\toprule
Order & Star & \multicolumn{1}{c}{\,\,$\varpi$} & $G$ & $A_{\rm G}$  & \multicolumn{1}{c}{$M_{\rm G}$}  & $G_{\rm BP}$ & $A_{\rm G_{\rm BP}}$ & \multicolumn{1}{c}{$M_{\rm G_{\rm BP}}$}  & $G_{\rm RP}$  & $A_{\rm G_{\rm RP}}$  & \multicolumn{1}{c}{$M_{\rm G_{\rm RP}}$}  \\
&  & \multicolumn{1}{c}{(mas)} &  \multicolumn{1}{c}{(mag)} &  \multicolumn{1}{c}{(mag)} &   \multicolumn{1}{c}{(mag)} &  \multicolumn{1}{c}{(mag)} &   \multicolumn{1}{c}{(mag)} &   \multicolumn{1}{c}{(mag)} &  \multicolumn{1}{c}{(mag)} &  \multicolumn{1}{c}{(mag)} &  \multicolumn{1}{c}{(mag)} \\
\midrule
\endhead

\midrule
\multicolumn{12}{r}{\normalsize{Continued on next page}} \\
\endfoot

\bottomrule
\endlastfoot
        01 & HD\,1279 & 2.8490 $\pm$ 0.0516 & 5.8352 $\pm$ 0.0028 & 0.0322 $\pm$ 0.0962 & -1.9200 $\pm$ 0.1039 & 5.7905 $\pm$ 0.0029 & 0.0369 $\pm$ 0.1102 & -1.9710 $\pm$ 0.1170 & 5.8697 $\pm$ 0.0040 & 0.0196 $\pm$ 0.0586 & -1.8757 $\pm$ 0.0707 \\ 
        02 & HD\,1404 & 23.2542 $\pm$ 0.1809 & 4.5015 $\pm$ 0.0011 & 0.0000 $\pm$ 0.0304 & 1.3340 $\pm$ 0.0348 & 4.5175 $\pm$ 0.0006 & 0.0000 $\pm$ 0.0356 & 1.3500 $\pm$ 0.0394 & 4.4224 $\pm$ 0.0012 & 0.0000 $\pm$ 0.0193 & 1.2549 $\pm$ 0.0257 \\ 
        03 & HD\,1439 & 6.6475 $\pm$ 0.0818 & 5.8741 $\pm$ 0.0005 & 0.0000 $\pm$ 0.0309 & -0.0126 $\pm$ 0.0409 & 5.8581 $\pm$ 0.0011 & 0.0000 $\pm$ 0.0359 & -0.0286 $\pm$ 0.0448 & 5.8649 $\pm$ 0.0015 & 0.0000 $\pm$ 0.0194 & -0.0219 $\pm$ 0.0331 \\ 
        04 & HD\,2729 & 3.8917 $\pm$ 0.0389 & 6.1576 $\pm$ 0.0004 & 0.0000 $\pm$ 0.0970 & -0.8917 $\pm$ 0.0994 & 6.1014 $\pm$ 0.0012 & 0.0000 $\pm$ 0.1108 & -0.9479 $\pm$ 0.1129 & 6.2255 $\pm$ 0.0008 & 0.0000 $\pm$ 0.0587 & -0.8238 $\pm$ 0.0626 \\ 
        05 & HD\,4778 & 9.6184 $\pm$ 0.0488 & 6.1255 $\pm$ 0.0007 & 0.0000 $\pm$ 0.0617 & 1.0410 $\pm$ 0.0627 & 6.1264 $\pm$ 0.0013 & 0.0000 $\pm$ 0.0718 & 1.0419 $\pm$ 0.0727 & 6.0977 $\pm$ 0.0031 & 0.0000 $\pm$ 0.0388 & 1.0132 $\pm$ 0.0405 \\ 
        06 & HD\,10362 & 2.4172 $\pm$ 0.0331 & 6.2951 $\pm$ 0.0002 & 0.4159 $\pm$ 0.2802 & -2.1617 $\pm$ 0.2818 & 6.3028 $\pm$ 0.0007 & 0.4775 $\pm$ 0.3257 & -2.2349 $\pm$ 0.3271 & 6.2260 $\pm$ 0.0006 & 0.2538 $\pm$ 0.1746 & -2.1025 $\pm$ 0.1771 \\ 
        07 & HD\,10700 & 273.8097 $\pm$ 0.1701 & 3.3004 $\pm$ 0.0030 & 0.0000 $\pm$ 0.1346 & 5.4876 $\pm$ 0.1346 & 3.7995 $\pm$ 0.0113 & 0.0000 $\pm$ 0.1659 & 5.9868 $\pm$ 0.1663 & 2.7483 $\pm$ 0.0053 & 0.0000 $\pm$ 0.0942 & 4.9355 $\pm$ 0.0944 \\ 
        08 & HD\,15318 & 16.5681 $\pm$ 0.4608 & 4.2554 $\pm$ 0.0025 & 0.0000 $\pm$ 0.0315 & 0.3517 $\pm$ 0.0682 & 4.2397 $\pm$ 0.0017 & 0.0000 $\pm$ 0.0363 & 0.3361 $\pm$ 0.0705 & 4.2806 $\pm$ 0.0051 & 0.0000 $\pm$ 0.0195 & 0.3769 $\pm$ 0.0637 \\ 
        09 & HD\,16440 & 1.4404 $\pm$ 0.0255 & 7.7021 $\pm$ 0.0003 & 2.2745 $\pm$ 0.1907 & -3.6595 $\pm$ 0.1945 & 8.0907 $\pm$ 0.0012 & 2.6799 $\pm$ 0.2363 & -3.7374 $\pm$ 0.2394 & 7.1108 $\pm$ 0.0010 & 1.4449 $\pm$ 0.1316 & -3.5105 $\pm$ 0.1371 \\ 
        10 & HD\,19736 & 4.5027 $\pm$ 0.0584 & 6.1280 $\pm$ 0.0006 & 0.2582 $\pm$ 0.0955 & -0.8344 $\pm$ 0.0996 & 6.0646 $\pm$ 0.0010 & 0.2953 $\pm$ 0.1100 & -0.9475 $\pm$ 0.1136 & 6.2028 $\pm$ 0.0013 & 0.1565 $\pm$ 0.0585 & -0.6806 $\pm$ 0.0649 \\ 
        11 & HD\,22879 & 38.3253 $\pm$ 0.0308 & 6.5358 $\pm$ 0.0028 & 0.0000 $\pm$ 0.0556 & 4.4532 $\pm$ 0.0557 & 6.8216 $\pm$ 0.0029 & 0.0000 $\pm$ 0.0676 & 4.7391 $\pm$ 0.0677 & 6.0750 $\pm$ 0.0039 & 0.0000 $\pm$ 0.0380 & 3.9924 $\pm$ 0.0382 \\ 
        12 & HD\,23300 & 4.6643 $\pm$ 0.0705 & 5.6315 $\pm$ 0.0008 & 0.0971 $\pm$ 0.1283 & -1.1107 $\pm$ 0.1325 & 5.5921 $\pm$ 0.0021 & 0.1109 $\pm$ 0.1471 & -1.1688 $\pm$ 0.1507 & 5.6686 $\pm$ 0.0025 & 0.0587 $\pm$ 0.0782 & -1.0440 $\pm$ 0.0848 \\ 
        13 & HD\,27295 & 11.9300 $\pm$ 0.0843 & 5.4806 $\pm$ 0.0005 & 0.0000 $\pm$ 0.0633 & 0.8638 $\pm$ 0.0651 & 5.4408 $\pm$ 0.0008 & 0.0000 $\pm$ 0.0729 & 0.8240 $\pm$ 0.0745 & 5.5324 $\pm$ 0.0019 & 0.0000 $\pm$ 0.0390 & 0.9156 $\pm$ 0.0420 \\ 
        14 & HD\,27563 & 3.7763 $\pm$ 0.0612 & 5.8299 $\pm$ 0.0007 & 0.0000 $\pm$ 0.0965 & -1.2847 $\pm$ 0.1027 & 5.7619 $\pm$ 0.0019 & 0.0000 $\pm$ 0.1105 & -1.3528 $\pm$ 0.1160 & 5.9192 $\pm$ 0.0031 & 0.0000 $\pm$ 0.0587 & -1.1955 $\pm$ 0.0685 \\ 
        15 & HD\,27778 & 4.7132 $\pm$ 0.0281 & 6.2934 $\pm$ 0.0004 & 0.8547 $\pm$ 0.2123 & -1.1155 $\pm$ 0.2127 & 6.3787 $\pm$ 0.0015 & 0.9864 $\pm$ 0.2499 & -1.1981 $\pm$ 0.2502 & 6.0896 $\pm$ 0.0014 & 0.5257 $\pm$ 0.1350 & -1.0527 $\pm$ 0.1356 \\ 
        16 & HD\,29138 & 0.4331 $\pm$ 0.0349 & 7.1554 $\pm$ 0.0005 & 0.4640 $\pm$ 0.0650 & -5.0637 $\pm$ 0.1867 & 7.0978 $\pm$ 0.0016 & 0.5257 $\pm$ 0.0743 & -5.2084 $\pm$ 0.1901 & 7.2052 $\pm$ 0.0012 & 0.2752 $\pm$ 0.0392 & -4.8758 $\pm$ 0.1793 \\ 
        17 & HD\,29335 & 7.0300 $\pm$ 0.3300 & 5.3124 $\pm$ 0.0012 & 0.1608 $\pm$ 0.1273 & -0.5965 $\pm$ 0.1631 & 5.2325 $\pm$ 0.0009 & 0.1842 $\pm$ 0.1464 & -0.7074 $\pm$ 0.1784 & 5.3852 $\pm$ 0.0015 & 0.0978 $\pm$ 0.0780 & -0.4742 $\pm$ 0.1284 \\ 
        18 & HD\,29589 & 8.2529 $\pm$ 0.1133 & 5.4138 $\pm$ 0.0007 & 0.0649 $\pm$ 0.0967 & -0.0606 $\pm$ 0.1012 & 5.3604 $\pm$ 0.0009 & 0.0740 $\pm$ 0.1107 & -0.1264 $\pm$ 0.1146 & 5.5033 $\pm$ 0.0016 & 0.0392 $\pm$ 0.0587 & 0.0487 $\pm$ 0.0659 \\ 
        19 & HD\,32040 & 5.8775 $\pm$ 0.0974 & 6.6405 $\pm$ 0.0003 & 0.0000 $\pm$ 0.0634 & 0.4864 $\pm$ 0.0729 & 6.5976 $\pm$ 0.0010 & 0.0000 $\pm$ 0.0730 & 0.4436 $\pm$ 0.0814 & 6.6815 $\pm$ 0.0007 & 0.0000 $\pm$ 0.0390 & 0.5275 $\pm$ 0.0531 \\ 
        20 & HD\,32115 & 20.3757 $\pm$ 0.0307 & 6.2522 $\pm$ 0.0002 & 0.0000 $\pm$ 0.0293 & 2.7977 $\pm$ 0.0295 & 6.3863 $\pm$ 0.0006 & 0.0000 $\pm$ 0.0348 & 2.9319 $\pm$ 0.0350 & 5.9910 $\pm$ 0.0006 & 0.0000 $\pm$ 0.0192 & 2.5366 $\pm$ 0.0195 \\ 
        21 & HD\,32309 & 16.5691 $\pm$ 0.0866 & 4.8712 $\pm$ 0.0007 & 0.0000 $\pm$ 0.0625 & 0.9676 $\pm$ 0.0635 & 4.8563 $\pm$ 0.0008 & 0.0000 $\pm$ 0.0723 & 0.9528 $\pm$ 0.0732 & 4.9002 $\pm$ 0.0024 & 0.0000 $\pm$ 0.0389 & 0.9967 $\pm$ 0.0406 \\ 
        22 & HD\,32612 & 2.4620 $\pm$ 0.0546 & 6.3798 $\pm$ 0.0005 & 0.0664 $\pm$ 0.0332 & -1.7212 $\pm$ 0.0585 & 6.2748 $\pm$ 0.0007 & 0.0751 $\pm$ 0.0375 & -1.8386 $\pm$ 0.0610 & 6.5529 $\pm$ 0.0006 & 0.0394 $\pm$ 0.0196 & -1.5284 $\pm$ 0.0520 \\ 
        23 & HD\,34078 & 2.5740 $\pm$ 0.0340 & 5.9397 $\pm$ 0.0018 & 1.3299 $\pm$ 0.0615 & -3.1848 $\pm$ 0.0679 & 6.0846 $\pm$ 0.0056 & 1.5223 $\pm$ 0.0721 & -3.2949 $\pm$ 0.0778 & 5.6447 $\pm$ 0.0036 & 0.8017 $\pm$ 0.0386 & -3.0751 $\pm$ 0.0482 \\ 
        24 & HD\,34310 & 11.9207 $\pm$ 0.0639 & 5.0412 $\pm$ 0.0007 & 0.0000 $\pm$ 0.0628 & 0.4227 $\pm$ 0.0639 & 5.0075 $\pm$ 0.0010 & 0.0000 $\pm$ 0.0726 & 0.3890 $\pm$ 0.0735 & 5.0982 $\pm$ 0.0016 & 0.0000 $\pm$ 0.0389 & 0.4797 $\pm$ 0.0406 \\ 
        25 & HD\,35299 & 2.7732 $\pm$ 0.0915 & 5.6543 $\pm$ 0.0006 & 0.0672 $\pm$ 0.1002 & -2.1883 $\pm$ 0.1232 & 5.5374 $\pm$ 0.0008 & 0.0757 $\pm$ 0.1131 & -2.3176 $\pm$ 0.1339 & 5.8460 $\pm$ 0.0012 & 0.0394 $\pm$ 0.0591 & -1.9768 $\pm$ 0.0929 \\ 
        26 & HD\,35693 & 6.6047 $\pm$ 0.0467 & 6.1635 $\pm$ 0.0003 & 0.0000 $\pm$ 0.0306 & 0.2628 $\pm$ 0.0342 & 6.1952 $\pm$ 0.0012 & 0.0000 $\pm$ 0.0357 & 0.2945 $\pm$ 0.0389 & 6.0689 $\pm$ 0.0019 & 0.0000 $\pm$ 0.0194 & 0.1682 $\pm$ 0.0248 \\ 
        27 & HD\,35912 & 2.7229 $\pm$ 0.0681 & 6.3665 $\pm$ 0.0005 & 0.1986 $\pm$ 0.0982 & -1.6307 $\pm$ 0.1123 & 6.2635 $\pm$ 0.0010 & 0.2249 $\pm$ 0.1119 & -1.7709 $\pm$ 0.1244 & 6.5235 $\pm$ 0.0014 & 0.1180 $\pm$ 0.0588 & -1.4144 $\pm$ 0.0801 \\ 
        28 & HD\,36285 & 2.8252 $\pm$ 0.0648 & 6.2967 $\pm$ 0.0004 & 0.2326 $\pm$ 0.0658 & -1.6491 $\pm$ 0.0825 & 6.1850 $\pm$ 0.0011 & 0.2631 $\pm$ 0.0748 & -1.8043 $\pm$ 0.0899 & 6.4809 $\pm$ 0.0008 & 0.1377 $\pm$ 0.0393 & -1.3958 $\pm$ 0.0634 \\ 
        29 & HD\,36430 & 2.5173 $\pm$ 0.0700 & 6.1883 $\pm$ 0.0004 & 0.0664 $\pm$ 0.0661 & -1.8645 $\pm$ 0.0895 & 6.0880 $\pm$ 0.0009 & 0.0751 $\pm$ 0.0749 & -1.9772 $\pm$ 0.0962 & 6.3465 $\pm$ 0.0009 & 0.0394 $\pm$ 0.0393 & -1.6865 $\pm$ 0.0721 \\ 
        30 & HD\,36512 & 2.4567 $\pm$ 0.1403 & 4.5462 $\pm$ 0.0012 & 0.0000 $\pm$ 0.0341 & -3.5020 $\pm$ 0.1286 & 4.4143 $\pm$ 0.0008 & 0.0000 $\pm$ 0.0382 & -3.6339 $\pm$ 0.1298 & 4.8075 $\pm$ 0.0022 & 0.0000 $\pm$ 0.0198 & -3.2408 $\pm$ 0.1256 \\ 
        31 & HD\,36591 & 2.3510 $\pm$ 0.1828 & 5.3038 $\pm$ 0.0010 & 0.1683 $\pm$ 0.1664 & -2.9836 $\pm$ 0.2371 & 5.2001 $\pm$ 0.0014 & 0.1894 $\pm$ 0.1883 & -3.1183 $\pm$ 0.2529 & 5.4816 $\pm$ 0.0028 & 0.0986 $\pm$ 0.0984 & -2.7564 $\pm$ 0.1954 \\ 
        32 & HD\,36960 & 2.6166 $\pm$ 0.1203 & 4.7165 $\pm$ 0.0010 & 0.0339 $\pm$ 0.0676 & -3.2235 $\pm$ 0.1206 & 4.5911 $\pm$ 0.0008 & 0.0381 $\pm$ 0.0759 & -3.3551 $\pm$ 0.1254 & 4.9639 $\pm$ 0.0016 & 0.0198 $\pm$ 0.0395 & -2.9662 $\pm$ 0.1074 \\ 
        33 & HD\,37077 & 8.7664 $\pm$ 0.0791 & 5.1747 $\pm$ 0.0006 & 0.0000 $\pm$ 0.0292 & -0.1112 $\pm$ 0.0352 & 5.3149 $\pm$ 0.0008 & 0.0000 $\pm$ 0.0347 & 0.0290 $\pm$ 0.0399 & 4.9314 $\pm$ 0.0013 & 0.0000 $\pm$ 0.0192 & -0.3545 $\pm$ 0.0275 \\ 
        34 & HD\,37356 & 2.1972 $\pm$ 0.0555 & 6.1585 $\pm$ 0.0003 & 0.6876 $\pm$ 0.0636 & -2.7353 $\pm$ 0.0840 & 6.1293 $\pm$ 0.0009 & 0.7830 $\pm$ 0.0734 & -2.8951 $\pm$ 0.0916 & 6.1504 $\pm$ 0.0007 & 0.4117 $\pm$ 0.0390 & -2.5361 $\pm$ 0.0673 \\ 
        35 & HD\,37744 & 2.5199 $\pm$ 0.0674 & 6.1772 $\pm$ 0.0004 & 0.1674 $\pm$ 0.0995 & -1.9594 $\pm$ 0.1152 & 6.0612 $\pm$ 0.0013 & 0.1888 $\pm$ 0.1127 & -2.1066 $\pm$ 0.1268 & 6.3614 $\pm$ 0.0008 & 0.0985 $\pm$ 0.0590 & -1.7259 $\pm$ 0.0828 \\ 
        36 & HD\,40967 & 3.4868 $\pm$ 0.1014 & 4.9612 $\pm$ 0.0010 & 0.1633 $\pm$ 0.1293 & -2.4702 $\pm$ 0.1439 & 4.8940 $\pm$ 0.0023 & 0.1859 $\pm$ 0.1479 & -2.5685 $\pm$ 0.1608 & 5.0205 $\pm$ 0.0035 & 0.0981 $\pm$ 0.0782 & -2.3616 $\pm$ 0.1006 \\ 
        37 & HD\,46189 & 2.9719 $\pm$ 0.0664 & 5.8876 $\pm$ 0.0005 & 0.0000 $\pm$ 0.0986 & -1.7472 $\pm$ 0.1099 & 5.8021 $\pm$ 0.0010 & 0.0000 $\pm$ 0.1119 & -1.8327 $\pm$ 0.1220 & 6.0195 $\pm$ 0.0017 & 0.0000 $\pm$ 0.0589 & -1.6153 $\pm$ 0.0763 \\ 
        38 & HD\,54764 & 0.7036 $\pm$ 0.0494 & 5.9994 $\pm$ 0.0006 & 0.5857 $\pm$ 0.3144 & -5.2809 $\pm$ 0.3495 & 6.0062 $\pm$ 0.0013 & 0.6684 $\pm$ 0.3644 & -5.3861 $\pm$ 0.3950 & 5.9217 $\pm$ 0.0015 & 0.3525 $\pm$ 0.1943 & -5.1810 $\pm$ 0.2470 \\ 
        39 & HD\,55856 & 1.3890 $\pm$ 0.0537 & 6.3312 $\pm$ 0.0003 & 0.1001 $\pm$ 0.0995 & -3.0415 $\pm$ 0.1302 & 6.2219 $\pm$ 0.0008 & 0.1130 $\pm$ 0.1127 & -3.1694 $\pm$ 0.1405 & 6.5070 $\pm$ 0.0006 & 0.0591 $\pm$ 0.0590 & -2.8360 $\pm$ 0.1026 \\ 
        40 & HD\,71155 & 25.3438 $\pm$ 0.2051 & 3.9023 $\pm$ 0.0018 & 0.0000 $\pm$ 0.1232 & 0.9216 $\pm$ 0.1245 & 3.8787 $\pm$ 0.0021 & 0.0000 $\pm$ 0.1435 & 0.8981 $\pm$ 0.1446 & 3.8679 $\pm$ 0.0023 & 0.0000 $\pm$ 0.0775 & 0.8872 $\pm$ 0.0795 \\ 
        41 & HD\,78556 & 4.8717 $\pm$ 0.1255 & 5.6115 $\pm$ 0.0006 & 0.0313 $\pm$ 0.0934 & -0.9788 $\pm$ 0.1089 & 5.5654 $\pm$ 0.0016 & 0.0362 $\pm$ 0.1083 & -1.0311 $\pm$ 0.1219 & 5.6056 $\pm$ 0.0031 & 0.0194 $\pm$ 0.0583 & -0.9748 $\pm$ 0.0809 \\ 
        42 & HD\,82106 & 78.1945 $\pm$ 0.0215 & 6.8855 $\pm$ 0.0028 & 0.0000 $\pm$ 0.2330 & 6.3514 $\pm$ 0.2330 & 7.4136 $\pm$ 0.0029 & 0.0000 $\pm$ 0.2923 & 6.8794 $\pm$ 0.2923 & 6.2027 $\pm$ 0.0038 & 0.0000 $\pm$ 0.1680 & 5.6686 $\pm$ 0.1680 \\ 
        43 & HD\,82621 & 12.3906 $\pm$ 0.1107 & 4.4408 $\pm$ 0.0009 & 0.0000 $\pm$ 0.0305 & -0.0937 $\pm$ 0.0362 & 4.4722 $\pm$ 0.0007 & 0.0000 $\pm$ 0.0356 & -0.0623 $\pm$ 0.0405 & 4.3925 $\pm$ 0.0021 & 0.0000 $\pm$ 0.0193 & -0.1420 $\pm$ 0.0274 \\ 
        44 & HD\,82943 & 36.1158 $\pm$ 0.0264 & 6.4022 $\pm$ 0.0028 & 0.0000 $\pm$ 0.1111 & 4.1907 $\pm$ 0.1111 & 6.7006 $\pm$ 0.0029 & 0.0000 $\pm$ 0.1351 & 4.4891 $\pm$ 0.1351 & 5.9357 $\pm$ 0.0039 & 0.0000 $\pm$ 0.0759 & 3.7242 $\pm$ 0.0760 \\ 
        45 & HD\,84937 & 13.4978 $\pm$ 0.0444 & 8.2067 $\pm$ 0.0028 & 0.0569 $\pm$ 0.1129 & 3.8006 $\pm$ 0.1132 & 8.4229 $\pm$ 0.0028 & 0.0685 $\pm$ 0.1363 & 4.0044 $\pm$ 0.1365 & 7.8167 $\pm$ 0.0038 & 0.0381 $\pm$ 0.0762 & 3.4303 $\pm$ 0.0766 \\ 
        46 & HD\,85503 & 26.0971 $\pm$ 0.1974 & 3.5302 $\pm$ 0.0029 & 0.0000 $\pm$ 0.1268 & 0.6131 $\pm$ 0.1279 & 4.2057 $\pm$ 0.0081 & 0.0000 $\pm$ 0.1605 & 1.2887 $\pm$ 0.1615 & 2.8111 $\pm$ 0.0055 & 0.0000 $\pm$ 0.0929 & -0.1059 $\pm$ 0.0945 \\ 
        47 & HD\,89021 & 14.4594 $\pm$ 1.0195 & 3.4462 $\pm$ 0.0035 & 0.0000 $\pm$ 0.0307 & -0.7530 $\pm$ 0.1562 & 3.5140 $\pm$ 0.0126 & 0.0000 $\pm$ 0.0358 & -0.6852 $\pm$ 0.1577 & 3.3666 $\pm$ 0.0046 & 0.0000 $\pm$ 0.0194 & -0.8326 $\pm$ 0.1544 \\ 
        48 & HD\,99922 & 8.6451 $\pm$ 0.0395 & 5.8188 $\pm$ 0.0004 & 0.0000 $\pm$ 0.0606 & 0.5026 $\pm$ 0.0614 & 5.8259 $\pm$ 0.0009 & 0.0000 $\pm$ 0.0710 & 0.5098 $\pm$ 0.0717 & 5.7724 $\pm$ 0.0019 & 0.0000 $\pm$ 0.0387 & 0.4563 $\pm$ 0.0400 \\ 
        49 & HD\,101364 & 16.7935 $\pm$ 0.0139 & 8.5225 $\pm$ 0.0028 & 0.0000 $\pm$ 0.0827 & 4.6482 $\pm$ 0.0828 & 8.8407 $\pm$ 0.0028 & 0.0000 $\pm$ 0.1009 & 4.9664 $\pm$ 0.1010 & 8.0326 $\pm$ 0.0038 & 0.0000 $\pm$ 0.0568 & 4.1583 $\pm$ 0.0570 \\ 
        50 & HD\,103095 & 109.0296 $\pm$ 0.0197 & 6.1985 $\pm$ 0.0028 & 0.0000 $\pm$ 0.1334 & 6.3862 $\pm$ 0.1334 & 6.6074 $\pm$ 0.0029 & 0.0000 $\pm$ 0.1651 & 6.7951 $\pm$ 0.1651 & 5.6058 $\pm$ 0.0039 & 0.0000 $\pm$ 0.0940 & 5.7935 $\pm$ 0.0941 \\ 
        51 & HD\,113226 & 30.2110 $\pm$ 0.1907 & 2.6191 $\pm$ 0.0032 & 0.0000 $\pm$ 0.1050 & 0.0199 $\pm$ 0.1059 & 3.4063 $\pm$ 0.0227 & 0.0000 $\pm$ 0.1309 & 0.8071 $\pm$ 0.1336 & 2.2496 $\pm$ 0.0141 & 0.0000 $\pm$ 0.0749 & -0.3496 $\pm$ 0.0774 \\ 
        52 & HD\,117176 & 55.2511 $\pm$ 0.0779 & 4.7790 $\pm$ 0.0028 & 0.0000 $\pm$ 0.1621 & 3.4907 $\pm$ 0.1622 & 5.1445 $\pm$ 0.0029 & 0.0000 $\pm$ 0.1995 & 3.8562 $\pm$ 0.1995 & 4.2316 $\pm$ 0.0049 & 0.0000 $\pm$ 0.1131 & 2.9433 $\pm$ 0.1132 \\ 
        53 & HD\,120198 & 10.7417 $\pm$ 0.0418 & 5.6709 $\pm$ 0.0005 & 0.0000 $\pm$ 0.0312 & 0.8262 $\pm$ 0.0323 & 5.6488 $\pm$ 0.0012 & 0.0000 $\pm$ 0.0361 & 0.8042 $\pm$ 0.0371 & 5.6854 $\pm$ 0.0015 & 0.0000 $\pm$ 0.0194 & 0.8407 $\pm$ 0.0212 \\ 
        54 & HD\,122563 & 3.0991 $\pm$ 0.0332 & 5.8750 $\pm$ 0.0028 & 0.1027 $\pm$ 0.1018 & -1.7838 $\pm$ 0.1044 & 6.3975 $\pm$ 0.0029 & 0.1293 $\pm$ 0.1287 & -1.2861 $\pm$ 0.1308 & 5.1787 $\pm$ 0.0039 & 0.0745 $\pm$ 0.0744 & -2.4406 $\pm$ 0.0781 \\ 
        55 & HD\,125924 & 0.3910 $\pm$ 0.0573 & 9.6133 $\pm$ 0.0005 & 0.0000 $\pm$ 0.0334 & -2.4258 $\pm$ 0.3200 & 9.5006 $\pm$ 0.0012 & 0.0000 $\pm$ 0.0377 & -2.5385 $\pm$ 0.3205 & 9.8146 $\pm$ 0.0005 & 0.0000 $\pm$ 0.0197 & -2.2245 $\pm$ 0.3188 \\ 
        56 & HD\,128167 & 63.4679 $\pm$ 0.1173 & 4.3420 $\pm$ 0.0028 & 0.0000 $\pm$ 0.1139 & 3.3547 $\pm$ 0.1140 & 4.5514 $\pm$ 0.0028 & 0.0000 $\pm$ 0.1370 & 3.5642 $\pm$ 0.1371 & 4.0114 $\pm$ 0.0045 & 0.0000 $\pm$ 0.0763 & 3.0241 $\pm$ 0.0765 \\ 
        57 & HD\,133208 & 13.8780 $\pm$ 0.1310 & 3.2450 $\pm$ 0.0010 & 0.0080 $\pm$ 0.3420 & -1.0520 $\pm$ 0.3426 & 3.8407 $\pm$ 0.0118 & 0.0099 $\pm$ 0.4261 & -0.4581 $\pm$ 0.4268 & 2.6520 $\pm$ 0.0072 & 0.0056 $\pm$ 0.2436 & -1.6420 $\pm$ 0.2446 \\ 
        58 & HD\,141714 & 19.4965 $\pm$ 0.0950 & 4.3783 $\pm$ 0.0005 & 0.0537 $\pm$ 0.0800 & 0.7706 $\pm$ 0.0807 & 4.7830 $\pm$ 0.0005 & 0.0663 $\pm$ 0.0990 & 1.1629 $\pm$ 0.0996 & 3.7979 $\pm$ 0.0013 & 0.0377 $\pm$ 0.0564 & 0.2100 $\pm$ 0.0574 \\ 
        59 & HD\,146233 & 70.7371 $\pm$ 0.0631 & 5.3277 $\pm$ 0.0028 & 0.0000 $\pm$ 0.0553 & 4.5759 $\pm$ 0.0554 & 5.6549 $\pm$ 0.0029 & 0.0000 $\pm$ 0.0673 & 4.9032 $\pm$ 0.0674 & 4.8379 $\pm$ 0.0057 & 0.0000 $\pm$ 0.0379 & 4.0861 $\pm$ 0.0384 \\ 
        60 & HD\,148379 & 0.6505 $\pm$ 0.0784 & 5.1298 $\pm$ 0.0013 & 2.1463 $\pm$ 0.3255 & -7.8431 $\pm$ 0.4177 & 5.4739 $\pm$ 0.0032 & 2.5312 $\pm$ 0.4040 & -7.9406 $\pm$ 0.4814 & 4.6032 $\pm$ 0.0035 & 1.3672 $\pm$ 0.2255 & -7.6689 $\pm$ 0.3455 \\ 
        61 & HD\,148688 & 0.7734 $\pm$ 0.1047 & 5.1761 $\pm$ 0.0015 & 1.9503 $\pm$ 0.5829 & -7.1912 $\pm$ 0.6528 & 5.3655 $\pm$ 0.0023 & 2.2772 $\pm$ 0.7162 & -7.3937 $\pm$ 0.7742 & 4.7866 $\pm$ 0.0077 & 1.2193 $\pm$ 0.3965 & -6.9589 $\pm$ 0.4936 \\ 
        62 & HD\,150117 & 7.8893 $\pm$ 0.1010 & 5.4018 $\pm$ 0.0008 & 0.0157 $\pm$ 0.0936 & -0.1273 $\pm$ 0.0977 & 5.3645 $\pm$ 0.0011 & 0.0181 $\pm$ 0.1085 & -0.1678 $\pm$ 0.1120 & 5.4409 $\pm$ 0.0022 & 0.0097 $\pm$ 0.0583 & -0.0833 $\pm$ 0.0646 \\ 
        63 & HD\,160762 & 6.5157 $\pm$ 0.3593 & 3.7906 $\pm$ 0.0028 & 0.0000 $\pm$ 0.1647 & -2.1395 $\pm$ 0.2036 & 3.6918 $\pm$ 0.0030 & 0.0000 $\pm$ 0.1869 & -2.2384 $\pm$ 0.2220 & 3.9072 $\pm$ 0.0018 & 0.0000 $\pm$ 0.0982 & -2.0230 $\pm$ 0.1549 \\ 
        64 & HD\,162570 & 10.6236 $\pm$ 0.0208 & 6.0869 $\pm$ 0.0002 & 0.0000 $\pm$ 0.0294 & 1.2183 $\pm$ 0.0297 & 6.2042 $\pm$ 0.0006 & 0.0000 $\pm$ 0.0349 & 1.3356 $\pm$ 0.0352 & 5.8508 $\pm$ 0.0008 & 0.0000 $\pm$ 0.0192 & 0.9822 $\pm$ 0.0197 \\ 
        65 & HD\,171301 & 9.3330 $\pm$ 0.0644 & 5.4560 $\pm$ 0.0008 & 0.0000 $\pm$ 0.0955 & 0.3061 $\pm$ 0.0967 & 5.4121 $\pm$ 0.0011 & 0.0000 $\pm$ 0.1097 & 0.2622 $\pm$ 0.1107 & 5.5049 $\pm$ 0.0024 & 0.0000 $\pm$ 0.0585 & 0.3550 $\pm$ 0.0604 \\ 
        66 & HD\,174959 & 2.9777 $\pm$ 0.0458 & 6.0769 $\pm$ 0.0007 & 0.0325 $\pm$ 0.0972 & -1.5824 $\pm$ 0.1027 & 6.0195 $\pm$ 0.0021 & 0.0371 $\pm$ 0.1109 & -1.6460 $\pm$ 0.1158 & 6.1465 $\pm$ 0.0019 & 0.0196 $\pm$ 0.0587 & -1.5029 $\pm$ 0.0676 \\ 
        67 & HD\,186427 & 47.3302 $\pm$ 0.0171 & 6.0734 $\pm$ 0.0028 & 0.0000 $\pm$ 0.1102 & 4.4491 $\pm$ 0.1102 & 6.4007 $\pm$ 0.0029 & 0.0000 $\pm$ 0.1345 & 4.7764 $\pm$ 0.1345 & 5.5746 $\pm$ 0.0039 & 0.0000 $\pm$ 0.0758 & 3.9503 $\pm$ 0.0759 \\ 
        68 & HD\,189319 & 11.3375 $\pm$ 0.3583 & 2.8594 $\pm$ 0.0015 & 0.2399 $\pm$ 0.2813 & -2.1552 $\pm$ 0.2896 & 3.9475 $\pm$ 0.0159 & 0.3116 $\pm$ 0.3697 & -1.1293 $\pm$ 0.3764 & 2.2505 $\pm$ 0.0167 & 0.1830 $\pm$ 0.2181 & -2.6654 $\pm$ 0.2292 \\ 
        69 & HD\,189957 & 0.4730 $\pm$ 0.0243 & 7.7510 $\pm$ 0.0005 & 0.8917 $\pm$ 0.3779 & -4.6501 $\pm$ 0.3940 & 7.7430 $\pm$ 0.0009 & 1.0126 $\pm$ 0.4384 & -4.8260 $\pm$ 0.4524 & 7.6951 $\pm$ 0.0007 & 0.5302 $\pm$ 0.2330 & -4.4396 $\pm$ 0.2583 \\ 
        70 & HD\,192263 & 50.9432 $\pm$ 0.0230 & 7.5130 $\pm$ 0.0028 & 0.0000 $\pm$ 0.2092 & 6.0484 $\pm$ 0.2092 & 7.9979 $\pm$ 0.0038 & 0.0000 $\pm$ 0.2612 & 6.5333 $\pm$ 0.2612 & 6.8673 $\pm$ 0.0043 & 0.0000 $\pm$ 0.1497 & 5.4028 $\pm$ 0.1498 \\ 
        71 & HD\,193183 & 0.4649 $\pm$ 0.0373 & 6.8625 $\pm$ 0.0005 & 1.8055 $\pm$ 0.2287 & -6.4665 $\pm$ 0.2875 & 7.0921 $\pm$ 0.0017 & 2.1028 $\pm$ 0.2770 & -6.5976 $\pm$ 0.3272 & 6.4514 $\pm$ 0.0023 & 1.1240 $\pm$ 0.1521 & -6.3051 $\pm$ 0.2313 \\ 
        72 & HD\,195556 & 2.5459 $\pm$ 0.0810 & 4.8946 $\pm$ 0.0012 & 0.2621 $\pm$ 0.0324 & -3.3059 $\pm$ 0.0763 & 4.8541 $\pm$ 0.0014 & 0.2981 $\pm$ 0.0370 & -3.3963 $\pm$ 0.0784 & 4.9495 $\pm$ 0.0026 & 0.1570 $\pm$ 0.0195 & -3.1721 $\pm$ 0.0718 \\ 
        73 & HD\,196740 & 6.0025 $\pm$ 0.0965 & 5.0209 $\pm$ 0.0009 & 0.0000 $\pm$ 0.1626 & -1.0875 $\pm$ 0.1663 & 4.9545 $\pm$ 0.0010 & 0.0000 $\pm$ 0.1854 & -1.1539 $\pm$ 0.1887 & 5.1344 $\pm$ 0.0022 & 0.0000 $\pm$ 0.0980 & -0.9739 $\pm$ 0.1041 \\ 
        74 & HD\,197512 & 0.7522 $\pm$ 0.0200 & 8.5212 $\pm$ 0.0004 & 0.8485 $\pm$ 0.4654 & -2.8438 $\pm$ 0.4689 & 8.5361 $\pm$ 0.0006 & 0.9679 $\pm$ 0.5431 & -2.9907 $\pm$ 0.5462 & 8.4428 $\pm$ 0.0004 & 0.5093 $\pm$ 0.2904 & -2.6657 $\pm$ 0.2961 \\ 
        75 & HD\,201091 & 285.9949 $\pm$ 0.0599 & 4.7667 $\pm$ 0.0028 & 0.0000 $\pm$ 0.2256 & 7.0485 $\pm$ 0.2256 & 5.4398 $\pm$ 0.0029 & 0.0000 $\pm$ 0.2872 & 7.7216 $\pm$ 0.2872 & 3.9772 $\pm$ 0.0047 & 0.0000 $\pm$ 0.1666 & 6.2590 $\pm$ 0.1667 \\ 
        76 & HD\,201092 & 286.0054 $\pm$ 0.0289 & 5.4506 $\pm$ 0.0028 & 0.0000 $\pm$ 0.3667 & 7.7325 $\pm$ 0.3667 & 6.2723 $\pm$ 0.0029 & 0.0000 $\pm$ 0.4723 & 8.5541 $\pm$ 0.4723 & 4.5569 $\pm$ 0.0044 & 0.0000 $\pm$ 0.2759 & 6.8388 $\pm$ 0.2759 \\ 
        77 & HD\,205139 & 1.1477 $\pm$ 0.0465 & 5.4582 $\pm$ 0.0007 & 1.0118 $\pm$ 0.4018 & -5.1330 $\pm$ 0.4113 & 5.5197 $\pm$ 0.0011 & 1.1545 $\pm$ 0.4699 & -5.2643 $\pm$ 0.4781 & 5.3013 $\pm$ 0.0020 & 0.6072 $\pm$ 0.2514 & -4.9839 $\pm$ 0.2664 \\ 
        78 & HD\,207538 & 1.1773 $\pm$ 0.0170 & 7.2152 $\pm$ 0.0004 & 1.5746 $\pm$ 0.6172 & -3.8377 $\pm$ 0.6180 & 7.3598 $\pm$ 0.0010 & 1.8100 $\pm$ 0.7392 & -3.9978 $\pm$ 0.7399 & 6.9265 $\pm$ 0.0005 & 0.9560 $\pm$ 0.4015 & -3.6427 $\pm$ 0.4027 \\ 
        79 & HD\,208266 & 1.0963 $\pm$ 0.0183 & 8.0506 $\pm$ 0.0003 & 1.4690 $\pm$ 0.2386 & -3.0708 $\pm$ 0.2413 & 8.1772 $\pm$ 0.0008 & 1.6923 $\pm$ 0.2836 & -3.2304 $\pm$ 0.2859 & 7.7799 $\pm$ 0.0004 & 0.8965 $\pm$ 0.1535 & -2.8876 $\pm$ 0.1577 \\ 
        80 & HD\,209419 & 2.9856 $\pm$ 0.0536 & 5.7601 $\pm$ 0.0005 & 0.0976 $\pm$ 0.0969 & -1.9509 $\pm$ 0.1045 & 5.6992 $\pm$ 0.0011 & 0.1112 $\pm$ 0.1108 & -2.0305 $\pm$ 0.1175 & 5.8338 $\pm$ 0.0010 & 0.0588 $\pm$ 0.0587 & -1.8476 $\pm$ 0.0705 \\ 
        81 & HD\,209975 & 1.0221 $\pm$ 0.0652 & 5.0328 $\pm$ 0.0008 & 1.2085 $\pm$ 0.4893 & -5.9824 $\pm$ 0.5085 & 5.0684 $\pm$ 0.0014 & 1.3792 $\pm$ 0.5753 & -6.1770 $\pm$ 0.5917 & 4.9251 $\pm$ 0.0017 & 0.7246 $\pm$ 0.3086 & -5.7249 $\pm$ 0.3383 \\ 
        82 & HD\,213087 & 0.9686 $\pm$ 0.0463 & 5.3906 $\pm$ 0.0011 & 1.5592 $\pm$ 0.2951 & -6.0860 $\pm$ 0.3129 & 5.5783 $\pm$ 0.0025 & 1.7991 $\pm$ 0.3525 & -6.2031 $\pm$ 0.3675 & 5.0496 $\pm$ 0.0043 & 0.9541 $\pm$ 0.1914 & -5.9433 $\pm$ 0.2178 \\ 
        83 & HD\,214263 & 1.9604 $\pm$ 0.0403 & 6.8144 $\pm$ 0.0003 & 0.2321 $\pm$ 0.2606 & -1.9249 $\pm$ 0.2644 & 6.7292 $\pm$ 0.0008 & 0.2628 $\pm$ 0.2973 & -2.0537 $\pm$ 0.3006 & 6.9334 $\pm$ 0.0006 & 0.1377 $\pm$ 0.1567 & -1.7368 $\pm$ 0.1629 \\ 
        84 & HD\,214994 & 11.6546 $\pm$ 0.1708 & 4.7996 $\pm$ 0.0010 & 0.0000 $\pm$ 0.0307 & 0.1321 $\pm$ 0.0442 & 4.7798 $\pm$ 0.0008 & 0.0000 $\pm$ 0.0358 & 0.1123 $\pm$ 0.0479 & 4.7606 $\pm$ 0.0037 & 0.0000 $\pm$ 0.0194 & 0.0931 $\pm$ 0.0375 \\ 
        85 & HD\,215191 & 2.1623 $\pm$ 0.0519 & 6.3920 $\pm$ 0.0003 & 0.3641 $\pm$ 0.1947 & -2.2494 $\pm$ 0.2015 & 6.3179 $\pm$ 0.0010 & 0.4126 $\pm$ 0.2226 & -2.3919 $\pm$ 0.2286 & 6.4780 $\pm$ 0.0009 & 0.2162 $\pm$ 0.1174 & -2.0548 $\pm$ 0.1285 \\ 
        86 & HD\,217014 & 64.4048 $\pm$ 0.0771 & 5.2832 $\pm$ 0.0028 & 0.0000 $\pm$ 0.0551 & 4.3278 $\pm$ 0.0552 & 5.6175 $\pm$ 0.0029 & 0.0000 $\pm$ 0.0672 & 4.6621 $\pm$ 0.0673 & 4.7889 $\pm$ 0.0041 & 0.0000 $\pm$ 0.0379 & 3.8335 $\pm$ 0.0382 \\ 
        87 & HD\,222173 & 6.4313 $\pm$ 0.1375 & 4.2537 $\pm$ 0.0008 & 0.0000 $\pm$ 0.0159 & -1.7048 $\pm$ 0.0491 & 4.2207 $\pm$ 0.0008 & 0.0000 $\pm$ 0.0183 & -1.7378 $\pm$ 0.0499 & 4.3099 $\pm$ 0.0012 & 0.0000 $\pm$ 0.0098 & -1.6486 $\pm$ 0.0475 \\ 
        88 & HD\,222762 & 1.9809 $\pm$ 0.0296 & 6.6077 $\pm$ 0.0003 & 0.2230 $\pm$ 0.1882 & -2.1090 $\pm$ 0.1909 & 6.5993 $\pm$ 0.0009 & 0.2564 $\pm$ 0.2178 & -2.1609 $\pm$ 0.2202 & 6.5739 $\pm$ 0.0006 & 0.1366 $\pm$ 0.1166 & -2.0737 $\pm$ 0.1210 \\
        \label{tab:3}

\end{longtable}

\end{landscape}


\newpage
\begin{landscape}
\scriptsize
\setlength{\tabcolsep}{3pt}
\renewcommand{\arraystretch}{0.55}
\begin{longtable}{ll@{\hspace{0.5cm}}rcc@{\hspace{0.5cm}}ccc@{\hspace{0.5cm}}rcc}
\caption{Photometric and spectroscopic data to calculate $C_2(\xi)$ the zero-point constant of $BC_\xi$ where $\xi$ stands for one of the  \textit{Gaia} filters, $G$, $G_{\rm BP}$, and $G_{\rm RP}$, respectively.} \\
\toprule
    Order & Star & \multicolumn{1}{c}{$BC_{\rm G}$} & $2.5\,\log{L_{\rm G}/L}$ & $C_{\rm G}$ & $BC_{\rm G_{\rm BP}}$ & $2.5\,\log{L_{\rm G_{\rm BP}}/L}$ & $C_{\rm G_{\rm RP}}$ & \multicolumn{1}{c}{$BC_{\rm G_{\rm RP}}$}  & $2.5\,\log{L_{\rm G_{\rm RP}}/L}$  & $C_{\rm G_{\rm RP}}$  \\
          &      &   \multicolumn{1}{c}{(mag)} &   \multicolumn{1}{c}{(mag)} &   \multicolumn{1}{c}{(mag)} &    \multicolumn{1}{c}{(mag)} &   \multicolumn{1}{c}{(mag)} &   \multicolumn{1}{c}{(mag)} &   \multicolumn{1}{c}{(mag)} &   \multicolumn{1}{c}{(mag)} &   \multicolumn{1}{c}{(mag)} \\
\midrule
\endfirsthead

\multicolumn{11}{l}%
{{\normalsize{\bf\tablename\ \thetable{}} -- continued from previous page}} \\
\toprule
    Order & Star & \multicolumn{1}{c}{$BC_{\rm G}$} & $2.5\,\log{L_{\rm G}/L}$ & $C_{\rm G}$ & $BC_{\rm G_{\rm BP}}$ & $2.5\,\log{L_{\rm G_{\rm BP}}/L}$ & $C_{\rm G_{\rm RP}}$ & \multicolumn{1}{c}{$BC_{\rm G_{\rm RP}}$}  & $2.5\,\log{L_{\rm G_{\rm RP}}/L}$  & $C_{\rm G_{\rm RP}}$  \\
          &      &   \multicolumn{1}{c}{(mag)} &   \multicolumn{1}{c}{(mag)} &   \multicolumn{1}{c}{(mag)} &   \multicolumn{1}{c}{(mag)} &   \multicolumn{1}{c}{(mag)} &   \multicolumn{1}{c}{(mag)} &   \multicolumn{1}{c}{(mag)} &   \multicolumn{1}{c}{(mag)} &   \multicolumn{1}{c}{(mag)} \\
\midrule
\endhead

\midrule
\multicolumn{11}{r}{\normalsize{Continued on next page}} \\
\endfoot

\bottomrule
\endlastfoot

        01 & HD1279 & -0.7561 $\pm$ 0.7621 & -1.5183 $\pm$ 0.0056 & 0.7621 $\pm$ 0.1610 & -0.7052 $\pm$ 0.1648 & -1.6327 $\pm$ 0.0056 & 0.9275 $\pm$ 0.1698 & -0.8005 $\pm$ 0.1358 & -2.8247 $\pm$ 0.0056 & 2.0242 $\pm$ 0.1358 \\ 
        02 & HD1404 & -0.0094 $\pm$ 0.9771 & -0.9865 $\pm$ 0.0072 & 0.9771 $\pm$ 0.1780 & -0.0254 $\pm$ 0.1570 & -1.2223 $\pm$ 0.0072 & 1.1969 $\pm$ 0.1790 & 0.0697 $\pm$ 0.1541 & -2.0261 $\pm$ 0.0072 & 2.0958 $\pm$ 0.1541 \\ 
        03 & HD1439 & -0.1753 $\pm$ 0.8804 & -1.0556 $\pm$ 0.0040 & 0.8804 $\pm$ 0.1335 & -0.1592 $\pm$ 0.1224 & -1.2564 $\pm$ 0.0040 & 1.0972 $\pm$ 0.1347 & -0.1660 $\pm$ 0.1186 & -2.1628 $\pm$ 0.0040 & 1.9969 $\pm$ 0.1186 \\ 
        04 & HD2729 & -0.8278 $\pm$ 0.7955 & -1.6233 $\pm$ 0.0048 & 0.7955 $\pm$ 0.4402 & -0.7717 $\pm$ 0.3310 & -1.7230 $\pm$ 0.0048 & 0.9513 $\pm$ 0.4435 & -0.8957 $\pm$ 0.3174 & -2.9636 $\pm$ 0.0048 & 2.0679 $\pm$ 0.3174 \\ 
        05 & HD4778 & -0.1284 $\pm$ 0.9392 & -1.0676 $\pm$ 0.0067 & 0.9392 $\pm$ 0.1495 & -0.1293 $\pm$ 0.1482 & -1.2713 $\pm$ 0.0067 & 1.1420 $\pm$ 0.1540 & -0.1006 $\pm$ 0.1354 & -2.1680 $\pm$ 0.0067 & 2.0673 $\pm$ 0.1354 \\ 
        06 & HD10362 & -0.9424 $\pm$ 0.7040 & -1.6464 $\pm$ 0.0077 & 0.7040 $\pm$ 0.3974 & -0.8692 $\pm$ 0.4209 & -1.7438 $\pm$ 0.0077 & 0.8746 $\pm$ 0.4306 & -1.0016 $\pm$ 0.3187 & -2.9959 $\pm$ 0.0077 & 1.9943 $\pm$ 0.3187 \\ 
        07 & HD10700 & 0.1662 $\pm$ 1.1086 & -0.9424 $\pm$ 0.0084 & 1.1086 $\pm$ 0.2417 & -0.3329 $\pm$ 0.2604 & -1.4926 $\pm$ 0.0084 & 1.1597 $\pm$ 0.2607 & 0.7183 $\pm$ 0.2215 & -1.4917 $\pm$ 0.0084 & 2.2100 $\pm$ 0.2215 \\ 
        08 & HD15318 & -0.2476 $\pm$ 0.9629 & -1.2105 $\pm$ 0.0046 & 0.9629 $\pm$ 0.3095 & -0.2319 $\pm$ 0.2376 & -1.3750 $\pm$ 0.0046 & 1.1431 $\pm$ 0.3101 & -0.2728 $\pm$ 0.2356 & -2.3990 $\pm$ 0.0046 & 2.1262 $\pm$ 0.2356 \\ 
        09 & HD16440 & -1.4236 $\pm$ 0.8184 & -2.2420 $\pm$ 0.0055 & 0.8184 $\pm$ 0.2731 & -1.3457 $\pm$ 0.3067 & -2.3001 $\pm$ 0.0055 & 0.9544 $\pm$ 0.3067 & -1.5726 $\pm$ 0.2357 & -3.7030 $\pm$ 0.0055 & 2.1305 $\pm$ 0.2357 \\ 
        10 & HD19736 & -0.8016 $\pm$ 0.9042 & -1.7058 $\pm$ 0.0069 & 0.9042 $\pm$ 0.1862 & -0.6885 $\pm$ 0.1848 & -1.7994 $\pm$ 0.0069 & 1.1109 $\pm$ 0.1940 & -0.9554 $\pm$ 0.1597 & -3.0645 $\pm$ 0.0069 & 2.1091 $\pm$ 0.1597 \\ 
        11 & HD22879 & 0.1833 $\pm$ 1.0143 & -0.8310 $\pm$ 0.0043 & 1.0143 $\pm$ 0.1512 & -0.1026 $\pm$ 0.1429 & -1.2696 $\pm$ 0.0043 & 1.1669 $\pm$ 0.1560 & 0.6441 $\pm$ 0.1315 & -1.5249 $\pm$ 0.0043 & 2.1690 $\pm$ 0.1315 \\ 
        12 & HD23300 & -0.8934 $\pm$ 0.7427 & -1.6361 $\pm$ 0.0051 & 0.7427 $\pm$ 0.2475 & -0.8353 $\pm$ 0.2224 & -1.7355 $\pm$ 0.0051 & 0.9002 $\pm$ 0.2577 & -0.9601 $\pm$ 0.1843 & -2.9764 $\pm$ 0.0051 & 2.0163 $\pm$ 0.1843 \\ 
        13 & HD27295 & -0.3948 $\pm$ 0.8985 & -1.2932 $\pm$ 0.0045 & 0.8985 $\pm$ 0.2154 & -0.3550 $\pm$ 0.1841 & -1.4403 $\pm$ 0.0045 & 1.0853 $\pm$ 0.2184 & -0.4465 $\pm$ 0.1735 & -2.5199 $\pm$ 0.0045 & 2.0734 $\pm$ 0.1735 \\ 
        14 & HD27563 & -0.5985 $\pm$ 0.9451 & -1.5436 $\pm$ 0.0059 & 0.9451 $\pm$ 0.1538 & -0.5305 $\pm$ 0.1597 & -1.6568 $\pm$ 0.0059 & 1.1263 $\pm$ 0.1630 & -0.6878 $\pm$ 0.1294 & -2.8537 $\pm$ 0.0059 & 2.1659 $\pm$ 0.1294 \\ 
        15 & HD27778 & -1.0157 $\pm$ 0.7688 & -1.7845 $\pm$ 0.0059 & 0.7688 $\pm$ 0.2935 & -0.9331 $\pm$ 0.3168 & -1.8724 $\pm$ 0.0059 & 0.9393 $\pm$ 0.3218 & -1.0785 $\pm$ 0.2369 & -3.1576 $\pm$ 0.0059 & 2.0791 $\pm$ 0.2369 \\ 
        16 & HD29138 & -2.1335 $\pm$ 0.6608 & -2.7943 $\pm$ 0.0096 & 0.6608 $\pm$ 0.2018 & -1.9888 $\pm$ 0.2003 & -2.8222 $\pm$ 0.0096 & 0.8334 $\pm$ 0.2050 & -2.3215 $\pm$ 0.1901 & -4.3402 $\pm$ 0.0096 & 2.0187 $\pm$ 0.1901 \\ 
        17 & HD29335 & -0.7658 $\pm$ 0.8110 & -1.5769 $\pm$ 0.0051 & 0.8110 $\pm$ 0.1882 & -0.6549 $\pm$ 0.1959 & -1.6857 $\pm$ 0.0051 & 1.0308 $\pm$ 0.2016 & -0.8881 $\pm$ 0.1518 & -2.8943 $\pm$ 0.0051 & 2.0062 $\pm$ 0.1518 \\ 
        18 & HD29589 & -0.8851 $\pm$ 0.7729 & -1.6579 $\pm$ 0.0051 & 0.7729 $\pm$ 0.1832 & -0.8192 $\pm$ 0.1811 & -1.7548 $\pm$ 0.0051 & 0.9355 $\pm$ 0.1909 & -0.9943 $\pm$ 0.1549 & -3.0077 $\pm$ 0.0051 & 2.0133 $\pm$ 0.1549 \\ 
        19 & HD32040 & -0.3877 $\pm$ 0.9268 & -1.3144 $\pm$ 0.0082 & 0.9268 $\pm$ 0.1421 & -0.3448 $\pm$ 0.1435 & -1.4620 $\pm$ 0.0082 & 1.1172 $\pm$ 0.1466 & -0.4288 $\pm$ 0.1295 & -2.5454 $\pm$ 0.0082 & 2.1167 $\pm$ 0.1295 \\ 
        20 & HD32115 & 0.0835 $\pm$ 0.9368 & -0.8533 $\pm$ 0.0056 & 0.9368 $\pm$ 0.1348 & -0.0507 $\pm$ 0.1222 & -1.1819 $\pm$ 0.0056 & 1.1312 $\pm$ 0.1361 & 0.3446 $\pm$ 0.1187 & -1.7226 $\pm$ 0.0056 & 2.0673 $\pm$ 0.1187 \\ 
        21 & HD32309 & -0.2493 $\pm$ 0.9160 & -1.1653 $\pm$ 0.0049 & 0.9160 $\pm$ 0.1501 & -0.2345 $\pm$ 0.1531 & -1.3452 $\pm$ 0.0049 & 1.1107 $\pm$ 0.1545 & -0.2784 $\pm$ 0.1405 & -2.3199 $\pm$ 0.0049 & 2.0415 $\pm$ 0.1405 \\ 
        22 & HD32612 & -1.5073 $\pm$ 0.8149 & -2.3222 $\pm$ 0.0044 & 0.8149 $\pm$ 0.1187 & -1.3899 $\pm$ 0.0978 & -2.3799 $\pm$ 0.0044 & 0.9900 $\pm$ 0.1200 & -1.7002 $\pm$ 0.0924 & -3.7813 $\pm$ 0.0044 & 2.0811 $\pm$ 0.0924 \\ 
        23 & HD34078 & -2.7815 $\pm$ 0.7008 & -3.4823 $\pm$ 0.0094 & 0.7008 $\pm$ 0.3471 & -2.6714 $\pm$ 0.2563 & -3.4987 $\pm$ 0.0094 & 0.8273 $\pm$ 0.3492 & -2.8912 $\pm$ 0.2489 & -5.0616 $\pm$ 0.0094 & 2.1704 $\pm$ 0.2489 \\ 
        24 & HD34310 & -0.2860 $\pm$ 0.9286 & -1.2146 $\pm$ 0.0068 & 0.9286 $\pm$ 0.1732 & -0.2523 $\pm$ 0.1743 & -1.3833 $\pm$ 0.0068 & 1.1309 $\pm$ 0.1769 & -0.3430 $\pm$ 0.1632 & -2.3954 $\pm$ 0.0068 & 2.0524 $\pm$ 0.1632 \\ 
        25 & HD35299 & -1.9609 $\pm$ 0.8256 & -2.7865 $\pm$ 0.0056 & 0.8256 $\pm$ 0.1445 & -1.8316 $\pm$ 0.1495 & -2.8168 $\pm$ 0.0056 & 0.9852 $\pm$ 0.1538 & -2.1724 $\pm$ 0.1142 & -4.3246 $\pm$ 0.0056 & 2.1522 $\pm$ 0.1142 \\ 
        26 & HD35693 & -0.1351 $\pm$ 0.8653 & -1.0004 $\pm$ 0.0065 & 0.8653 $\pm$ 0.1405 & -0.1668 $\pm$ 0.1167 & -1.2320 $\pm$ 0.0065 & 1.0653 $\pm$ 0.1417 & -0.0405 $\pm$ 0.1128 & -2.0552 $\pm$ 0.0065 & 2.0147 $\pm$ 0.1128 \\ 
        27 & HD35912 & -1.5097 $\pm$ 0.8378 & -2.3476 $\pm$ 0.0071 & 0.8378 $\pm$ 0.2746 & -1.3695 $\pm$ 0.2265 & -2.4038 $\pm$ 0.0071 & 1.0343 $\pm$ 0.2798 & -1.7260 $\pm$ 0.2055 & -3.8135 $\pm$ 0.0071 & 2.0875 $\pm$ 0.2055 \\ 
        28 & HD36285 & -1.7289 $\pm$ 0.8116 & -2.5405 $\pm$ 0.0050 & 0.8116 $\pm$ 0.1204 & -1.5736 $\pm$ 0.1102 & -2.5804 $\pm$ 0.0050 & 1.0068 $\pm$ 0.1255 & -1.9822 $\pm$ 0.0900 & -4.0489 $\pm$ 0.0050 & 2.0667 $\pm$ 0.0900 \\ 
        29 & HD36430 & -1.5219 $\pm$ 0.7646 & -2.2865 $\pm$ 0.0054 & 0.7646 $\pm$ 0.1189 & -1.4092 $\pm$ 0.1155 & -2.3470 $\pm$ 0.0054 & 0.9378 $\pm$ 0.1240 & -1.6999 $\pm$ 0.0964 & -3.7375 $\pm$ 0.0054 & 2.0377 $\pm$ 0.0964 \\ 
        30 & HD36512 & -2.9485 $\pm$ 0.7002 & -3.6486 $\pm$ 0.0039 & 0.7002 $\pm$ 0.1646 & -2.8166 $\pm$ 0.1522 & -3.6551 $\pm$ 0.0039 & 0.8385 $\pm$ 0.1655 & -3.2097 $\pm$ 0.1486 & -5.2604 $\pm$ 0.0039 & 2.0506 $\pm$ 0.1486 \\ 
        31 & HD36591 & -2.3838 $\pm$ 0.6946 & -3.0784 $\pm$ 0.0031 & 0.6946 $\pm$ 0.2646 & -2.2490 $\pm$ 0.2759 & -3.0959 $\pm$ 0.0031 & 0.8468 $\pm$ 0.2789 & -2.6110 $\pm$ 0.2244 & -4.6473 $\pm$ 0.0031 & 2.0363 $\pm$ 0.2244 \\ 
        32 & HD36960 & -2.5417 $\pm$ 0.7237 & -3.2655 $\pm$ 0.0031 & 0.7237 $\pm$ 0.1612 & -2.4101 $\pm$ 0.1563 & -3.2794 $\pm$ 0.0031 & 0.8693 $\pm$ 0.1649 & -2.7990 $\pm$ 0.1423 & -4.8501 $\pm$ 0.0031 & 2.0511 $\pm$ 0.1423 \\ 
        33 & HD37077 & 0.1347 $\pm$ 0.9803 & -0.8456 $\pm$ 0.0051 & 0.9803 $\pm$ 0.1006 & -0.0055 $\pm$ 0.0851 & -1.1773 $\pm$ 0.0051 & 1.1718 $\pm$ 0.1023 & 0.3780 $\pm$ 0.0800 & -1.7086 $\pm$ 0.0051 & 2.0866 $\pm$ 0.0800 \\ 
        34 & HD37356 & -1.8898 $\pm$ 0.7301 & -2.6198 $\pm$ 0.0055 & 0.7301 $\pm$ 0.1555 & -1.7299 $\pm$ 0.1338 & -2.6598 $\pm$ 0.0055 & 0.9299 $\pm$ 0.1598 & -2.0889 $\pm$ 0.1184 & -4.1309 $\pm$ 0.0055 & 2.0421 $\pm$ 0.1184 \\ 
        35 & HD37744 & -2.0359 $\pm$ 0.7691 & -2.8050 $\pm$ 0.0068 & 0.7691 $\pm$ 0.1678 & -1.8888 $\pm$ 0.1604 & -2.8431 $\pm$ 0.0068 & 0.9543 $\pm$ 0.1760 & -2.2695 $\pm$ 0.1285 & -4.3262 $\pm$ 0.0068 & 2.0567 $\pm$ 0.1285 \\ 
        36 & HD40967 & -1.1840 $\pm$ 0.7136 & -1.8976 $\pm$ 0.0041 & 0.7136 $\pm$ 0.1927 & -1.0858 $\pm$ 0.1977 & -1.9761 $\pm$ 0.0041 & 0.8904 $\pm$ 0.2057 & -1.2927 $\pm$ 0.1527 & -3.2983 $\pm$ 0.0041 & 2.0057 $\pm$ 0.1527 \\ 
        37 & HD46189 & -1.2549 $\pm$ 0.7470 & -2.0018 $\pm$ 0.0086 & 0.7470 $\pm$ 0.1693 & -1.1693 $\pm$ 0.1597 & -2.0726 $\pm$ 0.0086 & 0.9032 $\pm$ 0.1774 & -1.3867 $\pm$ 0.1283 & -3.4215 $\pm$ 0.0086 & 2.0348 $\pm$ 0.1283 \\ 
        38 & HD54764 & -1.5209 $\pm$ 0.7397 & -2.2606 $\pm$ 0.0054 & 0.7397 $\pm$ 0.5209 & -1.4157 $\pm$ 0.5030 & -2.3266 $\pm$ 0.0054 & 0.9109 $\pm$ 0.5525 & -1.6208 $\pm$ 0.3974 & -3.6992 $\pm$ 0.0054 & 2.0783 $\pm$ 0.3974 \\ 
        39 & HD55856 & -1.7626 $\pm$ 0.7659 & -2.5285 $\pm$ 0.0053 & 0.7659 $\pm$ 0.1982 & -1.6347 $\pm$ 0.1843 & -2.5658 $\pm$ 0.0053 & 0.9311 $\pm$ 0.2052 & -1.9680 $\pm$ 0.1573 & -4.0403 $\pm$ 0.0053 & 2.0723 $\pm$ 0.1573 \\ 
        40 & HD71155 & -0.2302 $\pm$ 0.8450 & -1.0752 $\pm$ 0.0051 & 0.8450 $\pm$ 0.2884 & -0.2066 $\pm$ 0.2949 & -1.2811 $\pm$ 0.0051 & 1.0745 $\pm$ 0.2976 & -0.1958 $\pm$ 0.2690 & -2.1751 $\pm$ 0.0051 & 1.9794 $\pm$ 0.2690 \\ 
        41 & HD78556 & -0.3763 $\pm$ 0.7837 & -1.1600 $\pm$ 0.0068 & 0.7837 $\pm$ 0.1206 & -0.3240 $\pm$ 0.1308 & -1.3374 $\pm$ 0.0068 & 1.0134 $\pm$ 0.1325 & -0.3803 $\pm$ 0.0938 & -2.3270 $\pm$ 0.0068 & 1.9467 $\pm$ 0.0938 \\ 
        42 & HD82106 & -0.0577 $\pm$ 0.9746 & -1.0323 $\pm$ 0.0079 & 0.9746 $\pm$ 0.3627 & -0.5858 $\pm$ 0.4031 & -1.7020 $\pm$ 0.0079 & 1.1162 $\pm$ 0.4034 & 0.6251 $\pm$ 0.3245 & -1.4649 $\pm$ 0.0079 & 2.0900 $\pm$ 0.3245 \\ 
        43 & HD82621 & -0.1587 $\pm$ 0.8268 & -0.9855 $\pm$ 0.0054 & 0.8268 $\pm$ 0.1544 & -0.1902 $\pm$ 0.1387 & -1.2186 $\pm$ 0.0054 & 1.0284 $\pm$ 0.1555 & -0.1105 $\pm$ 0.1354 & -2.0324 $\pm$ 0.0054 & 1.9220 $\pm$ 0.1354 \\ 
        44 & HD82943 & 0.0433 $\pm$ 0.8927 & -0.8494 $\pm$ 0.0060 & 0.8927 $\pm$ 0.1696 & -0.2551 $\pm$ 0.1861 & -1.2934 $\pm$ 0.0060 & 1.0383 $\pm$ 0.1862 & 0.5098 $\pm$ 0.1488 & -1.5374 $\pm$ 0.0060 & 2.0472 $\pm$ 0.1488 \\ 
        45 & HD84937 & 0.0286 $\pm$ 0.8349 & -0.8063 $\pm$ 0.0056 & 0.8349 $\pm$ 0.2078 & -0.1751 $\pm$ 0.2096 & -1.1773 $\pm$ 0.0056 & 1.0022 $\pm$ 0.2213 & 0.3989 $\pm$ 0.1766 & -1.5979 $\pm$ 0.0056 & 1.9968 $\pm$ 0.1766 \\ 
        46 & HD85503 & -0.1785 $\pm$ 0.9545 & -1.1330 $\pm$ 0.0079 & 0.9545 $\pm$ 0.2459 & -0.8541 $\pm$ 0.2644 & -1.8625 $\pm$ 0.0079 & 1.0084 $\pm$ 0.2650 & 0.5405 $\pm$ 0.2297 & -1.5076 $\pm$ 0.0079 & 2.0482 $\pm$ 0.2297 \\ 
        47 & HD89021 & -0.1062 $\pm$ 1.0355 & -1.1418 $\pm$ 0.0123 & 1.0355 $\pm$ 0.2362 & -0.1740 $\pm$ 0.2180 & -1.4014 $\pm$ 0.0123 & 1.2274 $\pm$ 0.2372 & -0.0266 $\pm$ 0.2156 & -2.1122 $\pm$ 0.0123 & 2.0856 $\pm$ 0.2156 \\ 
        48 & HD99922 & -0.0856 $\pm$ 0.8975 & -0.9830 $\pm$ 0.0068 & 0.8975 $\pm$ 0.1154 & -0.0927 $\pm$ 0.1144 & -1.2351 $\pm$ 0.0068 & 1.1424 $\pm$ 0.1212 & -0.0392 $\pm$ 0.0977 & -1.9917 $\pm$ 0.0068 & 1.9524 $\pm$ 0.0977 \\ 
        49 & HD101364 & 0.1254 $\pm$ 0.9692 & -0.8438 $\pm$ 0.0037 & 0.9692 $\pm$ 0.2361 & -0.1928 $\pm$ 0.2426 & -1.3071 $\pm$ 0.0037 & 1.1143 $\pm$ 0.2431 & 0.6153 $\pm$ 0.2278 & -1.5065 $\pm$ 0.0037 & 2.1218 $\pm$ 0.2278 \\ 
        50 & HD103095 & 0.0631 $\pm$ 0.9567 & -0.8935 $\pm$ 0.0044 & 0.9567 $\pm$ 0.3383 & -0.3458 $\pm$ 0.3517 & -1.4335 $\pm$ 0.0044 & 1.0878 $\pm$ 0.3521 & 0.6559 $\pm$ 0.3245 & -1.4578 $\pm$ 0.0044 & 2.1137 $\pm$ 0.3245 \\ 
        51 & HD113226 & -0.1106 $\pm$ 0.9189 & -1.0295 $\pm$ 0.0110 & 0.9189 $\pm$ 0.1432 & -0.8978 $\pm$ 0.1644 & -1.6846 $\pm$ 0.0110 & 0.7868 $\pm$ 0.1647 & 0.2589 $\pm$ 0.1233 & -1.4885 $\pm$ 0.0110 & 1.7475 $\pm$ 0.1233 \\ 
        52 & HD117176 & 0.0788 $\pm$ 0.9562 & -0.8775 $\pm$ 0.0029 & 0.9562 $\pm$ 0.1871 & -0.2867 $\pm$ 0.2196 & -1.3903 $\pm$ 0.0029 & 1.1035 $\pm$ 0.2203 & 0.6262 $\pm$ 0.1457 & -1.4771 $\pm$ 0.0029 & 2.1032 $\pm$ 0.1457 \\ 
        53 & HD120198 & -0.1636 $\pm$ 0.9692 & -1.1328 $\pm$ 0.0083 & 0.9692 $\pm$ 0.1402 & -0.1416 $\pm$ 0.1348 & -1.3203 $\pm$ 0.0083 & 1.1787 $\pm$ 0.1414 & -0.1781 $\pm$ 0.1313 & -2.2707 $\pm$ 0.0083 & 2.0927 $\pm$ 0.1313 \\ 
        54 & HD122563 & -0.0363 $\pm$ 0.9664 & -1.0028 $\pm$ 0.0051 & 0.9664 $\pm$ 0.1506 & -0.5340 $\pm$ 0.1668 & -1.6456 $\pm$ 0.0051 & 1.1116 $\pm$ 0.1699 & 0.6205 $\pm$ 0.1296 & -1.4522 $\pm$ 0.0051 & 2.0727 $\pm$ 0.1296 \\ 
        55 & HD125924 & -1.5893 $\pm$ 0.8793 & -2.4686 $\pm$ 0.0040 & 0.8793 $\pm$ 0.3437 & -1.4766 $\pm$ 0.3340 & -2.5153 $\pm$ 0.0040 & 1.0387 $\pm$ 0.3441 & -1.7906 $\pm$ 0.3325 & -3.9615 $\pm$ 0.0040 & 2.1709 $\pm$ 0.3325 \\ 
        56 & HD128167 & 0.0843 $\pm$ 0.9160 & -0.8317 $\pm$ 0.0029 & 0.9160 $\pm$ 0.2635 & -0.1251 $\pm$ 0.2708 & -1.2199 $\pm$ 0.0029 & 1.0948 $\pm$ 0.2742 & 0.4149 $\pm$ 0.2457 & -1.6059 $\pm$ 0.0029 & 2.0208 $\pm$ 0.2457 \\ 
        57 & HD133208 & 0.1333 $\pm$ 1.1574 & -1.0241 $\pm$ 0.0111 & 1.1574 $\pm$ 0.4475 & -0.4606 $\pm$ 0.5136 & -1.6597 $\pm$ 0.0111 & 1.1991 $\pm$ 0.5148 & 0.7233 $\pm$ 0.3761 & -1.4814 $\pm$ 0.0111 & 2.2048 $\pm$ 0.3761 \\ 
        58 & HD141714 & 0.0659 $\pm$ 0.9715 & -0.9056 $\pm$ 0.0035 & 0.9715 $\pm$ 0.1349 & -0.3264 $\pm$ 0.1455 & -1.4556 $\pm$ 0.0035 & 1.1293 $\pm$ 0.1470 & 0.6266 $\pm$ 0.1207 & -1.4628 $\pm$ 0.0035 & 2.0894 $\pm$ 0.1207 \\ 
        59 & HD146233 & 0.2175 $\pm$ 1.0740 & -0.8564 $\pm$ 0.0056 & 1.0740 $\pm$ 0.1288 & -0.1098 $\pm$ 0.1326 & -1.3255 $\pm$ 0.0056 & 1.2157 $\pm$ 0.1344 & 0.7073 $\pm$ 0.1204 & -1.5152 $\pm$ 0.0056 & 2.2225 $\pm$ 0.1204 \\ 
        60 & HD148379 & -1.0568 $\pm$ 0.9763 & -2.0332 $\pm$ 0.0054 & 0.9763 $\pm$ 0.6370 & -0.9593 $\pm$ 0.6307 & -2.1127 $\pm$ 0.0054 & 1.1533 $\pm$ 0.6805 & -1.2310 $\pm$ 0.5343 & -3.4254 $\pm$ 0.0054 & 2.1945 $\pm$ 0.5343 \\ 
        61 & HD148688 & -1.5993 $\pm$ 0.8625 & -2.4618 $\pm$ 0.0072 & 0.8625 $\pm$ 0.7667 & -1.3968 $\pm$ 0.8468 & -2.5129 $\pm$ 0.0072 & 1.1161 $\pm$ 0.8724 & -1.8316 $\pm$ 0.6011 & -3.9363 $\pm$ 0.0072 & 2.1047 $\pm$ 0.6011 \\ 
        62 & HD150117 & -0.4720 $\pm$ 0.7296 & -1.2016 $\pm$ 0.0075 & 0.7296 $\pm$ 0.1550 & -0.4315 $\pm$ 0.1631 & -1.3870 $\pm$ 0.0075 & 0.9555 $\pm$ 0.1644 & -0.5160 $\pm$ 0.1351 & -2.3400 $\pm$ 0.0075 & 1.8239 $\pm$ 0.1351 \\ 
        63 & HD160762 & -1.4002 $\pm$ 0.7653 & -2.1655 $\pm$ 0.0053 & 0.7653 $\pm$ 0.3798 & -1.3013 $\pm$ 0.3523 & -2.2366 $\pm$ 0.0053 & 0.9353 $\pm$ 0.3899 & -1.5168 $\pm$ 0.3144 & -3.5904 $\pm$ 0.0053 & 2.0736 $\pm$ 0.3144 \\ 
        64 & HD162570 & 0.0356 $\pm$ 0.9115 & -0.8759 $\pm$ 0.0054 & 0.9115 $\pm$ 0.1117 & -0.0818 $\pm$ 0.1093 & -1.2072 $\pm$ 0.0054 & 1.1254 $\pm$ 0.1133 & 0.2716 $\pm$ 0.1053 & -1.7491 $\pm$ 0.0054 & 2.0208 $\pm$ 0.1053 \\ 
        65 & HD171301 & -0.5535 $\pm$ 0.8517 & -1.4053 $\pm$ 0.0069 & 0.8517 $\pm$ 0.2073 & -0.5097 $\pm$ 0.2020 & -1.5537 $\pm$ 0.0069 & 1.0440 $\pm$ 0.2142 & -0.6025 $\pm$ 0.1794 & -2.6309 $\pm$ 0.0069 & 2.0284 $\pm$ 0.1794 \\ 
        66 & HD174959 & -0.9377 $\pm$ 0.7864 & -1.7241 $\pm$ 0.0073 & 0.7864 $\pm$ 0.1620 & -0.8741 $\pm$ 0.1600 & -1.8262 $\pm$ 0.0073 & 0.9521 $\pm$ 0.1706 & -1.0172 $\pm$ 0.1294 & -3.0642 $\pm$ 0.0073 & 2.0470 $\pm$ 0.1294 \\ 
        67 & HD186427 & -0.0869 $\pm$ 0.7695 & -0.8563 $\pm$ 0.0027 & 0.7695 $\pm$ 0.2166 & -0.4142 $\pm$ 0.2283 & -1.3304 $\pm$ 0.0027 & 0.9162 $\pm$ 0.2299 & 0.4120 $\pm$ 0.1995 & -1.5105 $\pm$ 0.0027 & 1.9225 $\pm$ 0.1995 \\ 
        68 & HD189319 & -0.1874 $\pm$ 1.1511 & -1.3385 $\pm$ 0.0092 & 1.1511 $\pm$ 0.5603 & -1.2134 $\pm$ 0.6088 & -2.2277 $\pm$ 0.0092 & 1.0143 $\pm$ 0.6097 & 0.3227 $\pm$ 0.5306 & -1.5737 $\pm$ 0.0092 & 1.8964 $\pm$ 0.5306 \\ 
        69 & HD189957 & -2.8211 $\pm$ 0.7358 & -3.5569 $\pm$ 0.0079 & 0.7358 $\pm$ 0.5028 & -2.6452 $\pm$ 0.5328 & -3.5689 $\pm$ 0.0079 & 0.9237 $\pm$ 0.5497 & -3.0316 $\pm$ 0.3821 & -5.1529 $\pm$ 0.0079 & 2.1212 $\pm$ 0.3821 \\ 
        70 & HD192263 & 0.0240 $\pm$ 1.0160 & -0.9921 $\pm$ 0.0071 & 1.0160 $\pm$ 0.3149 & -0.4609 $\pm$ 0.3514 & -1.6307 $\pm$ 0.0071 & 1.1698 $\pm$ 0.3516 & 0.6696 $\pm$ 0.2787 & -1.4583 $\pm$ 0.0071 & 2.1279 $\pm$ 0.2787 \\ 
        71 & HD193183 & -1.6796 $\pm$ 0.7365 & -2.4160 $\pm$ 0.0057 & 0.7365 $\pm$ 0.3697 & -1.5484 $\pm$ 0.4002 & -2.4619 $\pm$ 0.0057 & 0.9135 $\pm$ 0.4014 & -1.8409 $\pm$ 0.3264 & -3.8983 $\pm$ 0.0057 & 2.0574 $\pm$ 0.3264 \\ 
        72 & HD195556 & -1.3345 $\pm$ 0.7546 & -2.0891 $\pm$ 0.0083 & 0.7546 $\pm$ 0.5149 & -1.2442 $\pm$ 0.4474 & -2.1601 $\pm$ 0.0083 & 0.9159 $\pm$ 0.5152 & -1.4684 $\pm$ 0.4463 & -3.5195 $\pm$ 0.0083 & 2.0511 $\pm$ 0.4463 \\ 
        73 & HD196740 & -1.0346 $\pm$ 0.7656 & -1.8002 $\pm$ 0.0036 & 0.7656 $\pm$ 0.2542 & -0.9682 $\pm$ 0.2574 & -1.8906 $\pm$ 0.0036 & 0.9224 $\pm$ 0.2693 & -1.1482 $\pm$ 0.2037 & -3.1708 $\pm$ 0.0036 & 2.0226 $\pm$ 0.2037 \\ 
        74 & HD197512 & -2.0296 $\pm$ 0.7390 & -2.7687 $\pm$ 0.0057 & 0.7390 $\pm$ 0.6751 & -1.8827 $\pm$ 0.7099 & -2.8110 $\pm$ 0.0057 & 0.9283 $\pm$ 0.7309 & -2.2078 $\pm$ 0.5416 & -4.2785 $\pm$ 0.0057 & 2.0708 $\pm$ 0.5416 \\ 
        75 & HD201091 & -0.2222 $\pm$ 0.9358 & -1.1580 $\pm$ 0.0070 & 0.9358 $\pm$ 0.3242 & -0.8953 $\pm$ 0.3682 & -1.9480 $\pm$ 0.0070 & 1.0528 $\pm$ 0.3697 & 0.5673 $\pm$ 0.2843 & -1.4810 $\pm$ 0.0070 & 2.0483 $\pm$ 0.2843 \\ 
        76 & HD201092 & -0.2205 $\pm$ 1.0132 & -1.2337 $\pm$ 0.0071 & 1.0132 $\pm$ 0.6423 & -1.0421 $\pm$ 0.7062 & -2.0533 $\pm$ 0.0071 & 1.0111 $\pm$ 0.7079 & 0.6732 $\pm$ 0.5931 & -1.5252 $\pm$ 0.0071 & 2.1984 $\pm$ 0.5931 \\ 
        77 & HD205139 & -2.4334 $\pm$ 0.6823 & -3.1158 $\pm$ 0.0076 & 0.6823 $\pm$ 0.7024 & -2.3021 $\pm$ 0.6602 & -3.1566 $\pm$ 0.0076 & 0.8545 $\pm$ 0.7434 & -2.5825 $\pm$ 0.5275 & -4.6386 $\pm$ 0.0076 & 2.0560 $\pm$ 0.5275 \\ 
        78 & HD207538 & -2.7315 $\pm$ 0.7806 & -3.5121 $\pm$ 0.0068 & 0.7806 $\pm$ 0.7616 & -2.5714 $\pm$ 0.8607 & -3.5285 $\pm$ 0.0068 & 0.9571 $\pm$ 0.8635 & -2.9265 $\pm$ 0.5963 & -5.0861 $\pm$ 0.0068 & 2.1597 $\pm$ 0.5963 \\ 
        79 & HD208266 & -2.1307 $\pm$ 0.8055 & -2.9363 $\pm$ 0.0061 & 0.8055 $\pm$ 0.3904 & -1.9712 $\pm$ 0.3817 & -2.9768 $\pm$ 0.0061 & 1.0056 $\pm$ 0.4194 & -2.3140 $\pm$ 0.2980 & -4.4389 $\pm$ 0.0061 & 2.1248 $\pm$ 0.2980 \\ 
        80 & HD209419 & -0.9571 $\pm$ 0.8139 & -1.7710 $\pm$ 0.0043 & 0.8139 $\pm$ 0.1743 & -0.8775 $\pm$ 0.1698 & -1.8666 $\pm$ 0.0043 & 0.9891 $\pm$ 0.1823 & -1.0604 $\pm$ 0.1414 & -3.1125 $\pm$ 0.0043 & 2.0522 $\pm$ 0.1414 \\ 
        81 & HD209975 & -2.9733 $\pm$ 0.6497 & -3.6230 $\pm$ 0.0068 & 0.6497 $\pm$ 0.6412 & -2.7787 $\pm$ 0.6837 & -3.6302 $\pm$ 0.0068 & 0.8514 $\pm$ 0.7090 & -3.2307 $\pm$ 0.4813 & -5.2286 $\pm$ 0.0068 & 1.9978 $\pm$ 0.4813 \\ 
        82 & HD213087 & -2.1606 $\pm$ 0.7553 & -2.9159 $\pm$ 0.0057 & 0.7553 $\pm$ 0.4098 & -2.0435 $\pm$ 0.4523 & -2.9505 $\pm$ 0.0057 & 0.9071 $\pm$ 0.4529 & -2.3032 $\pm$ 0.3420 & -4.4265 $\pm$ 0.0057 & 2.1233 $\pm$ 0.3420 \\ 
        83 & HD214263 & -1.7003 $\pm$ 0.7545 & -2.4547 $\pm$ 0.0070 & 0.7545 $\pm$ 0.3503 & -1.5714 $\pm$ 0.3773 & -2.4998 $\pm$ 0.0070 & 0.9284 $\pm$ 0.3784 & -1.8883 $\pm$ 0.2802 & -3.9453 $\pm$ 0.0070 & 2.0571 $\pm$ 0.2802 \\ 
        84 & HD214994 & -0.0712 $\pm$ 0.9447 & -1.0159 $\pm$ 0.0054 & 0.9447 $\pm$ 0.5136 & -0.0514 $\pm$ 0.3728 & -1.2369 $\pm$ 0.0054 & 1.1855 $\pm$ 0.5139 & -0.0322 $\pm$ 0.3716 & -2.0896 $\pm$ 0.0054 & 2.0574 $\pm$ 0.3716 \\ 
        85 & HD215191 & -1.8613 $\pm$ 0.7188 & -2.5801 $\pm$ 0.0058 & 0.7188 $\pm$ 0.2652 & -1.7188 $\pm$ 0.2824 & -2.6181 $\pm$ 0.0058 & 0.8994 $\pm$ 0.2864 & -2.0558 $\pm$ 0.2097 & -4.0891 $\pm$ 0.0058 & 2.0333 $\pm$ 0.2097 \\ 
        86 & HD217014 & 0.1735 $\pm$ 1.0325 & -0.8590 $\pm$ 0.0050 & 1.0325 $\pm$ 0.1484 & -0.1608 $\pm$ 0.1434 & -1.3348 $\pm$ 0.0050 & 1.1740 $\pm$ 0.1533 & 0.6678 $\pm$ 0.1322 & -1.5075 $\pm$ 0.0050 & 2.1753 $\pm$ 0.1322 \\ 
        87 & HD222173 & -0.3910 $\pm$ 0.9441 & -1.3350 $\pm$ 0.0049 & 0.9441 $\pm$ 0.2695 & -0.3579 $\pm$ 0.1971 & -1.4824 $\pm$ 0.0049 & 1.1245 $\pm$ 0.2697 & -0.4471 $\pm$ 0.1965 & -2.5646 $\pm$ 0.0049 & 2.1175 $\pm$ 0.1965 \\ 
        88 & HD222762 & -0.6610 $\pm$ 0.8096 & -1.4706 $\pm$ 0.0057 & 0.8096 $\pm$ 0.2537 & -0.6091 $\pm$ 0.2744 & -1.5998 $\pm$ 0.0057 & 0.9907 $\pm$ 0.2764 & -0.6963 $\pm$ 0.2035 & -2.7458 $\pm$ 0.0057 & 2.0495 $\pm$ 0.2035 

        \label{tab:4}
\end{longtable}

\end{landscape}



\section{Results}

\subsection{Empirical Zero-Point Constants of $Gaia$ $BC$ Scales}

The bolometric correction ($BC_\xi$) from absolute bolometric (Table~\ref{tab:2}) and filtered magnitudes (Table~\ref{tab:3}) as $BC_{\xi} = M_{\rm Bol} - M_\xi$, and the logarithmic term obtained  from the sample spectra (Table~\ref{tab:1}) as $2.5\log{({L_\xi}/{L})}$ are the two parameters to calculate individual zero-point constants [$C_2(\xi)$] according to Equation~(\ref{eq:8}). The two input values and the calculated zero-point constants for each of the $Gaia$ passbands are given in Tables~\ref{tab:4} together with their propagated uncertainties, where the first two columns are for order and the star name. The next three columns are for the $G$ band, then the following three columns are for the $G_{\rm BP}$, and the last three columns are for the $G_{\rm RP}$ bands. 

\begin{figure}
    \centering
    \includegraphics[width=1\linewidth]{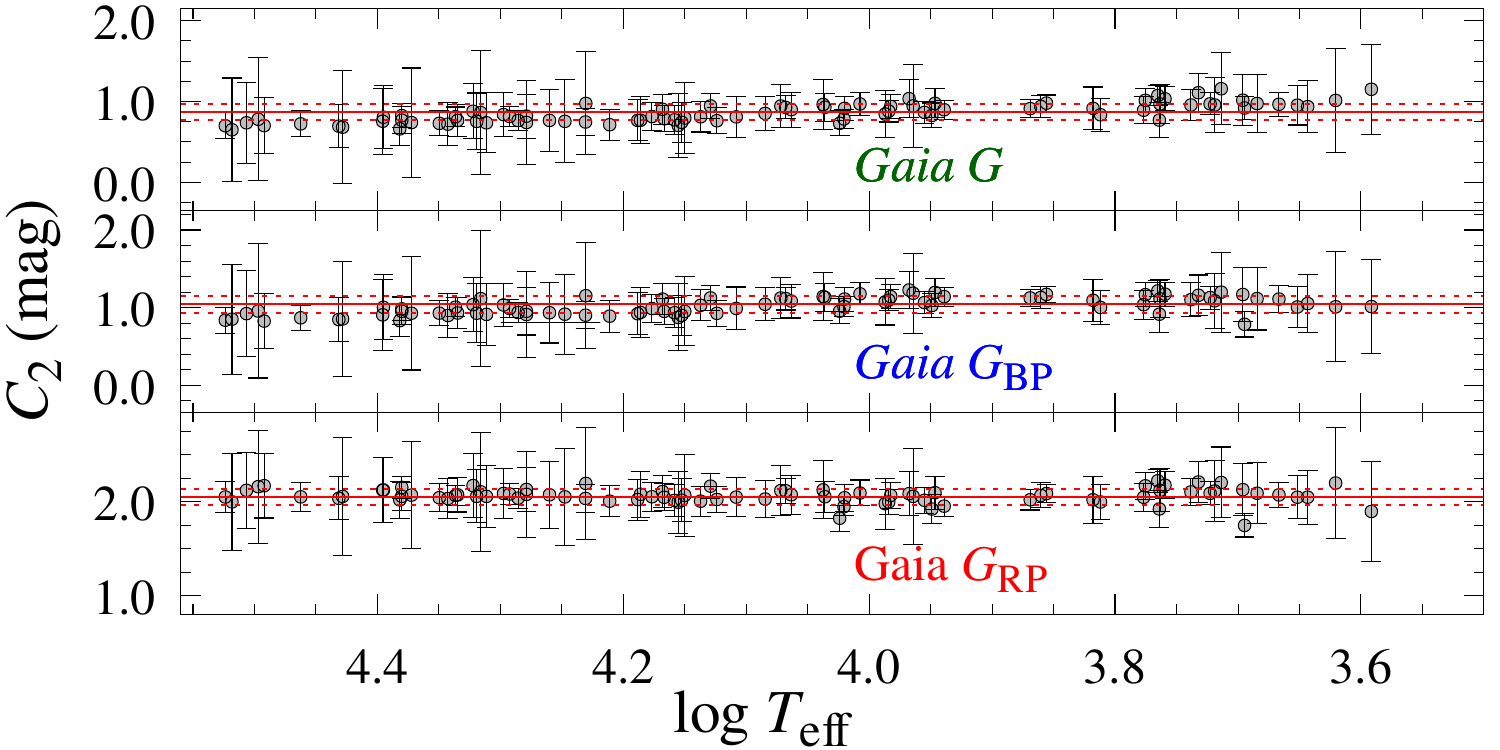}
    \caption{The zero-point constants of the $BC$ scales obtained from the sample spectra. Error bars are the propagated uncertainties. Horizontal lines mark values of weighted means: $0.8677$, $1.0449$, and $2.0510$ for \textit{Gaia G} (top), $G_{\rm BP}$ (middle), and $G_{\rm RP}$ (bottom), respectively. Dashed lines indicate standard deviations.}
    \label{fig:4}
\end{figure}

The arithmetic and weighted means of 88 zero-point $C_2(\xi)$ constants for each of the $BC{(\xi)}$ scales are given in Table~\ref{tab:5}. The standard deviations and standard errors of the mean values are indicated. Figure~\ref{fig:4} shows the weighted means   
and individual values distributed on the stellar effective temperature scale of the present study, which indicates no bias. Therefore, the weighted means $0.8677\pm0.0109$, $1.0449\pm0.0116$, and $2.0510\pm0.0087$ of $C_2$ for the $G$, $G_{\rm BP}$ and $G_{\rm RP}$ passbands could be used in future studies when computing {\it Gaia} spectroscopic $BC$ according to Equation~(\ref{eq:8}) from observed spectra until they are replaced by more precise and accurate constants.    

\begin{table*}
    \centering
    \caption{Statistics of the zero-point constant of the $BC$ scale $C_{\rm 2}$ estimated from the sample spectra and spectroscopic ($T_{\rm eff}$), photometric ($G, G_{\rm BP}, G_{\rm RP}$), and astrometric ($\varpi$) observations for \textit{Gaia} filters, $G$, $G_{\rm BP}$, and $G_{\rm RP}$.}
    \begin{tabular}{ccccccc}
    \toprule
        Arithmetic Mean of $C_{\rm 2}$ & S.D. & S.E. & Weighted Mean of $C_{\rm 2}$ & S.D. & S.E.  & Filter\\
        \midrule
         0.8560 & 0.1160 & 0.0124 & 0.8677 & 0.1015 & 0.0109 & $G$ \\
         1.0206 & 0.1105 & 0.0118 & 1.0449 & 0.1082 & 0.0116 & $G_{\rm BP}$\\
         2.0639 & 0.0780 & 0.0083 & 2.0510 & 0.0812 & 0.0087 & $G_{\rm RP}$\\
         \bottomrule
    \end{tabular}
    \label{tab:5}
\end{table*}

\subsection{Empirical Zero-Point Constants of absolute and apparent $Gaia$ Magnitudes}

After having the zero-point constants of the $BC(\xi)$ scales (Table~\ref{tab:5}), it is straightforward to calculate the empirical  zero-point constants of the absolute magnitudes of $Gaia$ ($M_{\rm G}$, $M_{G_{\rm BP}}$, $M_{G_{\rm RP}}$) as   
\begin{equation}
    C_{\xi} = (71.197425... - \langle C_{\rm 2}(\xi)\rangle) \pm \,\,\text{S.E., for the case if \(L_{\xi}\) is in SI units}
    \label{equ:9}
\end{equation}
using the zero-point constant $C_{bol} = 71.197425 ...$ of the  absolute bolometric magnitudes \citep{Mamajek2015}, and the following equation is to calculate empirical zero-point constants of the apparent magnitudes of $Gaia$ ($G$, $G_{\rm BP}$, $G_{\rm RP}$) as    
\begin{equation}
    c_{\rm \xi} = (-18.997351... - \langle C_{\rm 2}(\xi)\rangle) \pm \,\,\text{S.E., for the case if \(f_{\xi}\) is in SI units}
    \label{equ:14}
\end{equation}
using the zero-point constant $c_{\rm Bol} = -18.997351 ...$ of apparent bolometric magnitudes \citep{Mamajek2015}, where $\langle C_{\rm 2}(\xi)\rangle$ in both of the above equations would represent the weighted mean of the zero-point constant for the scales of $BC(\xi)$ estimated in this study. 

The zero-point constants of absolute and apparent magnitudes for the four bands, one for bolometric and three for filtered, are given in Table~\ref{tab:11}, where the luminosity corresponding to zero absolute magnitude ($L_0(\xi)$ if $M_\xi=0$) and the flux reaching above Earth's atmosphere (no extinction) from a star having zero apparent magnitude ($f_0(\xi)$ if $\xi =0$) are also indicated in both SI and cgs system of units. The zero-point constants ($C_\xi$) of the absolute filtered magnitudes are needed when calculating the filtered luminosities ($L_\xi$) from the absolute filtered magnitudes ($M_\xi$) according to Equation~(\ref{eq:5}). Similarly, zero-point constants ($c_\xi$) of the apparent filtered magnitudes are used when calculating the filtered fluxes ($f_\xi$) from the apparent filtered magnitudes ($\xi$) that arrive at Earth from the star if there are no extinctions.    
\begin{table*}
    \centering
    \caption{The zero-point (ZP) constant for absolute/apparent magnitudes, stellar luminosity ($L$) if $M_\xi=0$, and flux ($f$) at Earth (no extinction) if $\xi=0$, respectively, where $\xi$ is one of any of the four bands, one for bolometric and three for {\it Gaia}. }
   
    \resizebox{\textwidth}{!}{\begin{tabular}{ccccccc}
    \toprule
    & ZP of absolute mag & if $M_{\xi}=0$ & Unit & ZP of apparent mag & if $\xi=0$ & Unit \\
    & $C_{\rm Bol}$ & $L_0(\xi)$ &  & $c_{\rm Bol}$ & $f_0(\xi)$  &  \\
    \midrule
    SI  & 71.197425 & 3.0128E+28 & W            & -18.997351 & 2.5180E-08 & W m$^{-2}$             \\
    cgs & 88.697425 & 3.0128E+35 & erg s$^{-1}$ & -11.497351 & 2.5180E-05 & erg s$^{-1}$ cm$^{-2}$ \\
    \midrule
    \multicolumn{7}{c}{$C_{\rm G} = C_{\rm Bol} - 0.8677$ \qquad \qquad \qquad \qquad  $Gaia\,G$ \qquad \qquad \qquad \qquad $c_{\rm G} = c_{\rm Bol} - 0.8677$} \\
    SI  & 70.3297 & 1.3548E+28 & W              &  -19.8651  & 1.1328E-08 & W m$^{-2}$ \\
    cgs & 87.8297 & 1.3548E+35 & erg s$^{-1}$   &  -12.3651  & 1.1323E-05 & erg s$^{-1}$ cm$^{-2}$ \\
    \midrule
    \multicolumn{7}{c}{$C_{\rm G_{\rm BP}} = C_{\rm Bol} - 1.0449$ \qquad \qquad \qquad  $Gaia\,G_{\rm BP}$  \qquad \qquad \qquad $c_{\rm G_{\rm BP}} = c_{\rm Bol} -1.0449$} \\
    
    SI  & 70.1525 & 1.1508E+28 & W              & -20.0423   & 9.6177E-09 & W m$^{-2}$             \\
    cgs & 87.6525 & 1.1508E+35 & erg s$^{-1}$   & -12.5423   & 9.6177E-06 & erg s$^{-1}$ cm$^{-2}$ \\
    \midrule
    \multicolumn{7}{c}{$C_{\rm G_{\rm RP}} = C_{\rm Bol} - 2.0510$  \qquad \qquad \qquad  $Gaia\,G_{\rm RP}$ \qquad \qquad \qquad $c_{\rm G_{\rm RP}} = c_{\rm Bol} - 2.0510$} \\
    
    SI  & 69.1464 & 4.4449E+27 & W              & -21.0484   & 3.8074E-09 & W m$^{-2}$             \\
    cgs & 86.6464 & 4.5559E+34 & erg s$^{-1}$   & -13.5484   & 3.8074E-06 & erg s$^{-1}$ cm$^{-2}$ \\
    \bottomrule
    \end{tabular}}
    \label{tab:11}
\end{table*}


\subsection{Empirical Spectroscopic $BC$ from $Gaia$ Spectra}

Once the zero-point constants $C_2(\xi)$ of the $BC_\xi$ scales are ready (Table~\ref{tab:5}), it is straightforward to calculate the empirical spectroscopic $BC_\xi$ according to Equation~(\ref{eq:8}) using the fractional luminosities ($L_\xi / L$) given in Table~\ref{tab:1}. The calculated $BC_\xi$  values and associated errors in decreasing order of effective temperatures are listed in Table~\ref{tab:7}, where the columns are self-explanatory to indicate effective temperature ($T_{\rm eff}$), Spectroscopic $BC_{\rm G}$,  Spectroscopic $BC_{G_{\rm BP}}$, Spectroscopic $BC_{G_{\rm RP}}$, error in mag units, order of the star in Table~\ref{tab:1} and its name.

\begin{table*}[!ht]
    \centering
    \caption{Spectroscopic $BC$ and associated errors from {\it Gaia} spectra of the current sample of 88 stars for the {\it Gaia} photometric bands $G$, $G_{\rm BP}$ and $G_{\rm RP}$.}
    \resizebox{\textwidth}{!}{\begin{tabular}{crrrcclcrrrccl}
    \toprule
    $T_{\rm eff}$ & $BC_{\rm G}$ & $BC_{\rm G_{\rm BP}}$ & $BC_{\rm G_{\rm RP}}$ & err & Order & Star & $T_{\rm eff}$& $BC_{\rm G}$ & $BC_{\rm G_{\rm BP}}$ & $BC_{\rm G_{\rm RP}}$ & err & Order & Star\\
         (K) & (mag) & (mag) & (mag) & (mag) & & & (K) & (mag) & (mag) & (mag) & (mag) \\
    \midrule
        33400 & -2.781 & -2.610 & -3.209 & 0.004 & 30 &  HD\,36512 & 12828 & -0.603 & -0.555 & -0.695 & 0.006 & 88 &  HD\,222762  \\ 
        32983 & -2.755 & -2.585 & -3.178 & 0.007 & 81 &  HD\,209975   & 12153 & -0.538 & -0.509 & -0.580 & 0.007 & 65 &  HD\,171301  \\ 
        32100 & -2.689 & -2.524 & -3.102 & 0.008 & 69 &  HD\,189957   & 11800 & -0.467 & -0.437 & -0.514 & 0.005 & 87 &  HD\,222173  \\ 
        31400 & -2.644 & -2.484 & -3.035 & 0.007 & 78 &  HD\,207538   & 11695 & -0.447 & -0.417 & -0.494 & 0.008 & 19 &  HD\,32040  \\ 
        31057 & -2.615 & -2.454 & -3.011 & 0.009 & 23 &  HD\,34078   & 11572 & -0.426 & -0.395 & -0.469 & 0.004 & 13 &  HD\,27295  \\ 
        29000 & -2.398 & -2.235 & -2.799 & 0.003 & 32 &  HD\,36960   & 10900 & -0.343 & -0.330 & -0.348 & 0.005 & 8 &  HD\,15318  \\ 
        27000 & -2.211 & -2.051 & -2.596 & 0.003 & 31 &  HD\,36591   & 10864 & -0.347 & -0.338 & -0.344 & 0.007 & 24 &  HD\,34310  \\ 
        26810 & -2.248 & -2.112 & -2.588 & 0.008 & 77 &  HD\,205139   & 10568 & -0.334 & -0.342 & -0.289 & 0.008 & 62 &  HD\,150117  \\ 
        24870 & -2.048 & -1.906 & -2.376 & 0.006 & 82 &  HD\,213087   & 10484 & -0.292 & -0.293 & -0.276 & 0.007 & 41 &  HD\,78556  \\ 
        24840 & -2.069 & -1.932 & -2.388 & 0.006 & 79 &  HD\,208266   & 10471 & -0.298 & -0.300 & -0.269 & 0.005 & 21 &  HD\,32309  \\ 
        24090 & -1.927 & -1.777 & -2.289 & 0.010 & 16 &  HD\,29138   & 10174 & -0.265 & -0.275 & -0.220 & 0.008 & 53 &  HD\,120198  \\ 
        24000 & -1.919 & -1.772 & -2.274 & 0.006 & 25 &  HD\,35299   & 9705 & -0.208 & -0.236 & -0.124 & 0.005 & 40 &  HD\,71155  \\ 
        24000 & -1.937 & -1.798 & -2.275 & 0.007 & 35 &  HD\,37744   & 9640 & -0.188 & -0.212 & -0.112 & 0.004 & 3 &  HD\,1439  \\ 
        23570 & -1.901 & -1.766 & -2.228 & 0.006 & 74 &  HD\,197512   & 9605 & -0.200 & -0.226 & -0.117 & 0.007 & 5 &  HD\,4778  \\ 
        22370 & -1.752 & -1.615 & -2.080 & 0.006 & 34 &  HD\,37356   & 9280 & -0.274 & -0.357 & -0.061 & 0.012 & 47 &  HD\,89021  \\ 
        22010 & -1.712 & -1.573 & -2.038 & 0.006 & 85 &  HD\,215191   & 9209 & -0.148 & -0.192 & -0.039 & 0.005 & 84 &  HD\,214994  \\ 
        21700 & -1.673 & -1.536 & -1.998 & 0.005 & 28 &  HD\,36285   & 9016 & -0.133 & -0.187 & -0.004 & 0.007 & 26 &  HD\,35693  \\ 
        21616 & -1.661 & -1.521 & -1.989 & 0.005 & 39 &  HD\,55856   & 8892 & -0.118 & -0.174 & 0.019 & 0.005 & 43 &  HD\,82621  \\ 
        21000 & -1.601 & -1.470 & -1.911 & 0.004 & 55 &  HD\,125924   & 8840 & -0.119 & -0.177 & 0.025 & 0.007 & 2 &  HD\,1404  \\ 
        20860 & -1.587 & -1.455 & -1.894 & 0.007 & 83 &  HD\,214263   & 8689 & -0.115 & -0.190 & 0.059 & 0.007 & 48 &  HD\,99922  \\ 
        20700 & -1.594 & -1.468 & -1.885 & 0.007 & 61 &  HD\,148688   & 7396 & -0.008 & -0.162 & 0.302 & 0.005 & 64 &  HD\,162570  \\ 
        20480 & -1.548 & -1.417 & -1.847 & 0.006 & 71 &  HD\,193183   & 7250 & 0.014 & -0.137 & 0.328 & 0.006 & 20 &  HD\,32115  \\ 
        19833 & -1.480 & -1.359 & -1.763 & 0.007 & 27 &  HD\,35912   & 7174 & 0.022 & -0.132 & 0.342 & 0.005 & 33 &  HD\,37077  \\ 
        19612 & -1.454 & -1.335 & -1.730 & 0.004 & 22 &  HD\,32612   & 6575 & 0.036 & -0.175 & 0.445 & 0.003 & 56 &  HD\,128167  \\ 
        19300 & -1.419 & -1.302 & -1.687 & 0.005 & 29 &  HD\,36430   & 6484 & 0.061 & -0.132 & 0.453 & 0.006 & 45 &  HD\,84937  \\ 
        19000 & -1.374 & -1.255 & -1.652 & 0.006 & 9 &  HD\,16440   & 5981 & 0.018 & -0.248 & 0.514 & 0.006 & 44 &  HD\,82943  \\ 
        19000 & -1.393 & -1.282 & -1.648 & 0.005 & 38 &  HD\,54764   & 5962 & 0.037 & -0.225 & 0.526 & 0.004 & 11 &  HD\,22879  \\ 
        18200 & -1.298 & -1.192 & -1.539 & 0.005 & 63 &  HD\,160762   & 5824 & 0.011 & -0.281 & 0.536 & 0.006 & 59 &  HD\,146233  \\ 
        17680 & -1.221 & -1.115 & -1.468 & 0.008 & 72 &  HD\,195556   & 5807 & 0.011 & -0.285 & 0.541 & 0.003 & 67 &  HD\,186427  \\ 
        17010 & -1.134 & -1.028 & -1.371 & 0.009 & 37 &  HD\,46189   & 5800 & 0.024 & -0.262 & 0.545 & 0.004 & 49 &  HD\,101364  \\ 
        17000 & -1.166 & -1.068 & -1.374 & 0.005 & 60 &  HD\,148379   & 5746 & 0.009 & -0.290 & 0.543 & 0.005 & 86 &  HD\,217014  \\ 
        16258 & -1.030 & -0.931 & -1.247 & 0.004 & 36 &  HD\,40967   & 5473 & -0.010 & -0.345 & 0.574 & 0.003 & 52 &  HD\,117176  \\ 
        15427 & -0.933 & -0.846 & -1.120 & 0.004 & 73 &  HD\,196740   & 5397 & -0.075 & -0.448 & 0.559 & 0.008 & 7 &  HD\,10700  \\ 
        15345 & -0.917 & -0.827 & -1.107 & 0.006 & 15 &  HD\,27778   & 5280 & -0.038 & -0.411 & 0.588 & 0.004 & 58 &  HD\,141714  \\ 
        15020 & -0.903 & -0.822 & -1.062 & 0.004 & 80 &  HD\,209419   & 5235 & -0.026 & -0.389 & 0.593 & 0.004 & 50 &  HD\,103095  \\ 
        14727 & -0.838 & -0.755 & -1.014 & 0.007 & 10 &  HD\,19736   & 5171 & -0.156 & -0.615 & 0.570 & 0.011 & 57 &  HD\,133208  \\ 
        14681 & -0.856 & -0.781 & -1.013 & 0.007 & 66 &  HD\,174959   & 4967 & -0.124 & -0.586 & 0.593 & 0.007 & 70 &  HD\,192263  \\ 
        14400 & -0.790 & -0.710 & -0.957 & 0.005 & 18 &  HD\,29589   & 4950 & -0.162 & -0.640 & 0.562 & 0.011 & 51 &  HD\,113226  \\ 
        14300 & -0.779 & -0.699 & -0.945 & 0.008 & 6 &  HD\,10362   & 4836 & -0.165 & -0.657 & 0.586 & 0.008 & 42 &  HD\,82106  \\ 
        14207 & -0.768 & -0.691 & -0.925 & 0.005 & 12 &  HD\,23300   & 4642 & -0.135 & -0.601 & 0.599 & 0.005 & 54 &  HD\,122563  \\ 
        14125 & -0.756 & -0.678 & -0.913 & 0.005 & 4 &  HD\,2729   & 4484 & -0.265 & -0.818 & 0.543 & 0.008 & 46 &  HD\,85503  \\ 
        13711 & -0.709 & -0.641 & -0.843 & 0.005 & 17 &  HD\,29335   & 4398 & -0.290 & -0.903 & 0.570 & 0.007 & 75 &  HD\,201091  \\ 
        13461 & -0.676 & -0.612 & -0.803 & 0.006 & 14 &  HD\,27563   & 4174 & -0.366 & -1.008 & 0.526 & 0.007 & 76 &  HD\,201092  \\ 
        13300 & -0.651 & -0.588 & -0.774 & 0.006 & 1 &  HD\,1279   & 3904 & -0.471 & -1.183 & 0.477 & 0.009 & 68 &  HD\,189319 \\
        \bottomrule
    \end{tabular}}
    \label{tab:7}
\end{table*}

\subsection{Spectroscopic and Classical $BC_\xi -T_{\rm eff}$ Relations}

Spectroscopic and classical functions of $BC_{\xi}-T_{\rm eff}$ were calibrated by fitting the fourth-degree polynomials according to the least-squares method using already defined (\citetalias{Yucel2026}) effective temperature ($T_{\rm eff}$) values versus spectroscopic $BC_{\xi}$ data from the {\it Gaia} spectra of this study and the $T_{\rm eff}$ versus classical $BC_{\xi}$ data from absolute bolometric ($M_{\rm Bol}$) and filtered ($M_{\xi}$) {\it Gaia} magnitudes. The coefficients, errors in the coefficients, and related statistics [standard deviations, correlations, validity ranges, the two temperatures at which $BC=0$, the temperature at maximum, and $BC$ at the solar temperature (5772 K)] are given in Tables~\ref{tab:8} and \ref{tab:9} for the spectroscopic and classical relations, respectively.      
\begin{table*}[h]
\centering
    \caption{Spectroscopic $BC-T_{\rm eff}$ functions for the {\it Gaia} passbands, in which $BC$ data are from {\it Gaia} spectra.}
   \resizebox{\textwidth}{!}{\begin{tabular}{ccccccc}
    \toprule
        & \multicolumn{5}{c}{$BC = a + b \times (\log{T_{\rm eff}}) + c \times (\log{T_{\rm eff}})^2 + d \times (\log{T_{\rm eff}})^3 + e \times (\log{T_{\rm eff}})^4$} \\
        & $a$ & $b$ & $c$ & $d$ & $e$ \\
        \midrule
        \textit{Gaia G} & -827.678 & 714.86 & -229.83 & 32.764 & -1.7641 \\
        & ($\pm$183.100) & ($\pm$180.70) & ($\pm$66.76) & ($\pm$10.950) & ($\pm$0.6719) \\
        & \multicolumn{5}{c}{$\sigma = 0.024$, $R^2 = 0.9992$}\\
        & \multicolumn{5}{c}{valid in the range $3\,904 \leq T_{\rm eff} \leq 33\,400$ K}\\
        & & $BC = 0$ & $T_{\rm eff,1}= 5\,789$ K & $T_{\rm eff,2}= 7\,432$ K \\
        & & $BC_{\rm max} = 0.026$& $T_{\rm eff}= 6\,545$ K\\
        & & $BC_\odot = -0.001$ & $T_\odot = 5\,772$ K  \\
        \midrule
        \textit{Gaia} $G_{\rm BP}$ & -1350.93 & 1192.43  & -393.78 & 57.847 & -3.206 \\
        & ($\pm$279.80) & ($\pm$276.09) & ($\pm$102.02) & ($\pm$16.731) & ($\pm$1.027) \\
        & \multicolumn{5}{c}{$\sigma = 0.036$, $R^2 = 0.9974$}\\
        & \multicolumn{5}{c}{valid in the range $3\,904 \leq T_{\rm eff} \leq 33\,400$ K}\\
        & & $BC_{\rm max} = -0.013$& $T_{\rm eff}= 7\,694$ K\\ 
        & & $BC_\odot = -0.184$ & $T_\odot = 5\,772$ K  \\
        \midrule

        \textit{Gaia} $G_{\rm RP}$ & -1125.84 & 1030.58 & -351.882 & 53.332 & -3.046  \\
        & ($\pm$59.69) & ($\pm$58.92) & ($\pm$21.761) & ($\pm$3.568) & ($\pm$0.219) \\
        & \multicolumn{5}{c}{$\sigma = 0.008$, $R^2 = 0.9999$}\\
        & \multicolumn{5}{c}{valid in the range $3\,904 \leq T_{\rm eff} \leq 33\,400$ K}\\
        & & $BC = 0.00$ & $T_{\rm eff}= 8\,975$ K \\
        & & $BC_{\rm max} = 0.490$& $T_{\rm eff}= 4\,907$ K\\
        & & $BC_\odot = 0.526$ & $T_\odot = 5\,772$ K  \\
        \bottomrule
    \end{tabular}}
    \label{tab:8}
\end{table*}

\begin{table*}[h]
\centering
    \caption{Classical $BC-T_{\rm eff}$ functions for the {\it Gaia} passbands, in which $BC$ data are from bolometric and filtered magnitudes as $M_{bol}$ - $M_{\xi}$, where $\xi$ stands for one of the {\it Gaia} passbands.}
   \resizebox{\textwidth}{!}{\begin{tabular}{ccccccc}
    \toprule
        & \multicolumn{5}{c}{$BC = a + b \times (\log{T_{\rm eff}}) + c \times (\log{T_{\rm eff}})^2 + d \times (\log{T_{\rm eff}})^3 + e \times (\log{T_{\rm eff}})^4$} \\
        & $a$ & $b$ & $c$ & $d$ & $e$ \\
        \midrule
        \textit{Gaia G}  & -646.104 & 536.891 & -164.25 & 22.016 & -1.104 \\
        & ($\pm$527.498) & ($\pm$520.601) & ($\pm$192.30) & ($\pm$31.541) & ($\pm$1.936) \\
        & \multicolumn{5}{c}{$\sigma = 0.067$, $R^2 = 0.995$}\\
        & \multicolumn{5}{c}{valid in the range $3\,904 \leq T_{\rm eff} \leq 33\,400$ K}\\
        & & $BC = 0$ & $T_{\rm eff,1}= 4\,865$ K & $T_{\rm eff,2}= 8\,175$ K \\
        & & $BC_{\rm max} = 0.114$& $T_{\rm eff}= 6\,253$ K\\
        & & $BC_\odot = 0.103$ & $T_\odot = 5\,772$ K  \\
        \midrule
        \textit{Gaia} $G_{\rm BP}$ & -1913.860 & 1719.060 & -577.91 & 86.384 & -4.862 \\
        & ($\pm$630.997) & ($\pm$322.700) & ($\pm$230.10)  & ($\pm$37.721) & ($\pm$2.315) \\
        & \multicolumn{5}{c}{$\sigma = 0.080$, $R^2 = 0.989$}\\
        & \multicolumn{5}{c}{valid in the range $3\,904 \leq T_{\rm eff} \leq 33\,400$ K}\\
        & & $BC_{\rm max} = -0.048$& $T_{\rm eff}= 7\,543$ K\\ 
        & & $BC_\odot = -0.223$ & $T_\odot = 5\,772$ K  \\
        \midrule

        \textit{Gaia} $G_{\rm RP}$ & -2375.020 & 2265.900 & -808.95  & 128.326   & -7.650 \\
        & ($\pm$587.700) & ($\pm$217.10) & ($\pm$35.602) & ($\pm$595.501) & ($\pm$2.185) \\
        & \multicolumn{5}{c}{$\sigma = 0.076$, $R^2 = 0.996$}\\
        & \multicolumn{5}{c}{valid in the range $3\,904 \leq T_{\rm eff} \leq 33\,400$ K}\\
        & & $BC = 0.00$ & $T_{\rm eff}= 8\,814$ K \\
        & & $BC_{\rm max} = 0.582$ & $T_{\rm eff}= 5\,010$ K\\
        & & $BC_\odot = 0.537$ & $T_\odot = 5\,772$ K  \\
        \bottomrule
    \end{tabular}}
    \label{tab:9}
\end{table*}


\begin{figure}
    \centering
    \includegraphics[width=0.9\linewidth]{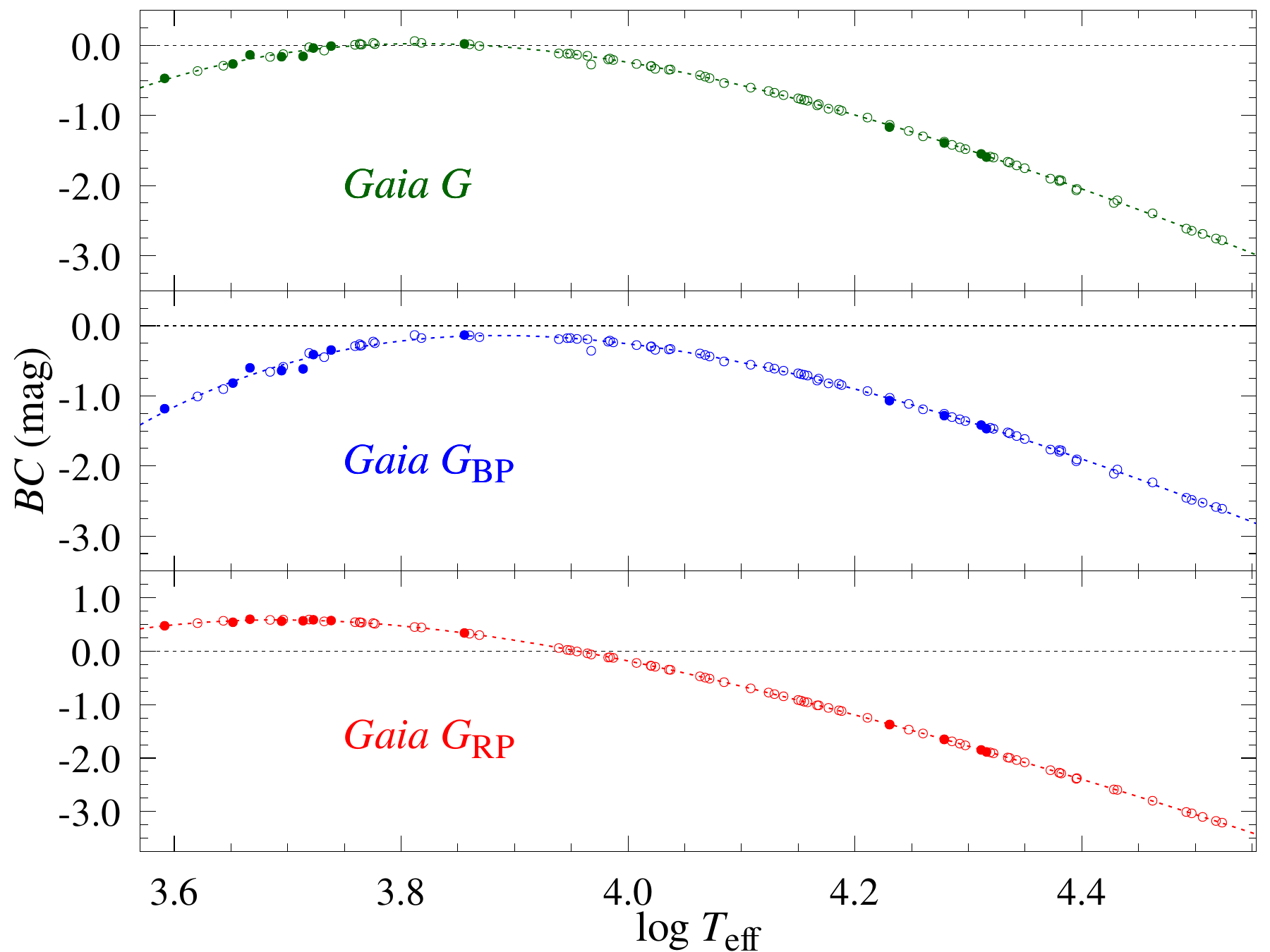}
    \caption{Spectroscopic $BC_\xi-T_{\rm eff}$ function with data.}
    \label{fig:spec_BCs}
\end{figure}


\begin{figure}
    \centering
    \includegraphics[width=0.9\linewidth]{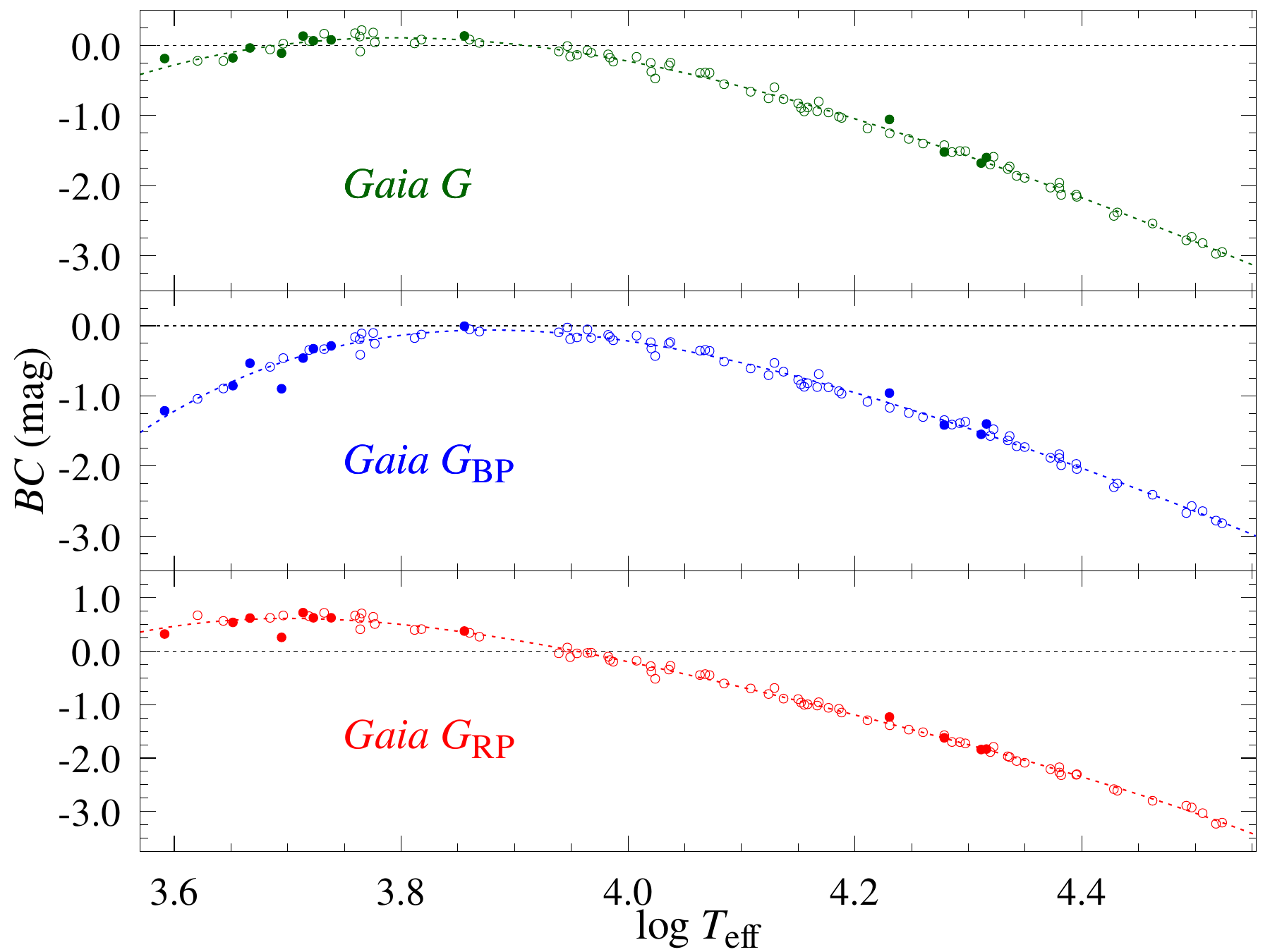}
    \caption{Classical $BC_\xi-T_{\rm eff}$ function with data.}
    \label{fig:phot_BCs}
\end{figure}

The spectroscopic and classical functions $BC_{\xi}-T_{\rm eff}$ are plotted in Figures~\ref{fig:spec_BCs} and~\ref{fig:phot_BCs}, respectively, together with the data. The smaller Standard Deviations of the spectroscopic data (Figure~\ref{fig:spec_BCs}) relative to the Standard Deviations of the classical data are obvious. Higher correlations of the spectroscopic function are also remarkable (see Tables~\ref{tab:8} and~\ref{tab:9}).


\section{Discussions}
\subsection{Why Spectroscopic $BCs$ are more accurate}

The about three times larger standard deviations of the data on the classical $BC-T_{\rm eff}$ functions with respect to the standard deviations on the spectroscopic $BC-T_{\rm eff}$ functions are given in Tables~\ref{tab:8} and~\ref{tab:9}, respectively. Why are $BC-T_{\rm eff}$ functions determined from the observed spectra much more accurate than classical functions?

Based on a single observed entity (spectrum), it is natural for a spectroscopic {\it BC} value to be more accurate than the classically computed $BC$ value, which requires at least four observational quantities, effective temperature and radius for calculating the absolute bolometric magnitude, apparent brightness, and trigonometric parallaxes for calculating the absolute filtered magnitude, which may also require another parameter, interstellar extinction, if the star is not sufficiently nearby. \cite{Eker2021b} have already discussed the great advantage of using spectroscopic {\it BC} values when computing standard stellar luminosities. 

Especially after \citetalias{Yucel2026}, which explained how to obtain a spectroscopic $BC$ from an observed spectrum, and this study, which used the same method to obtain \textit{Gaia} Spectroscopic $BC$s, we can say that today's technology has reached a level of measuring effective temperature and a bolometric correction of a star for a photometric band if its observed spectrum is wide enough to cover the wavelength range of the filter and has sufficient resolution to distinguish certain spectral lines useful for determining its effective temperature, and sufficiently high signal to noise ratio for minimizing errors in determining spectral continuum and normalization. If the star has an accurate trigonometric \textit{Gaia} parallax, one may even obtain the standard luminosity of the star from its apparent magnitude by using its spectroscopic $BC$ in an accuracy better than the Stefan-Boltzmann law could provide \citep{Eker2021b}. It is even possible to improve the accuracy and precision of the standard temperature of the star further by using multiple $BC$ values representing independent photometric observations at different wavelength ranges by the method described by \cite{Bakis2022}. Not only can its standard luminosity, but also its radius, be estimated with sufficient accuracy, from an accurately determined luminosity as shown by \cite{Eker2023} and \cite{Okuyan2026}, for the single stars as accurately as the radii of eclipsing binaries.

Therefore, we encourage improving the accuracy of the already determined zero-point constants by increasing the number of sample spectra. Having sufficiently large wavelength coverage and signal-to-noise ratio, the \textit{Gaia} spectrum is ideal for this purpose. Improving the accuracy of the zero-point constants will reduce the systematic errors in the measured $BC$ values. Introducing more photometric bands would be useful in further improving the accuracy of the predicted luminosities.

\subsection{Comparing $BC-T_{\rm eff}$ Functions}

\begin{figure}
    \centering
    \includegraphics[width=1\linewidth]{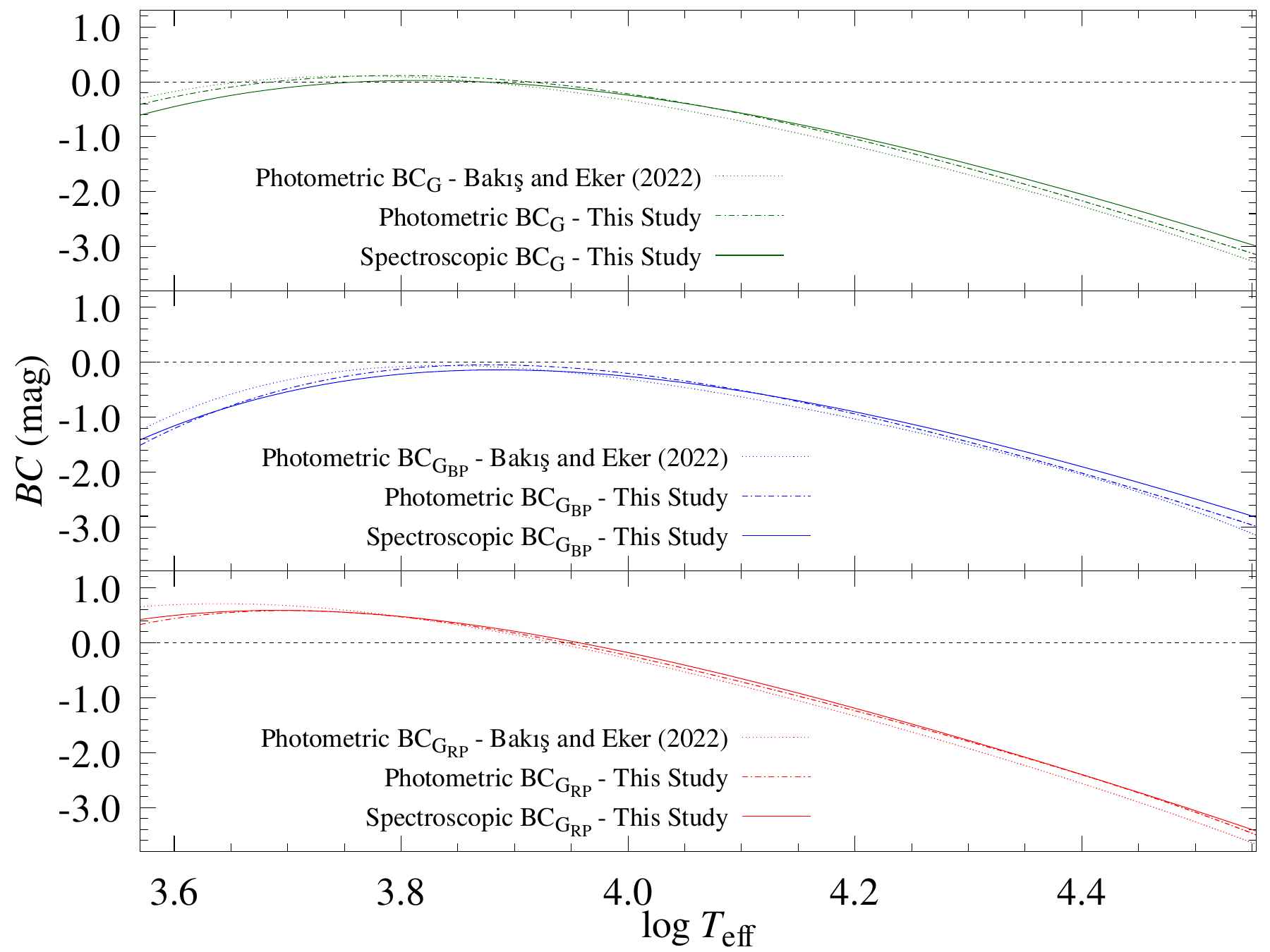}
    \caption{Spectroscopic and classical $BC_\xi-T_{\rm eff}$ of this study is compared the classical functions of $BC_\xi-T_{\rm eff}$ of \protect\cite{Bakis2022}.}
    \label{fig:comparison_g}
\end{figure}

The spectroscopic and classical curves $BC_\xi - T_{\rm eff}$ of this study are compared (Figure~\ref{fig:comparison_g} to the classical curves $BC_\xi - T_{\rm eff}$ of \cite{Bakis2022}, which are also for the \textit{Gaia} photometric bands, but extracted from detached eclipsing binaries. The agreement among them within the error limits is clear. Except for the curves of $G_{\rm BP}$, the curves of the other two bands, $G$ and $G_{\rm RP}$, cross over the horizontal axis. In other words, the functions related to $G$ and $G_{\rm RP}$ bands contain a limited amount of $BC$ values greater than zero. This would not have been acceptable before the paradigms ``{\it BC} of a star must always be negative'', ``the bolometric magnitude of a star ought to be brighter than its $V$-magnitude,'' and ``the zero point of bolometric corrections are arbitrary'' \citep{Torres2010, Eker2021, Eker2021b, Eker2022} discarded by the General Assembly resolution B2 announced after 2015 IAU meeting in Honolulu \citep{Mamajek2015}. The non-arbitrariness of the zero-point constants not only for the bolometric corrections of the visual band but also for the zero-point constants of the \textit{Gaia} bands is now undeniable.     

\subsection{Comparing distances}

The internal consistency of the stellar parameters in this study, which are reduced to absolute bolometric magnitude ($M_{\rm Bol}$) and bolometric correction ($BC_{\xi}$), is tested by comparing photometric and Gaia trigonometric distances. The Photometric distance of a star in parsec units is calculated using,   
\begin{equation}
    d_\xi = 10^{\left(\xi-M_\xi+5-A_\xi\right)/5
    },
    \label{eq:15}
\end{equation}
where ($\xi-A_\xi-M_\xi$) is the photometric distance modulus corrected for interstellar extinction, in which $M_\xi$ is the photometric absolute magnitude of a star calculated as $M_\xi=M_{\rm Bol}-BC_\xi$ from stellar parameters $R$ and $T_{\rm eff}$ and spectroscopic bolometric corrections.

\begin{figure*}
    \centering
    \includegraphics[width=0.32\linewidth]{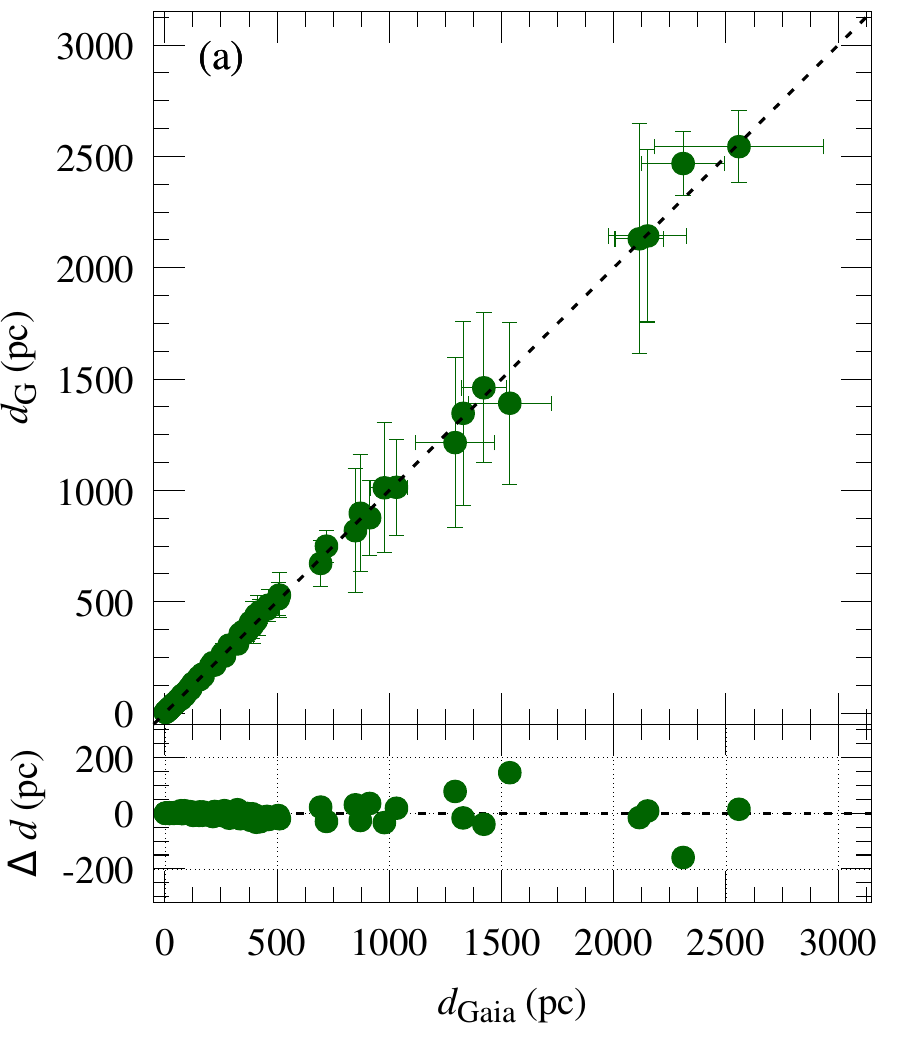}
    \includegraphics[width=0.32\linewidth]{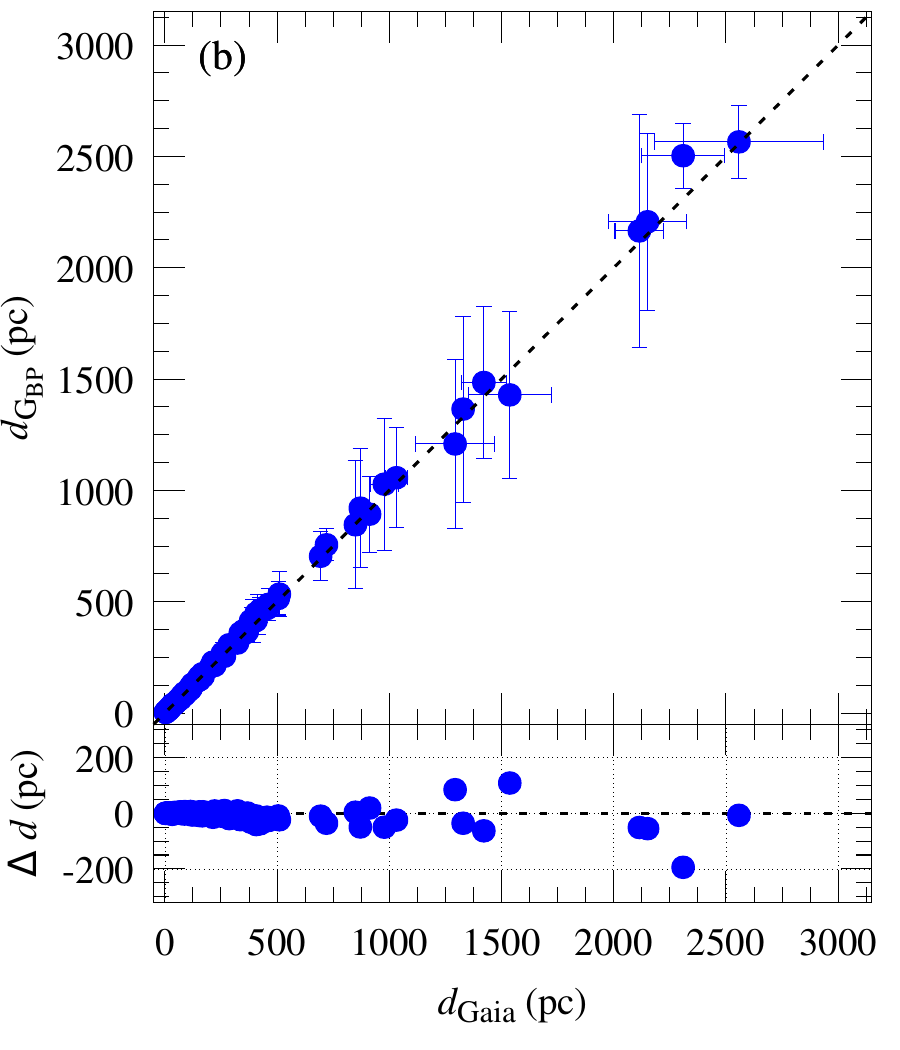}
    \includegraphics[width=0.32\linewidth]{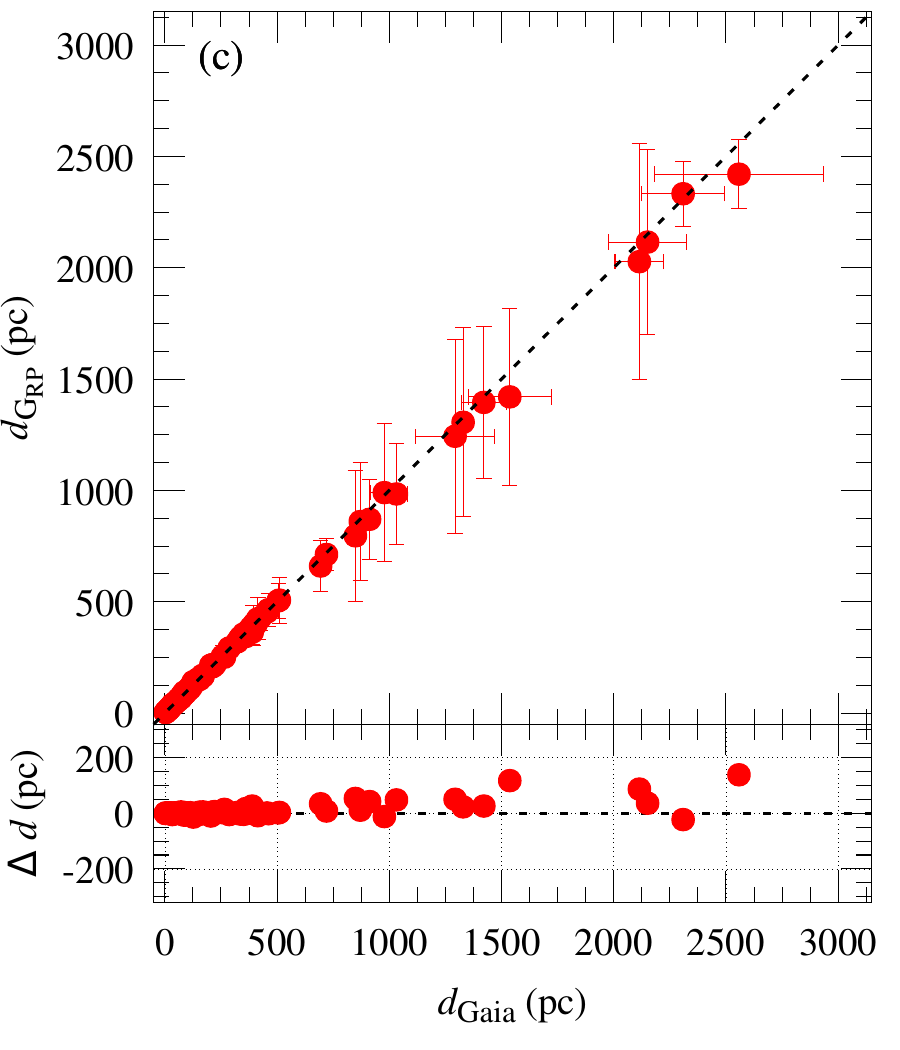} \\
    \includegraphics[width=0.32\linewidth]{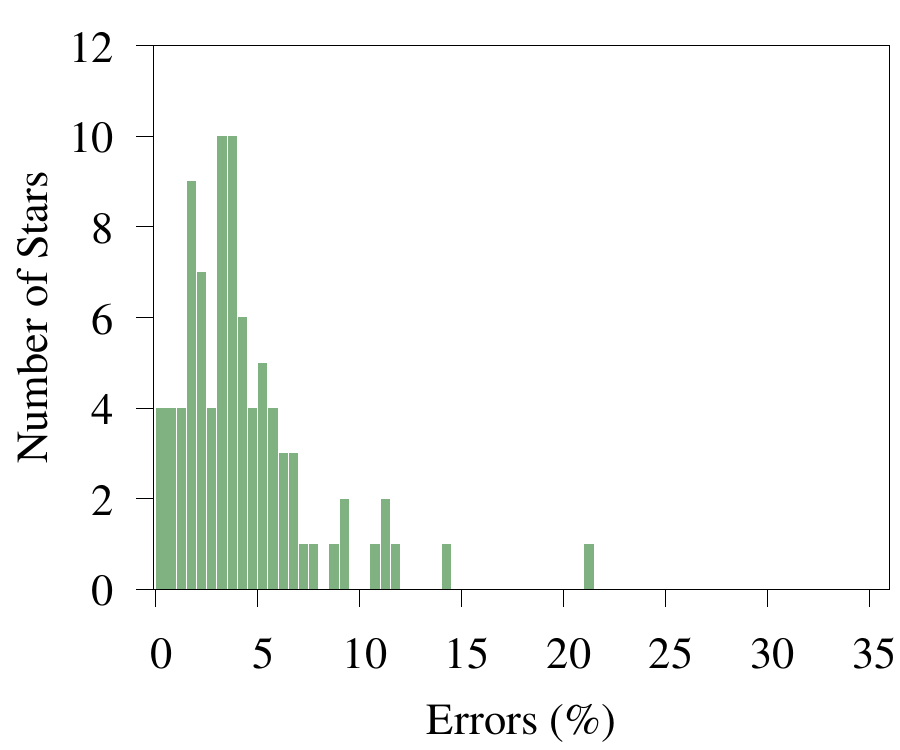}
    \includegraphics[width=0.32\linewidth]{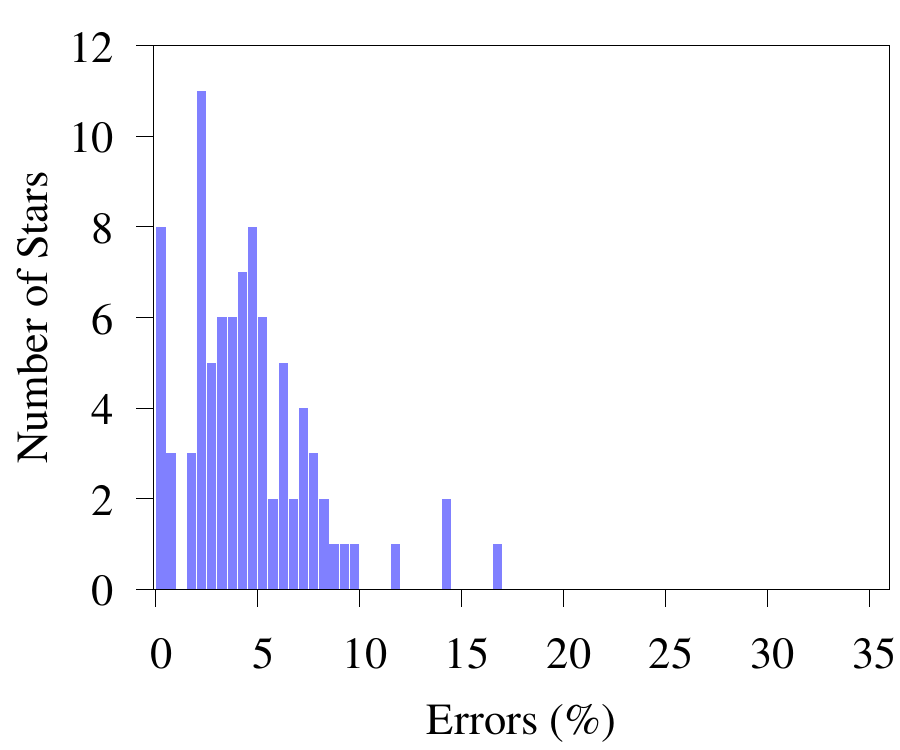}
    \includegraphics[width=0.32\linewidth]{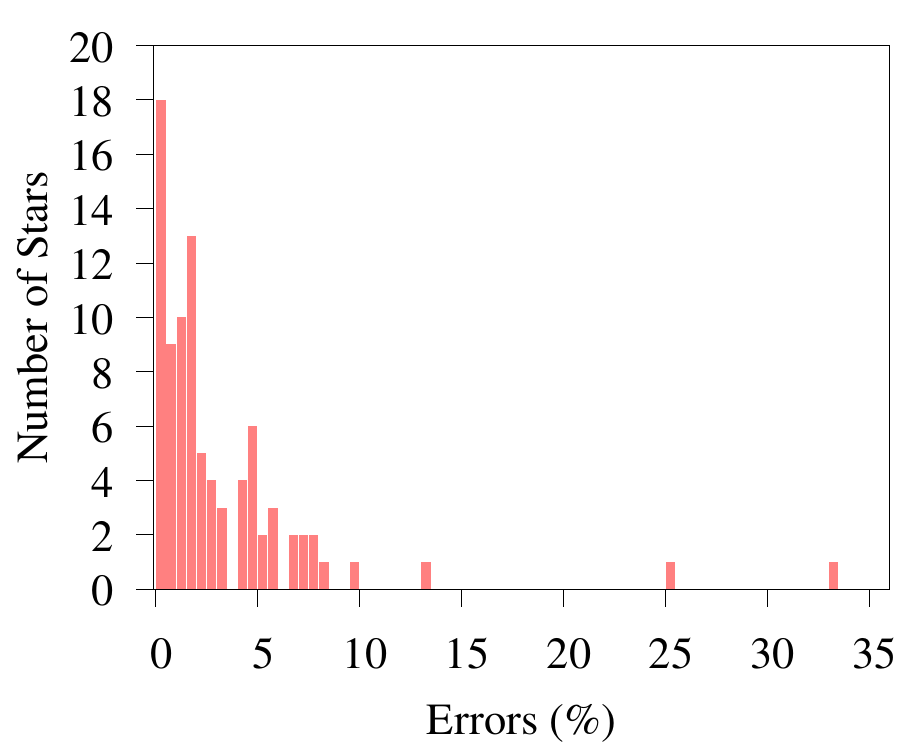}

    \caption{Comparison of the distances of 88 sample stars calculated using spectroscopic $BC$s with the \textit{Gaia}-based values for the \textit{Gaia} $G$ (upper left panel), \textit{Gaia} $G_{\rm BP}$ (upper middle panel), and \textit{Gaia} $G_{\rm RP}$ (upper right panel) bandpasses. The lower panels show the corresponding relative errors for each bandpass.}
    \label{fig:bc_distances}
\end{figure*}

The photometric distances at Gaia bandpasses are compared to the Gaia trigonometric distances in the upper panels of Figure~\ref{fig:bc_distances}, where the high correlation between the photometric and trigonometric distances is clearly seen. The lower panels show estimated errors in percent units for the photometric distances in Gaia bands $G$, $G_{\rm BP}$, and $G_{\rm RP}$, respectively. Therefore, we can conclude that the stellar parameters $R$ and $T_{\rm eff}$ and the spectroscopic bolometric corrections in this study are consistent with producing distances more accurate than five-percent error for 62 stars (70 percent of the sample) for the $G$ and $G_{\rm BP}$ bands and 72 stars (82 percent of the sample) for the $G_{\rm RP}$ band.

\section{Conclusion}
The results obtained in this study are listed below. 
\begin{enumerate}
    \item By analyzing 88 stars with {\it Gaia} XP spectra, the empirical zero-point constants of bolometric corrections, $C_2({\xi})$s, for the {\it Gaia} passbands were determined as $0.8677\pm0.0109$ mag, $1.0449\pm0.0116$ mag, and $2.0510\pm0.0087$ mag for $G$, $G_{\rm BP}$, and $G_{\rm RP}$, respectively.
    
    \item The zero-point constants of \textit{Gaia} passbands, $C_{\rm G}=70.1525\pm0.0109$ mag and $c_{\rm G}=-19.8651\pm0.0105$ mag, $C_{\rm G_{\rm BP}}=70.1525\pm0.0116$ mag, and $c_{\rm G_{\rm BP}}=-20.0423\pm0.0116$ mag, and $C_{\rm G_{\rm RP}}=69.1464\pm0.0087$ mag and $c_{\rm G_{\rm RP}}=-21.0484\pm0.0087$ mag for the absolute and apparent magnitudes were empirically obtained to be used, respectively, by $L_{\xi}$ and $f_{\xi}$ are in SI units to calculate absolute bolometric ($M_{bol}$) and filtered ($M_\xi$) magnitudes of stars.     
    
    \item Spectroscopic $BC_\xi$ of the sample stars for the three {\it Gaia} passbands have been calculated. Spectroscopic $BCs$ were found to be much more accurate than classical $BCs$ from absolute bolometric ($M_{bol}$) and filtered ($M_\xi$) magnitudes based on astrometric and photometric observations.    
    
    \item  Spectroscopic and classical functions $BC_\xi-T_{\rm eff}$ for {\it Gaia} passbands, $G$, $G_{\rm BP}$, and $G_{\rm RP}$, were calibrated by fitting a fourth degree polynomial according to the least squares method. The standard deviation of the spectroscopic $BC$ values from the functions is much smaller than the standard deviations of the classical data; thus, we suggest the spectroscopic functions to estimate an accurate $BC$ for a star with no spectra and distance from a single parameter, effective temperature.
     
    \item Observational stellar parameters $R$ and $T_{\rm eff}$ used in this study, and the spectroscopic bolometric corrections measured from stellar spectra are tested by comparing photometric and trigonometric distances and were found successful to produce a photometric distance for a star with an accuracy better than five percent error.   

\end{enumerate}

\section*{Acknowledgements}

We thank T\"{U}B{\.{I}}TAK for funding this research under project number 123C161. We thank Prof. Volkan Bak{\i}\c{s} for his valuable discussions to make this study better. This research has made use of NASA's Astrophysics Data System. The VizieR and Simbad databases at CDS, Strasbourg, France were invaluable for the project as were data from the European Space Agency (ESA) mission \emph{Gaia}\footnote{https://www.cosmos.esa.int/gaia}, processed by the \emph{Gaia} Data Processing and Analysis Consortium (DPAC)\footnote{https://www.cosmos.esa.int/web/gaia/dpac/consortium}. Funding for DPAC has been provided by national institutions, in particular, the institutions participating in the \emph{Gaia} Multilateral Agreement. This work made use of Astropy:\footnote{http://www.astropy.org}, a community-developed core Python package and an ecosystem of tools and resources for astronomy \citep{astro1, astro2, astro3}.


\bibliographystyle{mnras}
\bibliography{reference}

@ARTICLE{Aschenbrenner2023,
       author = {{Aschenbrenner}, P. and {Przybilla}, N. and {Butler}, K.},
      journal = {\aap},
         year = 2023,
        month = mar,
       volume = {671},
        title = "{Quantitative spectroscopy of late O-type main-sequence stars with a hybrid non-LTE method}",
          eid = {A36},
        pages = {A36},
          doi = {10.1051/0004-6361/202244906},
archivePrefix = {arXiv},
       eprint = {2301.09462},
 primaryClass = {astro-ph.SR},
       adsurl = {https://ui.adsabs.harvard.edu/abs/2023A&A...671A..36A}
}

@ARTICLE{Adelman2017,
       author = {{Adelman}, S.~J. and {Gulliver}, A.~F. and {Gucella}, L.~J.},
        title = "{Elemental abundance analyses with DAO spectrograms: XL}",
      journal = {Astronomische Nachrichten},
         year = 2017,
        month = jun,
       volume = {338},
       number = {5},
        pages = {584-597},
          doi = {10.1002/asna.201613214},
       adsurl = {https://ui.adsabs.harvard.edu/abs/2017AN....338..584A}
}

@ARTICLE{Arentsen2019,
       author = {{Arentsen}, Anke and {Prugniel}, Philippe and {Gonneau}, Anais and {Lan{\c{c}}on}, Ariane and {Trager}, Scott and {Peletier}, Reynier and {Lyubenova}, Mariya and {Chen}, Yan-Ping and {Falc{\'o}n Barroso}, Jes{\'u}s and {S{\'a}nchez Bl{\'a}zquez}, Patricia and {Vazdekis}, Alejandro},
        title = "{Stellar atmospheric parameters for 754 spectra from the X-shooter Spectral Library}",
      journal = {\aap},
         year = 2019,
        month = jul,
       volume = {627},
          eid = {A138},
        pages = {A138},
          doi = {10.1051/0004-6361/201834273},
archivePrefix = {arXiv},
       eprint = {1907.06391},
 primaryClass = {astro-ph.SR},
       adsurl = {https://ui.adsabs.harvard.edu/abs/2019A&A...627A.138A}
}

@ARTICLE{astro1,
       author = {{Astropy Collaboration} and {Robitaille}, Thomas P. and {Tollerud}, Erik J. and {Greenfield}, Perry and {Droettboom}, Michael and {Bray}, Erik and {Aldcroft}, Tom and {Davis}, Matt and {Ginsburg}, Adam and {Price-Whelan}, Adrian M. and {Kerzendorf}, Wolfgang E. and {Conley}, Alexander and {Crighton}, Neil and {Barbary}, Kyle and {Muna}, Demitri and {Ferguson}, Henry and {Grollier}, Fr{\'e}d{\'e}ric and {Parikh}, Madhura M. and {Nair}, Prasanth H. and {Unther}, Hans M. and {Deil}, Christoph and {Woillez}, Julien and {Conseil}, Simon and {Kramer}, Roban and {Turner}, James E.~H. and {Singer}, Leo and {Fox}, Ryan and {Weaver}, Benjamin A. and {Zabalza}, Victor and {Edwards}, Zachary I. and {Azalee Bostroem}, K. and {Burke}, D.~J. and {Casey}, Andrew R. and {Crawford}, Steven M. and {Dencheva}, Nadia and {Ely}, Justin and {Jenness}, Tim and {Labrie}, Kathleen and {Lim}, Pey Lian and {Pierfederici}, Francesco and {Pontzen}, Andrew and {Ptak}, Andy and {Refsdal}, Brian and {Servillat}, Mathieu and {Streicher}, Ole},
        title = "{Astropy: A community Python package for astronomy}",
      journal = {\aap},
         year = 2013,
        month = oct,
       volume = {558},
          eid = {A33},
        pages = {A33},
          doi = {10.1051/0004-6361/201322068},
archivePrefix = {arXiv},
       eprint = {1307.6212},
 primaryClass = {astro-ph.IM},
       adsurl = {https://ui.adsabs.harvard.edu/abs/2013A&A...558A..33A}
}

@ARTICLE{astro2,
       author = {{Astropy Collaboration} and {Price-Whelan}, A.~M. and {Sip{\H{o}}cz}, B.~M. and {G{\"u}nther}, H.~M. and {Lim}, P.~L. and {Crawford}, S.~M. and {Conseil}, S. and {Shupe}, D.~L. and {Craig}, M.~W. and {Dencheva}, N. and {Ginsburg}, A. and {VanderPlas}, J.~T. and {Bradley}, L.~D. and {P{\'e}rez-Su{\'a}rez}, D. and {de Val-Borro}, M. and {Aldcroft}, T.~L. and {Cruz}, K.~L. and {Robitaille}, T.~P. and {Tollerud}, E.~J. and {Ardelean}, C. and {Babej}, T. and {Bach}, Y.~P. and {Bachetti}, M. and {Bakanov}, A.~V. and {Bamford}, S.~P. and {Barentsen}, G. and {Barmby}, P. and {Baumbach}, A. and {Berry}, K.~L. and {Biscani}, F. and {Boquien}, M. and {Bostroem}, K.~A. and {Bouma}, L.~G. and {Brammer}, G.~B. and {Bray}, E.~M. and {Breytenbach}, H. and {Buddelmeijer}, H. and {Burke}, D.~J. and {Calderone}, G. and {Cano Rodr{\'\i}guez}, J.~L. and {Cara}, M. and {Cardoso}, J.~V.~M. and {Cheedella}, S. and {Copin}, Y. and {Corrales}, L. and {Crichton}, D. and {D'Avella}, D. and {Deil}, C. and {Depagne}, {\'E}. and {Dietrich}, J.~P. and {Donath}, A. and {Droettboom}, M. and {Earl}, N. and {Erben}, T. and {Fabbro}, S. and {Ferreira}, L.~A. and {Finethy}, T. and {Fox}, R.~T. and {Garrison}, L.~H. and {Gibbons}, S.~L.~J. and {Goldstein}, D.~A. and {Gommers}, R. and {Greco}, J.~P. and {Greenfield}, P. and {Groener}, A.~M. and {Grollier}, F. and {Hagen}, A. and {Hirst}, P. and {Homeier}, D. and {Horton}, A.~J. and {Hosseinzadeh}, G. and {Hu}, L. and {Hunkeler}, J.~S. and {Ivezi{\'c}}, {\v{Z}}. and {Jain}, A. and {Jenness}, T. and {Kanarek}, G. and {Kendrew}, S. and {Kern}, N.~S. and {Kerzendorf}, W.~E. and {Khvalko}, A. and {King}, J. and {Kirkby}, D. and {Kulkarni}, A.~M. and {Kumar}, A. and {Lee}, A. and {Lenz}, D. and {Littlefair}, S.~P. and {Ma}, Z. and {Macleod}, D.~M. and {Mastropietro}, M. and {McCully}, C. and {Montagnac}, S. and {Morris}, B.~M. and {Mueller}, M. and {Mumford}, S.~J. and {Muna}, D. and {Murphy}, N.~A. and {Nelson}, S. and {Nguyen}, G.~H. and {Ninan}, J.~P. and {N{\"o}the}, M. and {Ogaz}, S. and {Oh}, S. and {Parejko}, J.~K. and {Parley}, N. and {Pascual}, S. and {Patil}, R. and {Patil}, A.~A. and {Plunkett}, A.~L. and {Prochaska}, J.~X. and {Rastogi}, T. and {Reddy Janga}, V. and {Sabater}, J. and {Sakurikar}, P. and {Seifert}, M. and {Sherbert}, L.~E. and {Sherwood-Taylor}, H. and {Shih}, A.~Y. and {Sick}, J. and {Silbiger}, M.~T. and {Singanamalla}, S. and {Singer}, L.~P. and {Sladen}, P.~H. and {Sooley}, K.~A. and {Sornarajah}, S. and {Streicher}, O. and {Teuben}, P. and {Thomas}, S.~W. and {Tremblay}, G.~R. and {Turner}, J.~E.~H. and {Terr{\'o}n}, V. and {van Kerkwijk}, M.~H. and {de la Vega}, A. and {Watkins}, L.~L. and {Weaver}, B.~A. and {Whitmore}, J.~B. and {Woillez}, J. and {Zabalza}, V. and {Astropy Contributors}},
        title = "{The Astropy Project: Building an Open-science Project and Status of the v2.0 Core Package}",
      journal = {\aj},
         year = 2018,
        month = sep,
       volume = {156},
       number = {3},
          eid = {123},
        pages = {123},
          doi = {10.3847/1538-3881/aabc4f},
archivePrefix = {arXiv},
       eprint = {1801.02634},
 primaryClass = {astro-ph.IM},
       adsurl = {https://ui.adsabs.harvard.edu/abs/2018AJ....156..123A}
}

@ARTICLE{astro3,
       author = {{Astropy Collaboration} and {Price-Whelan}, Adrian M. and {Lim}, Pey Lian and {Earl}, Nicholas and {Starkman}, Nathaniel and {Bradley}, Larry and {Shupe}, David L. and {Patil}, Aarya A. and {Corrales}, Lia and {Brasseur}, C.~E. and {N{\"o}the}, Maximilian and {Donath}, Axel and {Tollerud}, Erik and {Morris}, Brett M. and {Ginsburg}, Adam and {Vaher}, Eero and {Weaver}, Benjamin A. and {Tocknell}, James and {Jamieson}, William and {van Kerkwijk}, Marten H. and {Robitaille}, Thomas P. and {Merry}, Bruce and {Bachetti}, Matteo and {G{\"u}nther}, H. Moritz and {Aldcroft}, Thomas L. and {Alvarado-Montes}, Jaime A. and {Archibald}, Anne M. and {B{\'o}di}, Attila and {Bapat}, Shreyas and {Barentsen}, Geert and {Baz{\'a}n}, Juanjo and {Biswas}, Manish and {Boquien}, M{\'e}d{\'e}ric and {Burke}, D.~J. and {Cara}, Daria and {Cara}, Mihai and {Conroy}, Kyle E. and {Conseil}, Simon and {Craig}, Matthew W. and {Cross}, Robert M. and {Cruz}, Kelle L. and {D'Eugenio}, Francesco and {Dencheva}, Nadia and {Devillepoix}, Hadrien A.~R. and {Dietrich}, J{\"o}rg P. and {Eigenbrot}, Arthur Davis and {Erben}, Thomas and {Ferreira}, Leonardo and {Foreman-Mackey}, Daniel and {Fox}, Ryan and {Freij}, Nabil and {Garg}, Suyog and {Geda}, Robel and {Glattly}, Lauren and {Gondhalekar}, Yash and {Gordon}, Karl D. and {Grant}, David and {Greenfield}, Perry and {Groener}, Austen M. and {Guest}, Steve and {Gurovich}, Sebastian and {Handberg}, Rasmus and {Hart}, Akeem and {Hatfield-Dodds}, Zac and {Homeier}, Derek and {Hosseinzadeh}, Griffin and {Jenness}, Tim and {Jones}, Craig K. and {Joseph}, Prajwel and {Kalmbach}, J. Bryce and {Karamehmetoglu}, Emir and {Ka{\l}uszy{\'n}ski}, Miko{\l}aj and {Kelley}, Michael S.~P. and {Kern}, Nicholas and {Kerzendorf}, Wolfgang E. and {Koch}, Eric W. and {Kulumani}, Shankar and {Lee}, Antony and {Ly}, Chun and {Ma}, Zhiyuan and {MacBride}, Conor and {Maljaars}, Jakob M. and {Muna}, Demitri and {Murphy}, N.~A. and {Norman}, Henrik and {O'Steen}, Richard and {Oman}, Kyle A. and {Pacifici}, Camilla and {Pascual}, Sergio and {Pascual-Granado}, J. and {Patil}, Rohit R. and {Perren}, Gabriel I. and {Pickering}, Timothy E. and {Rastogi}, Tanuj and {Roulston}, Benjamin R. and {Ryan}, Daniel F. and {Rykoff}, Eli S. and {Sabater}, Jose and {Sakurikar}, Parikshit and {Salgado}, Jes{\'u}s and {Sanghi}, Aniket and {Saunders}, Nicholas and {Savchenko}, Volodymyr and {Schwardt}, Ludwig and {Seifert-Eckert}, Michael and {Shih}, Albert Y. and {Jain}, Anany Shrey and {Shukla}, Gyanendra and {Sick}, Jonathan and {Simpson}, Chris and {Singanamalla}, Sudheesh and {Singer}, Leo P. and {Singhal}, Jaladh and {Sinha}, Manodeep and {Sip{\H{o}}cz}, Brigitta M. and {Spitler}, Lee R. and {Stansby}, David and {Streicher}, Ole and {{\v{S}}umak}, Jani and {Swinbank}, John D. and {Taranu}, Dan S. and {Tewary}, Nikita and {Tremblay}, Grant R. and {de Val-Borro}, Miguel and {Van Kooten}, Samuel J. and {Vasovi{\'c}}, Zlatan and {Verma}, Shresth and {de Miranda Cardoso}, Jos{\'e} Vin{\'\i}cius and {Williams}, Peter K.~G. and {Wilson}, Tom J. and {Winkel}, Benjamin and {Wood-Vasey}, W.~M. and {Xue}, Rui and {Yoachim}, Peter and {Zhang}, Chen and {Zonca}, Andrea and {Astropy Project Contributors}},
        title = "{The Astropy Project: Sustaining and Growing a Community-oriented Open-source Project and the Latest Major Release (v5.0) of the Core Package}",
      journal = {\apj},
         year = 2022,
        month = aug,
       volume = {935},
       number = {2},
          eid = {167},
        pages = {167},
          doi = {10.3847/1538-4357/ac7c74},
archivePrefix = {arXiv},
       eprint = {2206.14220},
 primaryClass = {astro-ph.IM},
       adsurl = {https://ui.adsabs.harvard.edu/abs/2022ApJ...935..167A}
}

@ARTICLE{Bailey2013,
       author = {{Bailey}, J.~D. and {Landstreet}, J.~D.},
        title = "{Abundances determined using Si ii and Si iii in B-type stars: evidence for stratification}",
      journal = {\aap},
         year = 2013,
        month = mar,
       volume = {551},
          eid = {A30},
        pages = {A30},
          doi = {10.1051/0004-6361/201220671},
archivePrefix = {arXiv},
       eprint = {1301.3050},
 primaryClass = {astro-ph.SR},
       adsurl = {https://ui.adsabs.harvard.edu/abs/2013A&A...551A..30B}
}

@ARTICLE{Bakis2022,
       author = {{Bak{\i}{\c{s}}}, V. and {Eker}, Z.},
        title = "{A Method of Improving Standard Stellar Luminosities with Multiband Standard Bolometric Corrections}",
      journal = {\actaa},
         year = 2022,
        month = dec,
       volume = {72},
       number = {3},
        pages = {195-232},
          doi = {10.32023/0001-5237/72.3.4},
archivePrefix = {arXiv},
       eprint = {2208.04110},
 primaryClass = {astro-ph.SR},
       adsurl = {https://ui.adsabs.harvard.edu/abs/2022AcA....72..195B}
}

@ARTICLE{Bessell1990,
       author = {{Bessell}, M.~S.},
        title = "{UBVRI passbands.}",
      journal = {\pasp},
         year = 1990,
        month = oct,
       volume = {102},
        pages = {1181-1199},
          doi = {10.1086/132749},
       adsurl = {https://ui.adsabs.harvard.edu/abs/1990PASP..102.1181B}
}

@ARTICLE{Borisov2023,
       author = {{Borisov}, Sviatoslav B. and {Chilingarian}, Igor V. and {Rubtsov}, Evgenii V. and {Ledoux}, C{\'e}dric and {Melo}, Claudio and {Grishin}, Kirill A. and {Katkov}, Ivan Yu. and {Goradzhanov}, Vladimir S. and {Afanasiev}, Anton V. and {Kasparova}, Anastasia V. and {Saburova}, Anna S.},
        title = "{New Generation Stellar Spectral Libraries in the Optical and Near-infrared. I. The Recalibrated UVES-POP Library for Stellar Population Synthesis}",
      journal = {\apjs},
         year = 2023,
        month = may,
       volume = {266},
       number = {1},
          eid = {11},
        pages = {11},
          doi = {10.3847/1538-4365/acc321},
archivePrefix = {arXiv},
       eprint = {2211.09130},
 primaryClass = {astro-ph.IM},
       adsurl = {https://ui.adsabs.harvard.edu/abs/2023ApJS..266...11B}
}

@ARTICLE{Bressan2012,
       author = {{Bressan}, Alessandro and {Marigo}, Paola and {Girardi}, L{\'e}o. and {Salasnich}, Bernardo and {Dal Cero}, Claudia and {Rubele}, Stefano and {Nanni}, Ambra},
        title = "{PARSEC: stellar tracks and isochrones with the PAdova and TRieste Stellar Evolution Code}",
      journal = {\mnras},
         year = 2012,
        month = nov,
       volume = {427},
       number = {1},
        pages = {127-145},
          doi = {10.1111/j.1365-2966.2012.21948.x},
archivePrefix = {arXiv},
       eprint = {1208.4498},
 primaryClass = {astro-ph.SR},
       adsurl = {https://ui.adsabs.harvard.edu/abs/2012MNRAS.427..127B}
}

@ARTICLE{Catanzaro1997,
       author = {{Catanzaro}, G.},
        title = "{High Resolution Spectral Atlas of Telluric Lines}",
      journal = {\apss},
         year = 1997,
        month = mar,
       volume = {257},
       number = {1},
        pages = {161-170},
          doi = {10.1023/A:1001197016750},
       adsurl = {https://ui.adsabs.harvard.edu/abs/1997Ap&SS.257..161C}
}

@ARTICLE{Cenarro2007,
       author = {{Cenarro}, A.~J. and {Peletier}, R.~F. and {S{\'a}nchez-Bl{\'a}zquez}, P. and {Selam}, S.~O. and {Toloba}, E. and {Cardiel}, N. and {Falc{\'o}n-Barroso}, J. and {Gorgas}, J. and {Jim{\'e}nez-Vicente}, J. and {Vazdekis}, A.},
        title = "{Medium-resolution Isaac Newton Telescope library of empirical spectra - II. The stellar atmospheric parameters}",
      journal = {\mnras},
         year = 2007,
        month = jan,
       volume = {374},
       number = {2},
        pages = {664-690},
          doi = {10.1111/j.1365-2966.2006.11196.x},
archivePrefix = {arXiv},
       eprint = {astro-ph/0611618},
 primaryClass = {astro-ph},
       adsurl = {https://ui.adsabs.harvard.edu/abs/2007MNRAS.374..664C}
}

@ARTICLE{Cunha1994,
       author = {{Cunha}, Katia and {Lambert}, David L.},
        title = "{Chemical Evolution of the Orion Association. II. The Carbon, Nitrogen, Oxygen, Silicon, and Iron Abundances of Main-Sequence B Stars}",
      journal = {\apj},
         year = 1994,
        month = may,
       volume = {426},
        pages = {170},
          doi = {10.1086/174053},
       adsurl = {https://ui.adsabs.harvard.edu/abs/1994ApJ...426..170C}
}

@ARTICLE{Daflon2003,
       author = {{Daflon}, S. and {Cunha}, K. and {Smith}, V.~V. and {Butler}, K.},
        title = "{Non-LTE abundances of magnesium, aluminum and sulfur in OB stars near the solar circle}",
      journal = {\aap},
         year = 2003,
        month = feb,
       volume = {399},
        pages = {525-530},
          doi = {10.1051/0004-6361:20021802},
archivePrefix = {arXiv},
       eprint = {astro-ph/0212171},
 primaryClass = {astro-ph},
       adsurl = {https://ui.adsabs.harvard.edu/abs/2003A&A...399..525D}
}

@ARTICLE{Daflon2007,
       author = {{Daflon}, Simone and {Cunha}, Katia and {de Ara{\'u}jo}, Francisco X. and {Wolff}, Sidney and {Przybilla}, Norbert},
        title = "{The Projected Rotational Velocity Distribution of a Sample of OB stars from a Calibration Based on Synthetic He I Lines}",
      journal = {\aj},
         year = 2007,
        month = oct,
       volume = {134},
       number = {4},
        pages = {1570-1578},
          doi = {10.1086/521707},
archivePrefix = {arXiv},
       eprint = {0707.3934},
 primaryClass = {astro-ph},
       adsurl = {https://ui.adsabs.harvard.edu/abs/2007AJ....134.1570D}
}

@ARTICLE{dasilva2015,
       author = {{da Silva}, Ronaldo and {Milone}, Andr{\'e} de C. and {Rocha-Pinto}, Helio J.},
        title = "{Homogeneous abundance analysis of FGK dwarf, subgiant, and giant stars with and without giant planets}",
      journal = {\aap},
         year = 2015,
        month = aug,
       volume = {580},
          eid = {A24},
        pages = {A24},
          doi = {10.1051/0004-6361/201525770},
       adsurl = {https://ui.adsabs.harvard.edu/abs/2015A&A...580A..24D}
}

@ARTICLE{DeAngeli2023,
       author = {{De Angeli}, F. and {Weiler}, M. and {Montegriffo}, P. and {Evans}, D.~W. and {Riello}, M. and {Andrae}, R. and {Carrasco}, J.~M. and {Busso}, G. and {Burgess}, P.~W. and {Cacciari}, C. and {Davidson}, M. and {Harrison}, D.~L. and {Hodgkin}, S.~T. and {Jordi}, C. and {Osborne}, P.~J. and {Pancino}, E. and {Altavilla}, G. and {Barstow}, M.~A. and {Bailer-Jones}, C.~A.~L. and {Bellazzini}, M. and {Brown}, A.~G.~A. and {Castellani}, M. and {Cowell}, S. and {Delchambre}, L. and {De Luise}, F. and {Diener}, C. and {Fabricius}, C. and {Fouesneau}, M. and {Fr{\'e}mat}, Y. and {Gilmore}, G. and {Giuffrida}, G. and {Hambly}, N.~C. and {Hidalgo}, S. and {Holland}, G. and {Kostrzewa-Rutkowska}, Z. and {van Leeuwen}, F. and {Lobel}, A. and {Marinoni}, S. and {Miller}, N. and {Pagani}, C. and {Palaversa}, L. and {Piersimoni}, A.~M. and {Pulone}, L. and {Ragaini}, S. and {Rainer}, M. and {Richards}, P.~J. and {Rixon}, G.~T. and {Ruz-Mieres}, D. and {Sanna}, N. and {Sarro}, L.~M. and {Rowell}, N. and {Sordo}, R. and {Walton}, N.~A. and {Yoldas}, A.},
        title = "{Gaia Data Release 3. Processing and validation of BP/RP low-resolution spectral data}",
      journal = {\aap},
         year = 2023,
        month = jun,
       volume = {674},
          eid = {A2},
        pages = {A2},
          doi = {10.1051/0004-6361/202243680},
archivePrefix = {arXiv},
       eprint = {2206.06143},
 primaryClass = {astro-ph.IM},
       adsurl = {https://ui.adsabs.harvard.edu/abs/2023A&A...674A...2D}
}

@BOOK{Eddington1926,
       author = {{Eddington}, A.~S.},
        title = "{The Internal Constitution of the Stars}",
         year = 1926,
       adsurl = {https://ui.adsabs.harvard.edu/abs/1926ics..book.....E}
}

@ARTICLE{Eker2021,
       author = {{Eker}, Z. and {Bak{\i}{\c{s}}}, V. and {Soydugan}, F. and {Bilir}, S.},
        title = "{On the zero point constant of the bolometric correction scale}",
      journal = {\mnras},
         year = 2021,
        month = may,
       volume = {503},
       number = {3},
        pages = {4231-4241},
          doi = {10.1093/mnras/stab684},
archivePrefix = {arXiv},
       eprint = {2103.03258},
 primaryClass = {astro-ph.SR},
       adsurl = {https://ui.adsabs.harvard.edu/abs/2021MNRAS.503.4231E}
}

@ARTICLE{Eker2021b,
       author = {{Eker}, Z. and {Soydugan}, F. and {Bilir}, S. and {Bak{\i}{\c{s}}}, V.},
        title = "{Standard stellar luminosities: what are typical and limiting accuracies in the era after Gaia?}",
      journal = {\mnras},
         year = 2021,
        month = nov,
       volume = {507},
       number = {3},
        pages = {3583-3592},
          doi = {10.1093/mnras/stab2302},
archivePrefix = {arXiv},
       eprint = {2108.01092},
 primaryClass = {astro-ph.SR},
       adsurl = {https://ui.adsabs.harvard.edu/abs/2021MNRAS.507.3583E}
}

@ARTICLE{Eker2022,
       author = {{Eker}, Zeki and {Soydugan}, Faruk and {Bak{\i}{\c{s}}}, Volkan and {Bilir}, Sel{\c{c}}uk and {Steer}, Ian},
        title = "{A Silent Revolution in Fundamental Astrophysics}",
      journal = {\aj},
         year = 2022,
        month = nov,
       volume = {164},
       number = {5},
          eid = {189},
        pages = {189},
          doi = {10.3847/1538-3881/ac9123},
archivePrefix = {arXiv},
       eprint = {2209.06254},
 primaryClass = {astro-ph.SR},
       adsurl = {https://ui.adsabs.harvard.edu/abs/2022AJ....164..189E}
}

@ARTICLE{Eker2023,
       author = {{Eker}, Z. and {Bak{\i}{\c{s}}}, V.},
        title = "{Testing multiband (G, G$_{BP}$, G$_{RP}$, B, V, and TESS) standard bolometric corrections by recovering luminosity and radii of 341 host stars}",
      journal = {\mnras},
         year = 2023,
        month = aug,
       volume = {523},
       number = {2},
        pages = {2440-2452},
          doi = {10.1093/mnras/stad1563},
archivePrefix = {arXiv},
       eprint = {2305.12538},
 primaryClass = {astro-ph.SR},
       adsurl = {https://ui.adsabs.harvard.edu/abs/2023MNRAS.523.2440E}
}

@ARTICLE{Eker2025,
       author = {{Eker}, Zeki and {Bak{\i}\c{s}}, Volkan},
        title = "{Fundamentals of Stars II: Revisiting Bolometric Corrections}",
      journal = {Physics and Astronomy Reports},
         year = 2025,
        month = jun,
       volume = {3},
        pages = {43-61},
          doi = {10.26650/PAR.2025.00005},
archivePrefix = {arXiv},
       eprint = {2503.19965},
 primaryClass = {astro-ph.SR},
       adsurl = {https://ui.adsabs.harvard.edu/abs/2025PARep...3...43E}
}

@ARTICLE{Espadons,
       author = {{Romanovskaya}, A.~M. and {Shulyak}, D.~V. and {Ryabchikova}, T.~A. and {Sitnova}, T.~M.},
        title = "{Fundamental parameters of the Ap-stars GO And, 84 UMa, and {\ensuremath{\kappa}} Psc}",
      journal = {\aap},
         year = 2021,
        month = nov,
       volume = {655},
          eid = {A106},
        pages = {A106},
          doi = {10.1051/0004-6361/202141740},
archivePrefix = {arXiv},
       eprint = {2111.03371},
 primaryClass = {astro-ph.SR},
       adsurl = {https://ui.adsabs.harvard.edu/abs/2021A&A...655A.106R}
}

@ARTICLE{Fossati2011,
       author = {{Fossati}, L. and {Ryabchikova}, T. and {Shulyak}, D.~V. and {Haswell}, C.~A. and {Elmasli}, A. and {Pandey}, C.~P. and {Barnes}, T.~G. and {Zwintz}, K.},
        title = "{The accuracy of stellar atmospheric parameter determinations: a case study with HD 32115 and HD 37594}",
      journal = {\mnras},
         year = 2011,
        month = oct,
       volume = {417},
       number = {1},
        pages = {495-507},
          doi = {10.1111/j.1365-2966.2011.19289.x},
archivePrefix = {arXiv},
       eprint = {1106.4406},
 primaryClass = {astro-ph.SR},
       adsurl = {https://ui.adsabs.harvard.edu/abs/2011MNRAS.417..495F}
}

@ARTICLE{Fraser2010,
       author = {{Fraser}, M. and {Dufton}, P.~L. and {Hunter}, I. and {Ryans}, R.~S.~I.},
        title = "{Atmospheric parameters and rotational velocities for a sample of Galactic B-type supergiants}",
      journal = {\mnras},
         year = 2010,
        month = may,
       volume = {404},
       number = {3},
        pages = {1306-1320},
          doi = {10.1111/j.1365-2966.2010.16392.x},
archivePrefix = {arXiv},
       eprint = {1001.3337},
 primaryClass = {astro-ph.SR},
       adsurl = {https://ui.adsabs.harvard.edu/abs/2010MNRAS.404.1306F}
}

@article{Gaia_DR3,
       author = {{Gaia Collaboration} and {Vallenari}, A. and {Brown}, A.~G.~A. and {Prusti}, T. and {de Bruijne}, J.~H.~J. and {Arenou}, F. and {Babusiaux}, C. and {Biermann}, M. and {Creevey}, O.~L. and {Ducourant}, C. and {Evans}, D.~W. and {Eyer}, L. and {Guerra}, R. and {Hutton}, A. and {Jordi}, C. and {Klioner}, S.~A. and {Lammers}, U.~L. and {Lindegren}, L. and {Luri}, X. and {Mignard}, F. and {Panem}, C. and {Pourbaix}, D. and {Randich}, S. and {Sartoretti}, P. and {Soubiran}, C. and {Tanga}, P. and {Walton}, N.~A. and {Bailer-Jones}, C.~A.~L. and {Bastian}, U. and {Drimmel}, R. and {Jansen}, F. and {Katz}, D. and {Lattanzi}, M.~G. and {van Leeuwen}, F. and {Bakker}, J. and {Cacciari}, C. and {Casta{\~n}eda}, J. and {De Angeli}, F. and {Fabricius}, C. and {Fouesneau}, M. and {Fr{\'e}mat}, Y. and {Galluccio}, L. and {Guerrier}, A. and {Heiter}, U. and {Masana}, E. and {Messineo}, R. and {Mowlavi}, N. and {Nicolas}, C. and {Nienartowicz}, K. and {Pailler}, F. and {Panuzzo}, P. and {Riclet}, F. and {Roux}, W. and {Seabroke}, G.~M. and {Sordo{\o}rcit}, R. and {Th{\'e}venin}, F. and {Gracia-Abril}, G. and {Portell}, J. and {Teyssier}, D. and {Altmann}, M. and {Andrae}, R. and {Audard}, M. and {Bellas-Velidis}, I. and {Benson}, K. and {Berthier}, J. and {Blomme}, R. and {Burgess}, P.~W. and {Busonero}, D. and {Busso}, G. and {C{\'a}novas}, H. and {Carry}, B. and {Cellino}, A. and {Cheek}, N. and {Clementini}, G. and {Damerdji}, Y. and {Davidson}, M. and {de Teodoro}, P. and {Nu{\~n}ez Campos}, M. and {Delchambre}, L. and {Dell'Oro}, A. and {Esquej}, P. and {Fern{\'a}ndez-Hern{\'a}ndez}, J. and {Fraile}, E. and {Garabato}, D. and {Garc{\'\i}a-Lario}, P. and {Gosset}, E. and {Haigron}, R. and {Halbwachs}, J. -L. and {Hambly}, N.~C. and {Harrison}, D.~L. and {Hern{\'a}ndez}, J. and {Hestroffer}, D. and {Hodgkin}, S.~T. and {Holl}, B. and {Jan{\ss}en}, K. and {Jevardat de Fombelle}, G. and {Jordan}, S. and {Krone-Martins}, A. and {Lanzafame}, A.~C. and {L{\"o}ffler}, W. and {Marchal}, O. and {Marrese}, P.~M. and {Moitinho}, A. and {Muinonen}, K. and {Osborne}, P. and {Pancino}, E. and {Pauwels}, T. and {Recio-Blanco}, A. and {Reyl{\'e}}, C. and {Riello}, M. and {Rimoldini}, L. and {Roegiers}, T. and {Rybizki}, J. and {Sarro}, L.~M. and {Siopis}, C. and {Smith}, M. and {Sozzetti}, A. and {Utrilla}, E. and {van Leeuwen}, M. and {Abbas}, U. and {{\'A}brah{\'a}m}, P. and {Abreu Aramburu}, A. and {Aerts}, C. and {Aguado}, J.~J. and {Ajaj}, M. and {Aldea-Montero}, F. and {Altavilla}, G. and {{\'A}lvarez}, M.~A. and {Alves}, J. and {Anders}, F. and {Anderson}, R.~I. and {Anglada Varela}, E. and {Antoja}, T. and {Baines}, D. and {Baker}, S.~G. and {Balaguer-N{\'u}{\~n}ez}, L. and {Balbinot}, E. and {Balog}, Z. and {Barache}, C. and {Barbato}, D. and {Barros}, M. and {Barstow}, M.~A. and {Bartolom{\'e}}, S. and {Bassilana}, J. -L. and {Bauchet}, N. and {Becciani}, U. and {Bellazzini}, M. and {Berihuete}, A. and {Bernet}, M. and {Bertone}, S. and {Bianchi}, L. and {Binnenfeld}, A. and {Blanco-Cuaresma}, S. and {Blazere}, A. and {Boch}, T. and {Bombrun}, A. and {Bossini}, D. and {Bouquillon}, S. and {Bragaglia}, A. and {Bramante}, L. and {Breedt}, E. and {Bressan}, A. and {Brouillet}, N. and {Brugaletta}, E. and {Bucciarelli}, B. and {Burlacu}, A. and {Butkevich}, A.~G. and {Buzzi}, R. and {Caffau}, E. and {Cancelliere}, R. and {Cantat-Gaudin}, T. and {Carballo}, R. and {Carlucci}, T. and {Carnerero}, M.~I. and {Carrasco}, J.~M. and {Casamiquela}, L. and {Castellani}, M. and {Castro-Ginard}, A. and {Chaoul}, L. and {Charlot}, P. and {Chemin}, L. and {Chiaramida}, V. and {Chiavassa}, A. and {Chornay}, N. and {Comoretto}, G. and {Contursi}, G. and {Cooper}, W.~J. and {Cornez}, T. and {Cowell}, S. and {Crifo}, F. and {Cropper}, M. and {Crosta}, M. and {Crowley}, C. and {Dafonte}, C. and {Dapergolas}, A. and {David}, M. and {David}, P. and {de Laverny}, P. and {De Luise}, F. and {De March}, R. and {De Ridder}, J. and {de Souza}, R. and {de Torres}, A. and {del Peloso}, E.~F. and {del Pozo}, E. and {Delbo}, M. and {Delgado}, A. and {Delisle}, J. -B. and {Demouchy}, C. and {Dharmawardena}, T.~E. and {Di Matteo}, P. and {Diakite}, S. and {Diener}, C. and {Distefano}, E. and {Dolding}, C. and {Edvardsson}, B. and {Enke}, H. and {Fabre}, C. and {Fabrizio}, M. and {Faigler}, S. and {Fedorets}, G. and {Fernique}, P. and {Fienga}, A. and {Figueras}, F. and {Fournier}, Y. and {Fouron}, C. and {Fragkoudi}, F. and {Gai}, M. and {Garcia-Gutierrez}, A. and {Garcia-Reinaldos}, M. and {Garc{\'\i}a-Torres}, M. and {Garofalo}, A. and {Gavel}, A. and {Gavras}, P. and {Gerlach}, E. and {Geyer}, R. and {Giacobbe}, P. and {Gilmore}, G. and {Girona}, S. and {Giuffrida}, G. and {Gomel}, R. and {Gomez}, A. and {Gonz{\'a}lez-N{\'u}{\~n}ez}, J. and {Gonz{\'a}lez-Santamar{\'\i}a}, I. and {Gonz{\'a}lez-Vidal}, J.~J. and {Granvik}, M. and {Guillout}, P. and {Guiraud}, J. and {Guti{\'e}rrez-S{\'a}nchez}, R. and {Guy}, L.~P. and {Hatzidimitriou}, D. and {Hauser}, M. and {Haywood}, M. and {Helmer}, A. and {Helmi}, A. and {Sarmiento}, M.~H. and {Hidalgo}, S.~L. and {Hilger}, T. and {H{\l}adczuk}, N. and {Hobbs}, D. and {Holland}, G. and {Huckle}, H.~E. and {Jardine}, K. and {Jasniewicz}, G. and {Jean-Antoine Piccolo}, A. and {Jim{\'e}nez-Arranz}, {\'O}. and {Jorissen}, A. and {Juaristi Campillo}, J. and {Julbe}, F. and {Karbevska}, L. and {Kervella}, P. and {Khanna}, S. and {Kontizas}, M. and {Kordopatis}, G. and {Korn}, A.~J. and {K{\'o}sp{\'a}l}, {\'A} and {Kostrzewa-Rutkowska}, Z. and {Kruszy{\'n}ska}, K. and {Kun}, M. and {Laizeau}, P. and {Lambert}, S. and {Lanza}, A.~F. and {Lasne}, Y. and {Le Campion}, J. -F. and {Lebreton}, Y. and {Lebzelter}, T. and {Leccia}, S. and {Leclerc}, N. and {Lecoeur-Taibi}, I. and {Liao}, S. and {Licata}, E.~L. and {Lindstr{\o}m}, H.~E.~P. and {Lister}, T.~A. and {Livanou}, E. and {Lobel}, A. and {Lorca}, A. and {Loup}, C. and {Madrero Pardo}, P. and {Magdaleno Romeo}, A. and {Managau}, S. and {Mann}, R.~G. and {Manteiga}, M. and {Marchant}, J.~M. and {Marconi}, M. and {Marcos}, J. and {Marcos Santos}, M.~M.~S. and {Mar{\'\i}n Pina}, D. and {Marinoni}, S. and {Marocco}, F. and {Marshall}, D.~J. and {Polo}, L. Martin and {Mart{\'\i}n-Fleitas}, J.~M. and {Marton}, G. and {Mary}, N. and {Masip}, A. and {Massari}, D. and {Mastrobuono-Battisti}, A. and {Mazeh}, T. and {McMillan}, P.~J. and {Messina}, S. and {Michalik}, D. and {Millar}, N.~R. and {Mints}, A. and {Molina}, D. and {Molinaro}, R. and {Moln{\'a}r}, L. and {Monari}, G. and {Mongui{\'o}}, M. and {Montegriffo}, P. and {Montero}, A. and {Mor}, R. and {Mora}, A. and {Morbidelli}, R. and {Morel}, T. and {Morris}, D. and {Muraveva}, T. and {Murphy}, C.~P. and {Musella}, I. and {Nagy}, Z. and {Noval}, L. and {Oca{\~n}a}, F. and {Ogden}, A. and {Ordenovic}, C. and {Osinde}, J.~O. and {Pagani}, C. and {Pagano}, I. and {Palaversa}, L. and {Palicio}, P.~A. and {Pallas-Quintela}, L. and {Panahi}, A. and {Payne-Wardenaar}, S. and {Pe{\~n}alosa Esteller}, X. and {Penttil{\"a}}, A. and {Pichon}, B. and {Piersimoni}, A.~M. and {Pineau}, F. -X. and {Plachy}, E. and {Plum}, G. and {Poggio}, E. and {Pr{\v{s}}a}, A. and {Pulone}, L. and {Racero}, E. and {Ragaini}, S. and {Rainer}, M. and {Raiteri}, C.~M. and {Rambaux}, N. and {Ramos}, P. and {Ramos-Lerate}, M. and {Re Fiorentin}, P. and {Regibo}, S. and {Richards}, P.~J. and {Rios Diaz}, C. and {Ripepi}, V. and {Riva}, A. and {Rix}, H. -W. and {Rixon}, G. and {Robichon}, N. and {Robin}, A.~C. and {Robin}, C. and {Roelens}, M. and {Rogues}, H.~R.~O. and {Rohrbasser}, L. and {Romero-G{\'o}mez}, M. and {Rowell}, N. and {Royer}, F. and {Ruz Mieres}, D. and {Rybicki}, K.~A. and {Sadowski}, G. and {S{\'a}ez N{\'u}{\~n}ez}, A. and {Sagrist{\`a} Sell{\'e}s}, A. and {Sahlmann}, J. and {Salguero}, E. and {Samaras}, N. and {Sanchez Gimenez}, V. and {Sanna}, N. and {Santove{\~n}a}, R. and {Sarasso}, M. and {Schultheis}, M. and {Sciacca}, E. and {Segol}, M. and {Segovia}, J.~C. and {S{\'e}gransan}, D. and {Semeux}, D. and {Shahaf}, S. and {Siddiqui}, H.~I. and {Siebert}, A. and {Siltala}, L. and {Silvelo}, A. and {Slezak}, E. and {Slezak}, I. and {Smart}, R.~L. and {Snaith}, O.~N. and {Solano}, E. and {Solitro}, F. and {Souami}, D. and {Souchay}, J. and {Spagna}, A. and {Spina}, L. and {Spoto}, F. and {Steele}, I.~A. and {Steidelm{\"u}ller}, H. and {Stephenson}, C.~A. and {S{\"u}veges}, M. and {Surdej}, J. and {Szabados}, L. and {Szegedi-Elek}, E. and {Taris}, F. and {Taylo}, M.~B. and {Teixeira}, R. and {Tolomei}, L. and {Tonello}, N. and {Torra}, F. and {Torra}, J. and {Torralba Elipe}, G. and {Trabucchi}, M. and {Tsounis}, A.~T. and {Turon}, C. and {Ulla}, A. and {Unger}, N. and {Vaillant}, M.~V. and {van Dillen}, E. and {van Reeven}, W. and {Vanel}, O. and {Vecchiato}, A. and {Viala}, Y. and {Vicente}, D. and {Voutsinas}, S. and {Weiler}, M. and {Wevers}, T. and {Wyrzykowski}, L. and {Yoldas}, A. and {Yvard}, P. and {Zhao}, H. and {Zorec}, J. and {Zucker}, S. and {Zwitter}, T.},
       title = "{Gaia Data Release 3. Summary of the content and survey properties}",
      journal = {\aap},
         year = 2023,
        month = jun,
       volume = {674},
          eid = {A1},
        pages = {A1},
          doi = {10.1051/0004-6361/202243940},
archivePrefix = {arXiv},
       eprint = {2012.01533},
 primaryClass = {astro-ph.GA},
       adsurl = {https://ui.adsabs.harvard.edu/abs/2021A&A...649A...1G}
}

@ARTICLE{Gebran2016,
       author = {{Gebran}, M. and {Farah}, W. and {Paletou}, F. and {Monier}, R. and {Watson}, V.},
        title = "{A new method for the inversion of atmospheric parameters of A/Am stars}",
      journal = {\aap},
         year = 2016,
        month = may,
       volume = {589},
          eid = {A83},
        pages = {A83},
          doi = {10.1051/0004-6361/201528052},
archivePrefix = {arXiv},
       eprint = {1603.01146},
 primaryClass = {astro-ph.IM},
       adsurl = {https://ui.adsabs.harvard.edu/abs/2016A&A...589A..83G}
}

@ARTICLE{Gordon2019,
       author = {{Gordon}, Kathryn D. and {Gies}, Douglas R. and {Schaefer}, Gail H. and {Huber}, Daniel and {Ireland}, Michael},
        title = "{Angular Sizes, Radii, and Effective Temperatures of B-type Stars from Optical Interferometry with the CHARA Array}",
      journal = {\apj},
         year = 2019,
        month = mar,
       volume = {873},
       number = {1},
          eid = {91},
        pages = {91},
          doi = {10.3847/1538-4357/ab04b2},
       adsurl = {https://ui.adsabs.harvard.edu/abs/2019ApJ...873...91G}
}

@ARTICLE{Griffin1973,
       author = {{Griffin}, R. and {Griffin}, R.},
        title = "{Accurate wavelengths of stellar and telluric absorption lines near lambda 7000 Angstroms}",
      journal = {\mnras},
         year = 1973,
        month = jan,
       volume = {162},
        pages = {255},
          doi = {10.1093/mnras/162.3.255},
       adsurl = {https://ui.adsabs.harvard.edu/abs/1973MNRAS.162..255G}
}

@ARTICLE{Griffin2000,
       author = {{Griffin}, R.~F.},
        title = "{The New `High Resolution Spectral Atlas of Telluric Lines'}",
      journal = {\apss},
         year = 2000,
        month = mar,
       volume = {271},
       number = {2},
        pages = {205-208},
          doi = {10.1023/A:1002095404664},
       adsurl = {https://ui.adsabs.harvard.edu/abs/2000Ap&SS.271..205G}
}

@ARTICLE{Hertzsprung1923,
       author = {{Hertzsprung}, E.},
        title = "{On the relation between mass and absolute brightness of components of double stars}",
      journal = {\bain},
         year = 1923,
        month = jul,
       volume = {2},
        pages = {15},
       adsurl = {https://ui.adsabs.harvard.edu/abs/1923BAN.....2...15H}
}

@ARTICLE{Hill1995,
       author = {{Hill}, G.~M.},
        title = "{Compositional differences among the A-type stars. II. Spectrum synthesis up to V sin i = 110 km/s.}",
      journal = {\aap},
         year = 1995,
        month = feb,
       volume = {294},
        pages = {536-546},
       adsurl = {https://ui.adsabs.harvard.edu/abs/1995A&A...294..536H}
}

@ARTICLE{Hillen2012,
       author = {{Hillen}, M. and {Verhoelst}, T. and {Degroote}, P. and {Acke}, B. and {van Winckel}, H.},
        title = "{The dynamic atmospheres of Mira stars: comparing the CODEX models to PTI time series of TU Andromedae}",
      journal = {\aap},
         year = 2012,
        month = feb,
       volume = {538},
          eid = {L6},
        pages = {L6},
          doi = {10.1051/0004-6361/201118653},
archivePrefix = {arXiv},
       eprint = {1201.5815},
 primaryClass = {astro-ph.SR},
       adsurl = {https://ui.adsabs.harvard.edu/abs/2012A&A...538L...6H}
}

@ARTICLE{Hourihane2023,
       author = {{Hourihane}, A. and {Fran{\c{c}}ois}, P. and {Worley}, C.~C. and {Magrini}, L. and {Gonneau}, A. and {Casey}, A.~R. and {Gilmore}, G. and {Randich}, S. and {Sacco}, G.~G. and {Recio-Blanco}, A. and {Korn}, A.~J. and {Allende Prieto}, C. and {Smiljanic}, R. and {Blomme}, R. and {Bragaglia}, A. and {Walton}, N.~A. and {Van Eck}, S. and {Bensby}, T. and {Lanzafame}, A. and {Frasca}, A. and {Franciosini}, E. and {Damiani}, F. and {Lind}, K. and {Bergemann}, M. and {Bonifacio}, P. and {Hill}, V. and {Lobel}, A. and {Montes}, D. and {Feuillet}, D.~K. and {Tautvai{\v{s}}ien{\.{e}}}, G. and {Guiglion}, G. and {Tabernero}, H.~M. and {Gonz{\'a}lez Hern{\'a}ndez}, J.~I. and {Gebran}, M. and {Van der Swaelmen}, M. and {Mikolaitis}, {\v{S}}. and {Daflon}, S. and {Merle}, T. and {Morel}, T. and {Lewis}, J.~R. and {Gonz{\'a}lez Solares}, E.~A. and {Murphy}, D.~N.~A. and {Jeffries}, R.~D. and {Jackson}, R.~J. and {Feltzing}, S. and {Prusti}, T. and {Carraro}, G. and {Biazzo}, K. and {Prisinzano}, L. and {Jofr{\'e}}, P. and {Zaggia}, S. and {Drazdauskas}, A. and {Stonkut{\'e}}, E. and {Marfil}, E. and {Jim{\'e}nez-Esteban}, F. and {Mahy}, L. and {Guti{\'e}rrez Albarr{\'a}n}, M.~L. and {Berlanas}, S.~R. and {Santos}, W. and {Morbidelli}, L. and {Spina}, L. and {Minkevi{\v{c}}i{\={u}}t{\.{e}}}, R.},
        title = "{The Gaia-ESO Survey: Homogenisation of stellar parameters and elemental abundances}",
      journal = {\aap},
         year = 2023,
        month = aug,
       volume = {676},
          eid = {A129},
        pages = {A129},
          doi = {10.1051/0004-6361/202345910},
archivePrefix = {arXiv},
       eprint = {2304.07720},
 primaryClass = {astro-ph.SR},
       adsurl = {https://ui.adsabs.harvard.edu/abs/2023A&A...676A.129H}
}

@ARTICLE{Holgado2022,
       author = {{Holgado}, G. and {Sim{\'o}n-D{\'\i}az}, S. and {Herrero}, A. and {Barb{\'a}}, R.~H.},
        title = "{The IACOB project. VII. The rotational properties of Galactic massive O-type stars revisited}",
      journal = {\aap},
         year = 2022,
        month = sep,
       volume = {665},
          eid = {A150},
        pages = {A150},
          doi = {10.1051/0004-6361/202243851},
archivePrefix = {arXiv},
       eprint = {2207.12776},
 primaryClass = {astro-ph.SR},
       adsurl = {https://ui.adsabs.harvard.edu/abs/2022A&A...665A.150H}
}

@ARTICLE{Huang2010,
       author = {{Huang}, Wenjin and {Gies}, D.~R. and {McSwain}, M.~V.},
        title = "{A Stellar Rotation Census of B Stars: From ZAMS to TAMS}",
      journal = {\apj},
         year = 2010,
        month = oct,
       volume = {722},
       number = {1},
        pages = {605-619},
          doi = {10.1088/0004-637X/722/1/605},
archivePrefix = {arXiv},
       eprint = {1008.1761},
 primaryClass = {astro-ph.SR},
       adsurl = {https://ui.adsabs.harvard.edu/abs/2010ApJ...722..605H}
}

@ARTICLE{Huang2024,
       author = {{Huang}, Bowen and {Yuan}, Haibo and {Xiang}, Maosheng and {Huang}, Yang and {Xiao}, Kai and {Xu}, Shuai and {Zhang}, Ruoyi and {Yang}, Lin and {Niu}, Zexi and {Gu}, Hongrui},
        title = "{A Comprehensive Correction of the Gaia DR3 XP Spectra}",
      journal = {\apjs},
         year = 2024,
        month = mar,
       volume = {271},
       number = {1},
          eid = {13},
        pages = {13},
          doi = {10.3847/1538-4365/ad18b1},
archivePrefix = {arXiv},
       eprint = {2401.12006},
 primaryClass = {astro-ph.IM},
       adsurl = {https://ui.adsabs.harvard.edu/abs/2024ApJS..271...13H}
}

@ARTICLE{Koleva2012,
       author = {{Koleva}, M. and {Vazdekis}, A.},
        title = "{Stellar population models in the UV. I. Characterisation of the New Generation Stellar Library}",
      journal = {\aap},
         year = 2012,
        month = feb,
       volume = {538},
          eid = {A143},
        pages = {A143},
          doi = {10.1051/0004-6361/201118065},
archivePrefix = {arXiv},
       eprint = {1111.5449},
 primaryClass = {astro-ph.CO},
       adsurl = {https://ui.adsabs.harvard.edu/abs/2012A&A...538A.143K}
}

@ARTICLE{Kuiper1938,
       author = {{Kuiper}, G.~P.},
        title = "{The Magnitude of the Sun, the Stellar Temperature Scale, and Bolometric Corrections.}",
      journal = {\apj},
         year = 1938,
        month = nov,
       volume = {88},
        pages = {429},
          doi = {10.1086/143998},
       adsurl = {https://ui.adsabs.harvard.edu/abs/1938ApJ....88..429K}
}

@ARTICLE{Landolt1992,
       author = {{Landolt}, Arlo U.},
        title = "{Broadband UBVRI Photometry of the Baldwin-Stone Southern Hemisphere Spectrophotometric Standards}",
      journal = {\aj},
         year = 1992,
        month = jul,
       volume = {104},
        pages = {372},
          doi = {10.1086/116243},
       adsurl = {https://ui.adsabs.harvard.edu/abs/1992AJ....104..372L}
}

@ARTICLE{Lefever2007,
       author = {{Lefever}, K. and {Puls}, J. and {Aerts}, C.},
        title = "{Statistical properties of a sample of periodically variable B-type supergiants. Evidence for opacity-driven gravity-mode oscillations}",
      journal = {\aap},
         year = 2007,
        month = mar,
       volume = {463},
       number = {3},
        pages = {1093-1109},
          doi = {10.1051/0004-6361:20066038},
archivePrefix = {arXiv},
       eprint = {astro-ph/0611484},
 primaryClass = {astro-ph},
       adsurl = {https://ui.adsabs.harvard.edu/abs/2007A&A...463.1093L}
}

@ARTICLE{Levenhagen2006,
       author = {{Levenhagen}, R.~S. and {Leister}, N.~V.},
        title = "{Spectroscopic analysis of southern B and Be stars}",
      journal = {\mnras},
         year = 2006,
        month = sep,
       volume = {371},
       number = {1},
        pages = {252-262},
          doi = {10.1111/j.1365-2966.2006.10655.x},
archivePrefix = {arXiv},
       eprint = {astro-ph/0606149},
 primaryClass = {astro-ph},
       adsurl = {https://ui.adsabs.harvard.edu/abs/2006MNRAS.371..252L}
}

@ARTICLE{Lindegren2021,
       author = {{Lindegren}, L. and {Klioner}, S.~A. and {Hern{\'a}ndez}, J. and {Bombrun}, A. and {Ramos-Lerate}, M. and {Steidelm{\"u}ller}, H. and {Bastian}, U. and {Biermann}, M. and {de Torres}, A. and {Gerlach}, E. and {Geyer}, R. and {Hilger}, T. and {Hobbs}, D. and {Lammers}, U. and {McMillan}, P.~J. and {Stephenson}, C.~A. and {Casta{\~n}eda}, J. and {Davidson}, M. and {Fabricius}, C. and {Gracia-Abril}, G. and {Portell}, J. and {Rowell}, N. and {Teyssier}, D. and {Torra}, F. and {Bartolom{\'e}}, S. and {Clotet}, M. and {Garralda}, N. and {Gonz{\'a}lez-Vidal}, J.~J. and {Torra}, J. and {Abbas}, U. and {Altmann}, M. and {Anglada Varela}, E. and {Balaguer-N{\'u}{\~n}ez}, L. and {Balog}, Z. and {Barache}, C. and {Becciani}, U. and {Bernet}, M. and {Bertone}, S. and {Bianchi}, L. and {Bouquillon}, S. and {Brown}, A.~G.~A. and {Bucciarelli}, B. and {Busonero}, D. and {Butkevich}, A.~G. and {Buzzi}, R. and {Cancelliere}, R. and {Carlucci}, T. and {Charlot}, P. and {Cioni}, M.-R.~L. and {Crosta}, M. and {Crowley}, C. and {del Peloso}, E.~F. and {del Pozo}, E. and {Drimmel}, R. and {Esquej}, P. and {Fienga}, A. and {Fraile}, E. and {Gai}, M. and {Garcia-Reinaldos}, M. and {Guerra}, R. and {Hambly}, N.~C. and {Hauser}, M. and {Jan{\ss}en}, K. and {Jordan}, S. and {Kostrzewa-Rutkowska}, Z. and {Lattanzi}, M.~G. and {Liao}, S. and {Licata}, E. and {Lister}, T.~A. and {L{\"o}ffler}, W. and {Marchant}, J.~M. and {Masip}, A. and {Mignard}, F. and {Mints}, A. and {Molina}, D. and {Mora}, A. and {Morbidelli}, R. and {Murphy}, C.~P. and {Pagani}, C. and {Panuzzo}, P. and {Pe{\~n}alosa Esteller}, X. and {Poggio}, E. and {Re Fiorentin}, P. and {Riva}, A. and {Sagrist{\`a} Sell{\'e}s}, A. and {Sanchez Gimenez}, V. and {Sarasso}, M. and {Sciacca}, E. and {Siddiqui}, H.~I. and {Smart}, R.~L. and {Souami}, D. and {Spagna}, A. and {Steele}, I.~A. and {Taris}, F. and {Utrilla}, E. and {van Reeven}, W. and {Vecchiato}, A.},
        title = "{Gaia Early Data Release 3. The astrometric solution}",
      journal = {\aap},
         year = 2021,
        month = may,
       volume = {649},
          eid = {A2},
        pages = {A2},
          doi = {10.1051/0004-6361/202039709},
archivePrefix = {arXiv},
       eprint = {2012.03380},
 primaryClass = {astro-ph.IM},
       adsurl = {https://ui.adsabs.harvard.edu/abs/2021A&A...649A...2L}
}

@ARTICLE{Mamajek2015,
       author = {{Mamajek}, E.~E. and {Torres}, G. and {Prsa}, A. and {Harmanec}, P. and {Asplund}, M. and {Bennett}, P.~D. and {Capitaine}, N. and {Christensen-Dalsgaard}, J. and {Depagne}, E. and {Folkner}, W.~M. and {Haberreiter}, M. and {Hekker}, S. and {Hilton}, J.~L. and {Kostov}, V. and {Kurtz}, D.~W. and {Laskar}, J. and {Mason}, B.~D. and {Milone}, E.~F. and {Montgomery}, M.~M. and {Richards}, M.~T. and {Schou}, J. and {Stewart}, S.~G.},
        title = "{IAU 2015 Resolution B2 on Recommended Zero Points for the Absolute and Apparent Bolometric Magnitude Scales}",
      journal = {arXiv e-prints},
         year = 2015,
        month = oct,
          eid = {arXiv:1510.06262},
        pages = {arXiv:1510.06262},
          doi = {10.48550/arXiv.1510.06262},
archivePrefix = {arXiv},
       eprint = {1510.06262},
 primaryClass = {astro-ph.SR},
       adsurl = {https://ui.adsabs.harvard.edu/abs/2015arXiv151006262M}
}

@ARTICLE{Martin2018,
       author = {{Martin}, A.~J. and {Neiner}, C. and {Oksala}, M.~E. and {Wade}, G.~A. and {Keszthelyi}, Z. and {Fossati}, L. and {Marcolino}, W. and {Mathis}, S. and {Georgy}, C.},
        title = "{First results from the LIFE project: discovery of two magnetic hot evolved stars}",
      journal = {\mnras},
         year = 2018,
        month = apr,
       volume = {475},
       number = {2},
        pages = {1521-1536},
          doi = {10.1093/mnras/stx3264},
archivePrefix = {arXiv},
       eprint = {1712.07403},
 primaryClass = {astro-ph.SR},
       adsurl = {https://ui.adsabs.harvard.edu/abs/2018MNRAS.475.1521M}
}

@ARTICLE{Melendez2012,
       author = {{Mel{\'e}ndez}, J. and {Bergemann}, M. and {Cohen}, J.~G. and {Endl}, M. and {Karakas}, A.~I. and {Ram{\'\i}rez}, I. and {Cochran}, W.~D. and {Yong}, D. and {MacQueen}, P.~J. and {Kobayashi}, C. and {Asplund}, M.},
        title = "{The remarkable solar twin HIP 56948: a prime target in the quest for other Earths}",
      journal = {\aap},
         year = 2012,
        month = jul,
       volume = {543},
          eid = {A29},
        pages = {A29},
          doi = {10.1051/0004-6361/201117222},
archivePrefix = {arXiv},
       eprint = {1204.2766},
 primaryClass = {astro-ph.SR},
       adsurl = {https://ui.adsabs.harvard.edu/abs/2012A&A...543A..29M}
}

@ARTICLE{Monier2023,
       author = {{Monier}, Richard and {Niemczura}, E. and {Kurtz}, D.~W. and {Rappaport}, S. and {Bowman}, D.~M. and {Murphy}, Simon J. and {Lebreton}, Yveline and {Stuik}, Remko and {Deal}, Morgan and {Merle}, Thibault and {K{\i}l{\i}{\c{c}}o{\u{g}}lu}, T. and {Gebran}, M. and {Le Ster}, Ewen},
        title = "{The Surface Composition of Six Newly Discovered Chemically Peculiar Stars. Comparison to the HgMn Stars mu Lep and beta Scl and the Superficially Normal B Star nu Cap}",
      journal = {\aj},
         year = 2023,
        month = aug,
       volume = {166},
       number = {2},
          eid = {54},
        pages = {54},
          doi = {10.3847/1538-3881/acdb50},
archivePrefix = {arXiv},
       eprint = {2306.01601},
 primaryClass = {astro-ph.SR},
       adsurl = {https://ui.adsabs.harvard.edu/abs/2023AJ....166...54M}
}

@ARTICLE{Montegriffo2023,
       author = {{Montegriffo}, P. and {De Angeli}, F. and {Andrae}, R. and {Riello}, M. and {Pancino}, E. and {Sanna}, N. and {Bellazzini}, M. and {Evans}, D.~W. and {Carrasco}, J.~M. and {Sordo}, R. and {Busso}, G. and {Cacciari}, C. and {Jordi}, C. and {van Leeuwen}, F. and {Vallenari}, A. and {Altavilla}, G. and {Barstow}, M.~A. and {Brown}, A.~G.~A. and {Burgess}, P.~W. and {Castellani}, M. and {Cowell}, S. and {Davidson}, M. and {De Luise}, F. and {Delchambre}, L. and {Diener}, C. and {Fabricius}, C. and {Fr{\'e}mat}, Y. and {Fouesneau}, M. and {Gilmore}, G. and {Giuffrida}, G. and {Hambly}, N.~C. and {Harrison}, D.~L. and {Hidalgo}, S. and {Hodgkin}, S.~T. and {Holland}, G. and {Marinoni}, S. and {Osborne}, P.~J. and {Pagani}, C. and {Palaversa}, L. and {Piersimoni}, A.~M. and {Pulone}, L. and {Ragaini}, S. and {Rainer}, M. and {Richards}, P.~J. and {Rowell}, N. and {Ruz-Mieres}, D. and {Sarro}, L.~M. and {Walton}, N.~A. and {Yoldas}, A.},
        title = "{Gaia Data Release 3. External calibration of BP/RP low-resolution spectroscopic data}",
      journal = {\aap},
         year = 2023,
        month = jun,
       volume = {674},
          eid = {A3},
        pages = {A3},
          doi = {10.1051/0004-6361/202243880},
archivePrefix = {arXiv},
       eprint = {2206.06205},
 primaryClass = {astro-ph.IM},
       adsurl = {https://ui.adsabs.harvard.edu/abs/2023A&A...674A...3M}
}

@ARTICLE{Mugnes2015,
       author = {{Mugnes}, J. -M. and {Robert}, C.},
        title = "{Bayesian statistics as a new tool for spectral analysis - I. Application for the determination of basic parameters of massive stars}",
      journal = {\mnras},
         year = 2015,
        month = nov,
       volume = {454},
       number = {1},
        pages = {28-52},
          doi = {10.1093/mnras/stv1889},
archivePrefix = {arXiv},
       eprint = {1507.02928},
 primaryClass = {astro-ph.IM},
       adsurl = {https://ui.adsabs.harvard.edu/abs/2015MNRAS.454...28M}
}

@ARTICLE{Niemczura2022,
       author = {{Niemczura}, E. and {Walczak}, P. and {Miko{\l}ajczyk}, P. and {Sch{\"o}ller}, M. and {Hummel}, C.~A. and {Hubrig}, S. and {R{\'o}{\.z}a{\'n}ski}, T.},
        title = "{Seismic modelling of the pulsating mercury-manganese star HD 29589}",
      journal = {\mnras},
         year = 2022,
        month = aug,
       volume = {514},
       number = {4},
        pages = {5640-5658},
          doi = {10.1093/mnras/stac1632},
       adsurl = {https://ui.adsabs.harvard.edu/abs/2022MNRAS.514.5640N}
}

@ARTICLE{Nieva2011,
       author = {{Nieva}, M. -F. and {Sim{\'o}n-D{\'\i}az}, S.},
        title = "{The chemical composition of the Orion star forming region. III. C, N, Ne, Mg, and Fe abundances in B-type stars revisited}",
      journal = {\aap},
         year = 2011,
        month = aug,
       volume = {532},
          eid = {A2},
        pages = {A2},
          doi = {10.1051/0004-6361/201116478},
archivePrefix = {arXiv},
       eprint = {1104.3154},
 primaryClass = {astro-ph.SR},
       adsurl = {https://ui.adsabs.harvard.edu/abs/2011A&A...532A...2N}
}

@ARTICLE{Okuyan2026,
       author = {{Okuyan}, O{\u{g}}uzhan and {Bak{\i}{\textcommabelow s}}, Volkan and {Eker}, Zeki},
        title = "{Determining accurate radii and luminosities using multiband BC for LAMOST stars within 500 pc}",
      journal = {Advances in Space Research},
         year = 2026,
        month = may,
       volume = {77},
       number = {9},
        pages = {9868-9882},
          doi = {10.1016/j.asr.2026.03.029},
       adsurl = {https://ui.adsabs.harvard.edu/abs/2026AdSpR..77.9868O}
}

@ARTICLE{Paunzen2005,
       author = {{Paunzen}, E. and {Schnell}, A. and {Maitzen}, H.~M.},
        title = "{An empirical temperature calibration for the {\ensuremath{\Delta}} a photometric system . I. The B-type stars}",
      journal = {\aap},
         year = 2005,
        month = dec,
       volume = {444},
       number = {3},
        pages = {941-946},
          doi = {10.1051/0004-6361:20053546},
archivePrefix = {arXiv},
       eprint = {astro-ph/0509049},
 primaryClass = {astro-ph},
       adsurl = {https://ui.adsabs.harvard.edu/abs/2005A&A...444..941P}
}

@ARTICLE{Paunzen2006,
       author = {{Paunzen}, E. and {Schnell}, A. and {Maitzen}, H.~M.},
        title = "{An empirical temperature calibration for the {\ensuremath{\Delta}} a photometric system. II. The A-type and mid F-type stars}",
      journal = {\aap},
         year = 2006,
        month = oct,
       volume = {458},
       number = {1},
        pages = {293-296},
          doi = {10.1051/0004-6361:20064889},
archivePrefix = {arXiv},
       eprint = {astro-ph/0607567},
 primaryClass = {astro-ph},
       adsurl = {https://ui.adsabs.harvard.edu/abs/2006A&A...458..293P}
}

@ARTICLE{Prugniel2011,
       author = {{Prugniel}, Ph. and {Vauglin}, I. and {Koleva}, M.},
        title = "{The atmospheric parameters and spectral interpolator for the MILES stars}",
      journal = {\aap},
         year = 2011,
        month = jul,
       volume = {531},
          eid = {A165},
        pages = {A165},
          doi = {10.1051/0004-6361/201116769},
archivePrefix = {arXiv},
       eprint = {1104.4952},
 primaryClass = {astro-ph.CO},
       adsurl = {https://ui.adsabs.harvard.edu/abs/2011A&A...531A.165P}
}

@ARTICLE{Riello2021,
       author = {{Riello}, M. and {De Angeli}, F. and {Evans}, D.~W. and {Montegriffo}, P. and {Carrasco}, J.~M. and {Busso}, G. and {Palaversa}, L. and {Burgess}, P.~W. and {Diener}, C. and {Davidson}, M. and {Rowell}, N. and {Fabricius}, C. and {Jordi}, C. and {Bellazzini}, M. and {Pancino}, E. and {Harrison}, D.~L. and {Cacciari}, C. and {van Leeuwen}, F. and {Hambly}, N.~C. and {Hodgkin}, S.~T. and {Osborne}, P.~J. and {Altavilla}, G. and {Barstow}, M.~A. and {Brown}, A.~G.~A. and {Castellani}, M. and {Cowell}, S. and {De Luise}, F. and {Gilmore}, G. and {Giuffrida}, G. and {Hidalgo}, S. and {Holland}, G. and {Marinoni}, S. and {Pagani}, C. and {Piersimoni}, A.~M. and {Pulone}, L. and {Ragaini}, S. and {Rainer}, M. and {Richards}, P.~J. and {Sanna}, N. and {Walton}, N.~A. and {Weiler}, M. and {Yoldas}, A.},
        title = "{Gaia Early Data Release 3. Photometric content and validation}",
      journal = {\aap},
         year = 2021,
        month = may,
       volume = {649},
          eid = {A3},
        pages = {A3},
          doi = {10.1051/0004-6361/202039587},
archivePrefix = {arXiv},
       eprint = {2012.01916},
 primaryClass = {astro-ph.IM},
       adsurl = {https://ui.adsabs.harvard.edu/abs/2021A&A...649A...3R}
}

@ARTICLE{Royer2014,
       author = {{Royer}, F. and {Gebran}, M. and {Monier}, R. and {Adelman}, S. and {Smalley}, B. and {Pintado}, O. and {Reiners}, A. and {Hill}, G. and {Gulliver}, A.},
        title = "{Normal A0-A1 stars with low rotational velocities. I. Abundance determination and classification}",
      journal = {\aap},
         year = 2014,
        month = feb,
       volume = {562},
          eid = {A84},
        pages = {A84},
          doi = {10.1051/0004-6361/201322762},
archivePrefix = {arXiv},
       eprint = {1401.2372},
 primaryClass = {astro-ph.SR},
       adsurl = {https://ui.adsabs.harvard.edu/abs/2014A&A...562A..84R}
}

@ARTICLE{Russell1923,
       author = {{Russell}, H.~N. and {Adams}, W.~S. and {Joy}, A.~H.},
        title = "{A Comparison of Spectroscopic and Dynamical Parallaxes}",
      journal = {\pasp},
         year = 1923,
        month = aug,
       volume = {35},
       number = {206},
        pages = {189},
          doi = {10.1086/123303},
       adsurl = {https://ui.adsabs.harvard.edu/abs/1923PASP...35..189R}
}

@ARTICLE{scipy,
  author  = {Virtanen, Pauli and Gommers, Ralf and Oliphant, Travis E. and
            Haberland, Matt and Reddy, Tyler and Cournapeau, David and
            Burovski, Evgeni and Peterson, Pearu and Weckesser, Warren and
            Bright, Jonathan and {van der Walt}, St{\'e}fan J. and
            Brett, Matthew and Wilson, Joshua and Millman, K. Jarrod and
            Mayorov, Nikolay and Nelson, Andrew R. J. and Jones, Eric and
            Kern, Robert and Larson, Eric and Carey, C J and
            Polat, {\.I}lhan and Feng, Yu and Moore, Eric W. and
            {VanderPlas}, Jake and Laxalde, Denis and Perktold, Josef and
            Cimrman, Robert and Henriksen, Ian and Quintero, E. A. and
            Harris, Charles R. and Archibald, Anne M. and
            Ribeiro, Ant{\^o}nio H. and Pedregosa, Fabian and
            {van Mulbregt}, Paul and {SciPy 1.0 Contributors}},
  title   = {{{SciPy} 1.0: Fundamental Algorithms for Scientific
            Computing in Python}},
  journal = {Nature Methods},
  year    = {2020},
  volume  = {17},
  pages   = {261--272},
  adsurl  = {https://rdcu.be/b08Wh},
  doi     = {10.1038/s41592-019-0686-2},
}

@ARTICLE{SimonDiaz2017,
       author = {{Sim{\'o}n-D{\'\i}az}, S. and {Godart}, M. and {Castro}, N. and {Herrero}, A. and {Aerts}, C. and {Puls}, J. and {Telting}, J. and {Grassitelli}, L.},
        title = "{The IACOB project . III. New observational clues to understand macroturbulent broadening in massive O- and B-type stars}",
      journal = {\aap},
         year = 2017,
        month = jan,
       volume = {597},
          eid = {A22},
        pages = {A22},
          doi = {10.1051/0004-6361/201628541},
archivePrefix = {arXiv},
       eprint = {1608.05508},
 primaryClass = {astro-ph.SR},
       adsurl = {https://ui.adsabs.harvard.edu/abs/2017A&A...597A..22S}
}

@ARTICLE{Soubiran2022,
       author = {{Soubiran}, C. and {Brouillet}, N. and {Casamiquela}, L.},
        title = "{Assessment of [Fe/H] determinations for FGK stars in spectroscopic surveys}",
      journal = {\aap},
         year = 2022,
        month = jul,
       volume = {663},
          eid = {A4},
        pages = {A4},
          doi = {10.1051/0004-6361/202142409},
archivePrefix = {arXiv},
       eprint = {2112.07545},
 primaryClass = {astro-ph.SR},
       adsurl = {https://ui.adsabs.harvard.edu/abs/2022A&A...663A...4S}
}

@ARTICLE{Soubiran2024,
       author = {{Soubiran}, C. and {Creevey}, O.~L. and {Lagarde}, N. and {Brouillet}, N. and {Jofr{\'e}}, P. and {Casamiquela}, L. and {Heiter}, U. and {Aguilera-G{\'o}mez}, C. and {Vitali}, S. and {Worley}, C. and {de Brito Silva}, D.},
        title = "{Gaia FGK benchmark stars: Fundamental T$_{eff}$ and log g of the third version}",
      journal = {\aap},
         year = 2024,
        month = feb,
       volume = {682},
          eid = {A145},
        pages = {A145},
          doi = {10.1051/0004-6361/202347136},
archivePrefix = {arXiv},
       eprint = {2310.11302},
 primaryClass = {astro-ph.SR},
       adsurl = {https://ui.adsabs.harvard.edu/abs/2024A&A...682A.145S}
}

@ARTICLE{Tabernero2022,
       author = {{Tabernero}, H.~M. and {Marfil}, E. and {Montes}, D. and {Gonz{\'a}lez Hern{\'a}ndez}, J.~I.},
        title = "{STEPARSYN: A Bayesian code to infer stellar atmospheric parameters using spectral synthesis}",
      journal = {\aap},
         year = 2022,
        month = jan,
       volume = {657},
          eid = {A66},
        pages = {A66},
          doi = {10.1051/0004-6361/202141763},
archivePrefix = {arXiv},
       eprint = {2110.00444},
 primaryClass = {astro-ph.SR},
       adsurl = {https://ui.adsabs.harvard.edu/abs/2022A&A...657A..66T}
}

@ARTICLE{Takeda2010,
       author = {{Takeda}, Yoichi and {Kambe}, Eiji and {Sadakane}, Kozo and {Masada}, Seiji},
        title = "{Oxygen and Neon Abundances of B-Type Stars in Comparison with the Sun}",
      journal = {\pasj},
         year = 2010,
        month = oct,
       volume = {62},
        pages = {1239-1248},
          doi = {10.1093/pasj/62.5.1239},
archivePrefix = {arXiv},
       eprint = {1008.1220},
 primaryClass = {astro-ph.SR},
       adsurl = {https://ui.adsabs.harvard.edu/abs/2010PASJ...62.1239T}
}

@ARTICLE{Torres2010,
       author = {{Torres}, Guillermo},
        title = "{On the Use of Empirical Bolometric Corrections for Stars}",
      journal = {\aj},
         year = 2010,
        month = nov,
       volume = {140},
       number = {5},
        pages = {1158-1162},
          doi = {10.1088/0004-6256/140/5/1158},
archivePrefix = {arXiv},
       eprint = {1008.3913},
 primaryClass = {astro-ph.SR},
       adsurl = {https://ui.adsabs.harvard.edu/abs/2010AJ....140.1158T}
}

@ARTICLE{Wu2011,
       author = {{Wu}, Yue and {Singh}, H.~P. and {Prugniel}, P. and {Gupta}, R. and {Koleva}, M.},
        title = "{Coud{\'e}-feed stellar spectral library - atmospheric parameters}",
      journal = {\aap},
         year = 2011,
        month = jan,
       volume = {525},
          eid = {A71},
        pages = {A71},
          doi = {10.1051/0004-6361/201015014},
archivePrefix = {arXiv},
       eprint = {1009.1491},
 primaryClass = {astro-ph.SR},
       adsurl = {https://ui.adsabs.harvard.edu/abs/2011A&A...525A..71W}
}

@ARTICLE{Yucel2026,
       author = {{Y{\"u}cel}, G{\"o}khan and {Bilir}, Sel{\c{c}}uk and {Bak{\i}{\textcommabelow s}}, Volkan and {Eker}, Zeki},
        title = "{Empirical Bolometric Correction and Zero-point Constants of Visual Magnitudes from High-resolution Spectra}",
      journal = {\aj},
         year = 2026,
        month = jan,
       volume = {171},
       number = {1},
          eid = {49},
        pages = {49},
          doi = {10.3847/1538-3881/ae1f8c},
archivePrefix = {arXiv},
       eprint = {2511.07522},
 primaryClass = {astro-ph.SR},
       adsurl = {https://ui.adsabs.harvard.edu/abs/2026AJ....171...49Y}
}

@ARTICLE{Zorec2009,
       author = {{Zorec}, J. and {Cidale}, L. and {Arias}, M.~L. and {Fr{\'e}mat}, Y. and {Muratore}, M.~F. and {Torres}, A.~F. and {Martayan}, C.},
        title = "{Fundamental parameters of B supergiants from the BCD system. I. Calibration of the ({\ensuremath{\lambda}}\_1, D) parameters into T$_{eff}$}",
      journal = {\aap},
         year = 2009,
        month = jul,
       volume = {501},
       number = {1},
        pages = {297-320},
          doi = {10.1051/0004-6361/200811147},
archivePrefix = {arXiv},
       eprint = {0903.5134},
 primaryClass = {astro-ph.SR},
       adsurl = {https://ui.adsabs.harvard.edu/abs/2009A&A...501..297Z}
}

@ARTICLE{Zorec2012,
       author = {{Zorec}, J. and {Royer}, F.},
        title = "{Rotational velocities of A-type stars. IV. Evolution of rotational velocities}",
      journal = {\aap},
         year = 2012,
        month = jan,
       volume = {537},
          eid = {A120},
        pages = {A120},
          doi = {10.1051/0004-6361/201117691},
archivePrefix = {arXiv},
       eprint = {1201.2052},
 primaryClass = {astro-ph.SR},
       adsurl = {https://ui.adsabs.harvard.edu/abs/2012A&A...537A.120Z}
}


\bsp	
\label{lastpage}
\end{document}